\documentclass[twocolumn]{aastex63}

\usepackage{amsmath}
\usepackage{amssymb}
\usepackage[T1]{fontenc}
\usepackage{gensymb}
\usepackage{enumitem}
\usepackage{url}

\newcommand{\M}{\, M}

\newenvironment{myfont}{\fontfamily{lmtt}\selectfont}{\par}
\DeclareTextFontCommand{\textmyfont}{\myfont}

\received{February 15, 2021}
\revised{\today}
\accepted{..... .., 2021}
\submitjournal{ApJ}

\shorttitle{Stellar transits through accretion flow}
\shortauthors{P. Sukov\'a et al.}

\begin{document}

\title{Stellar transits across a magnetized accretion torus as a mechanism for plasmoid ejection}


\correspondingauthor{Petra Sukov\'a}
\email{petra.sukova@asu.cas.cz}

\author[0000-0002-4779-5635]{Petra Sukov\'a}
\affiliation{Astronomical Institute of the Czech Academy of Sciences, Bo\v{c}n\'{\i} II 1401, 141 00 Prague, Czech Republic}

\author[0000-0001-6450-1187]{Michal Zaja\v{c}ek}
\affiliation{Center for Theoretical Physics, Polish Academy of Sciences, Al. Lotnikow 32/46, 02-668 Warsaw, Poland} 
\affil{Department of Theoretical Physics and Astrophysics, Faculty of Science, Masaryk University, Kotl\'a\v{r}sk\'a 2, Brno, 611 37, Czech Republic}

\author[0000-0002-9209-5355]{Vojt\v{e}ch Witzany}
\affiliation{School of Mathematics and Statistics, University College Dublin, Belfield, Dublin 4, D04 V1W8, Ireland}

\author[0000-0002-5760-0459]{Vladim\'{\i}r Karas}
\affiliation{Astronomical Institute of the Czech Academy of Sciences, Bo\v{c}n\'{\i} II 1401, 141 00 Prague, Czech Republic}


\begin{abstract}
\noindent The close neighbourhood of a supermassive black hole contains not only the accreting gas and dust but also stellar-sized objects, such as late-type and early-type stars and compact remnants that belong to the nuclear star cluster. When passing through the accretion flow, these objects perturb it by the direct action of stellar winds, as well as their magnetic and gravitational effects. By performing General-Relativistic Magnetohydrodynamic (GRMHD) simulations, we investigate how the passages of a star can influence the supermassive black hole gaseous environment. We focus on the changes in the accretion rate and the emergence of blobs of plasma in the funnel of an accretion torus. We compare results from 2D and 3D numerical computations that have been started with comparable initial conditions. We find that a quasi-stationary inflow can be temporarily inhibited by a transiting star, and the plasmoids can be ejected along the magnetic field lines near the rotation axis. We observe the characteristic signatures of the perturbing motion in the power spectrum of the accretion variability, which provides an avenue for a multi-messenger detection of these transient events. Finally, we discuss the connection of our results to multi-wavelength observations of galactic nuclei, with the emphasis on ten promising sources (Sgr~A*, OJ 287, J0849$+$5108, RE J1034$+$396, 1ES 1927+65, ESO 253--G003, GSN 069, RX J1301.9$+$2747, eRO-QPE1, eRO-QPE2).
\end{abstract}


\keywords{accretion, accretion disks --- 
black hole physics --- magnetohydrodynamics --- methods: numerical -- gravitational waves }


\section{Introduction} \label{sec:intro}


Active galactic nuclei (hereafter AGN) harbour supermassive black holes (hereafter SMBH) in their cores. The SMBHs are surrounded by dilute plasma in rotating, toroidal configurations known as accretion disks. The process of gradual accretion of the plasma onto SMBH and the corresponding conversion of the gravitational binding energy of the plasma into radiation is the main source of the intense, non-thermal signal that defines AGN \citep{1997iagn.book.....P,1999agnc.book.....K,2006eac..book.....S}. Stars of various types are also present around the nucleus where they often form a dense Nuclear Star Cluster (NSC) and contribute to the observed spectrum. Furthermore, magnetic fields are generated by electric currents flowing in the plasma; they exhibit a component ordered on large scales, as well as loops entangled on small scales with respect to the size of the SMBH horizon \citep[see, e.g.,][]{1990agn..conf.....B,2019ARA&A..57..467B,2012rjag.book.....B}. 

The material of the torus gradually sinks to the centre and a fraction of it eventually accretes onto the black hole. In the inner region near the event horizon, plasma overflows across the relativistic cusp into the plunging region and then it falls rapidly towards the event horizon. However, some material is ejected in a form of a highly collimated jet or a less focused outflow. This is because there are electromagnetic forces that accelerate some of the plasma against gravity, and the plasma coupled with strong electromagnetic fields can even use the SMBH rotation as an energy source for this acceleration \citep{1977MNRAS.179..433B,2001bhgh.book.....P}. As a result, a part of the infalling plasma becomes accelerated outwards and escapes the attraction of the black hole.  

Although stars and NSCs are typically not considered in the standard unified model of AGN \citep{1993ARA&A..31..473A,1995PASP..107..803U}, they are certainly present in most galactic nuclei \citep{2014CQGra..31x4007S}. We can directly detect them in the closest galactic nucleus -- the Galactic center \citep[see][for reviews]{2010RvMP...82.3121G,2012RAA....12..995M,2017FoPh...47..553E,2019JPhCS1258a2019E}. The bright star S2 (or S0-2) orbits the compact and variable radio source Sgr~A* every $\sim 16$ years \citep{2002MNRAS.331..917E,2002Natur.419..694S,2003ApJ...586L.127G,2017ApJ...845...22P,2018A&A...615L..15G, 2019Sci...365..664D}, with the closest approach of $\sim 2944$ gravitational radii, at which the pericentre velocity is $\sim 2.5\%$ of the light speed. The continuum emission of S2 does not vary significantly and the type of environment, through which it passes along its orbit, is consistent with the hot and diluted accretion flow \citep{2020A&A...644A.105H}, typical of low luminous galactic nuclei \citep{2014ARA&A..52..529Y}. In addition, a group of fainter stars has been detected, some of which can approach Sgr~A* at distances one order of magnitude closer than S2 \citep{2020ApJ...889...61P,2020ApJ...899...50P}. Recently, the first interaction between the S star and the black hole environment, which involves the comet-shaped source X7 with the host star S50, was resolved thanks to the collection of 20 years of near-infrared data \citep{2021arXiv210102077P}. Therefore, it is natural to ask if and on what timescales the interaction of orbiting stars with the hot accretion flow can affect the accretion state of the central SMBH.

In a more general context, variable accretion onto SMBHs and associated outflows are considered to be the main drivers for the observed nuclear variability. For AGN, the brightness variability appears to be largely stochastic and can be interpreted and modelled as a red-noise and a damped random-walk process \citep{1995A&A...300..707T,2009ApJ...698..895K,2010ApJ...708..927K,2010ApJ...721.1014M,2013ApJ...765..106Z}. Deviations from a simple and a broken power-law profile of the power spectral densities of AGN light curves, in particular in the form of narrow periodicity and quasiperiodicity peaks, are generally associated with the presence of bound perturbers \citep[see e.g.][]{2006MmSAI..77..733K}, such as secondary supermassive and intermediate-mass black holes, stars and stellar compact remnants, or partial tidal disruption events. 

Specifically, the repetitive perturbing events occur on the time-scale of the orbital period \citep[see][and further references cited therein]{1991MNRAS.250..505S,2001A&A...376..686K,2020ApJ...889...94M}. The outcome of the interaction between the perturber is then sensitive both to the nature of the perturber but also to the orbital parameters. This can generate quasi-periodicity in observable properties of the source \citep{1999PASJ...51..571S,2010MNRAS.402.1614D} and it may cause the perturber to inspiral into the SMBH due to hydrodynamical drag \citep{2000ApJ...536..663N,2016MNRAS.460..240K}. Depending on the position of the critical radius for tidal disruption, the star can eventually get partially or completely damaged by tidal force, thus contributing in an intermittent manner to the gas environment \citep{2019GReGr..51...30S,2020ApJ...899...36M}. The remnant material from the disruption event gradually disperses and enriches the torus and the outflow. On the other hand, a sufficiently compact object such as a white dwarf, a neutron star, or a stellar-mass black hole, will never get tidally disrupted by a SMBH above its horizon. The compact object can then approach sufficiently close to significantly emit gravitational waves and thus embark on an extreme mass ratio inspiral into the SMBH \citep{2017PhRvD..95j3012B}. Identifying the host galaxy of such a gravitational-wave event through matching frequencies in the AGN variability would make it a multi-messenger standard siren for cosmology \citep{holz2005using}.  

Hence, from a number of perspectives, it is timely and relevant to study the effect a stellar-type perturber can have on an accretion disk near a SMBH, which is the subject of the current paper. In Section \ref{sec:Theory}, we consider various scenarios for the perturber, in particular a star with an outflow characterized by a terminal velocity and a mass-loss rate, a magnetized and a rotating pulsar characterized by a spin-down energy, and a heavy, purely gravitationally interacting compact object. In all three scenarios, the effect of the perturber can be captured by a certain cross-sectional radius wherein the motion of the accretion-disk gas is synchronized with the perturber motion. 

We then proceed to describe our general-relativistic magneto-hydrodynamics (GRMHD) simulations of the perturbed accretion disk in Section~\ref{sec:num}. In the simulations the perturbers with various choices of cross-sectional radii orbit the SMBH at various distances, inclinations and eccentricities. Since we are interested only in the dynamics of the accretion disk, the star is assumed to orbit the SMBH eternally with no back-reaction on its orbit or cross-sectional radius. One of the important technical points is that most of our simulations are done in a 2D idealization and two were then carried out on a full 3D grid for reference.  

The results of the simulations are discussed in Sections~\ref{sec:2D} and~\ref{sec:3D}. We observe that stars inclined with respect to the disk often kick out matter into an outflow along the rotational axis of the SMBH. In other cases, we observe that the stars can quench or regulate the accretion rate onto the SMBH. We then study the induced periodicities emerging in the flow due to the motion of the perturber in Section~\ref{subsec_periodicity_psd}. We find that the induced periodicity strongly depends both on the orbital parameters but also on the 2D/3D setup.    

We deal with the possible drawbacks of our GRMHD approach in Sections~\ref{subsec:MADdiscuss} and \ref{subsec:radiative-cooling}, while the comparison of the results with the observational data is given in Sections \ref{subsec:discussion-observation-I} (Sgr~A*), \ref{subsec:discussion-observation-II} (OJ287) and \ref{subsec:discussion-TDE} (RE J1034+396, 1ES 1927+65, ESO 253-G003). The results are summarised in Section~\ref{sec:conclusions}.


\section{Star--flow interaction} \label{sec:Theory}


Since NSCs are found at the centers of $\sim 75\%$ of all the galaxies in the local Universe \citep{2002AJ....123.1389B,1998AJ....116...68C,2006ApJS..165...57C,2011MNRAS.413.1875N}, interactions of stars with the accretion flow are expected to be common phenomena during the galaxy evolution. The only case where this can directly be detected and resolved is the Milky Way NSC, where comet-shaped sources X3 and X7 \citep{2010A&A...521A..13M,2021arXiv210102077P} as well as the bow-shock source X8 \citep{2019A&A...624A..97P} appear to interact with the nuclear wind or a low surface-brightness jet \citep{2020MNRAS.499.3909Y}. For the comet-shaped source X7, \citet{2021arXiv210102077P}\footnote{See also the ATEL announcement: \citet{2021ATel14306....1P}.} revealed the significant structural change between the years 1999 and 2018 from a bow-shock-like source to a source with a potential detachment of the envelope from the host star associated with S50. This transition in the source morphology could be induced by a nuclear outflow or a weak jet from the direction of Sgr~A*.

In this section, we discuss three astrophysical scenarios of the interaction of a stellar-type perturber with the accretion flow near the SMBH with the hope of finding a cross-sectional radius of interaction that can be used in the simulations.

In Subsection~\ref{sec:bowshock}, we derive the effective cross-section of wind-blowing stars, which is motivated by the set-up observed in the Galactic center. In Subsection~\ref{sec:pulsar}, we treat another possibility of forming shocks and cavities in the accretion flow, which involves a young magnetized neutron stars -- pulsar with a strong electromagnetic outflow. This set-up is less likely since neutron stars are remnants of the most massive stars \citep{2017CoSka..47..124K}, but occasionally they can plunge towards the SMBH via supernova kicks \citep{2017MNRAS.469.1510B} and dynamical scattering \citep{2007ApJ...656..709P}. Currently, only one magnetar is detected in the Galactic center orbiting the SMBH at the larger distance comparable to the Bondi radius \citep[at $3"\sim 0.12\,{\rm pc}$, projected; ][]{2013Natur.501..391E}. Last but not least, in Subsection \ref{sec:CO} we describe the effective cross-section of the purely gravitational interaction between the perturbing object and the disk, which applies in the case of dark compact objects.


\subsection{Case of a massive star with strong wind}\label{sec:bowshock}

We consider a wind-blowing star characterized by the mass-loss rate $\dot{m}_{\rm w}$ and the terminal wind velocity $v_{\rm w}$, moving with a Keplerian velocity around the SMBH, $v_{\star}\sim \sqrt{GM/r}$, at the distance $r$ expressed in gravitational radii, $r_{\rm g}=GM/c^2$. The hot ADAF-like flow is also assumed to have a $\phi$-component of the velocity field close to Keplerian, $v_{\rm f}\sim \sqrt{GM/r}$. The relative velocity between the star and the flow can be expressed as ${\bf v_{\rm rel}}={\bf v_{\star}}-{\bf v_{\rm f}}$. The magnitude of the relative velocity is 

\begin{align}
    v_{\rm rel} \sim \sqrt{\frac{2GM}{r} (1 - \cos i)} \,,
\end{align}
where $i$ is the inclination angle of the orbit with respect to the disk. Inclination $i=0$ corresponds to a star embedded in the disk and $v_{\rm rel} \sim 0$; $i=90^\circ$ to perpendicular crossings, $v_{\rm rel} \sim \sqrt{2GM/r}$; and $i = 180^\circ$ to counter-rotation such that $v_{\rm rel} = 2 \sqrt{GM/r}$.

\begin{figure*}
    \centering
    \includegraphics[width=\columnwidth]{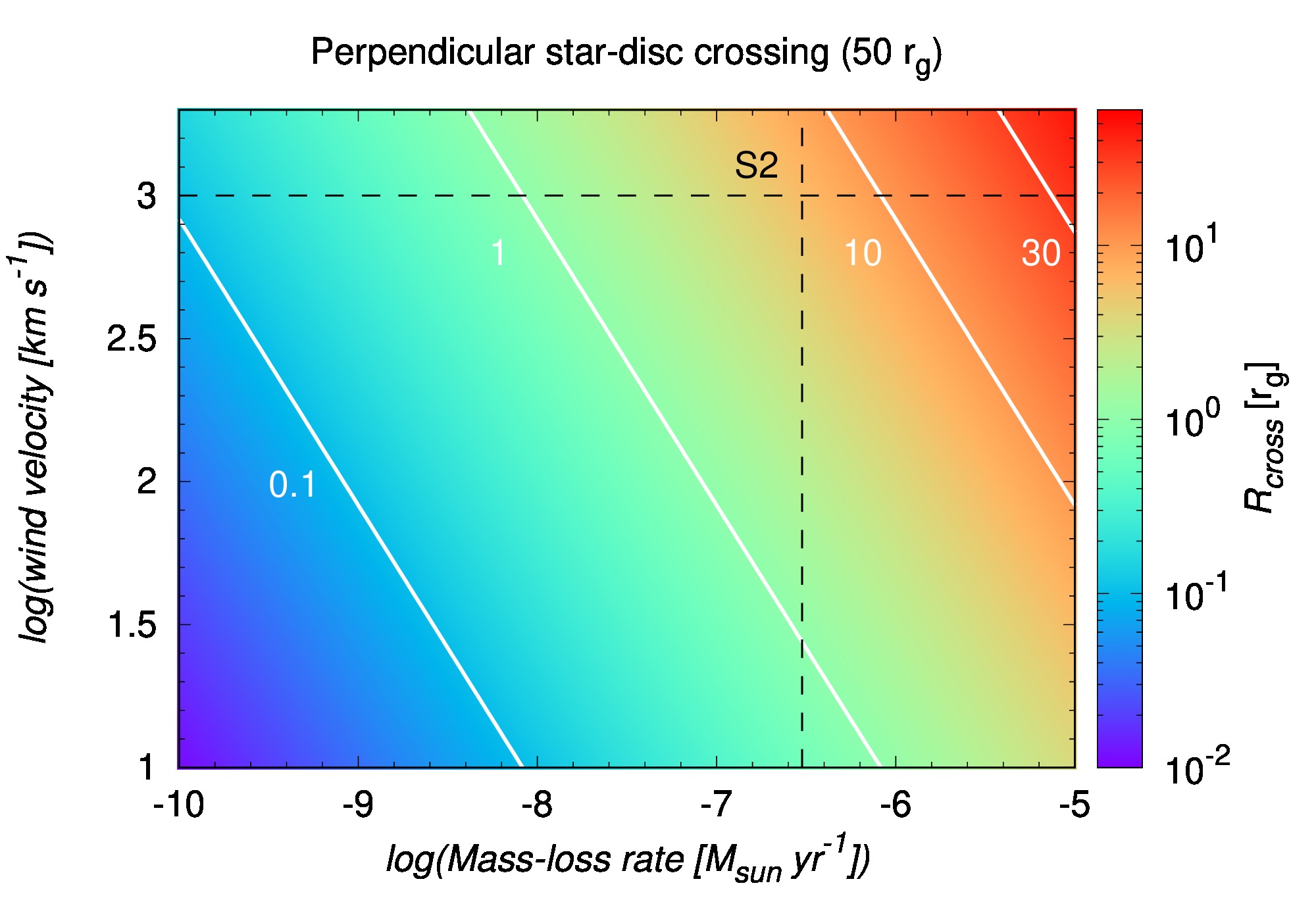}
    \includegraphics[width=\columnwidth]{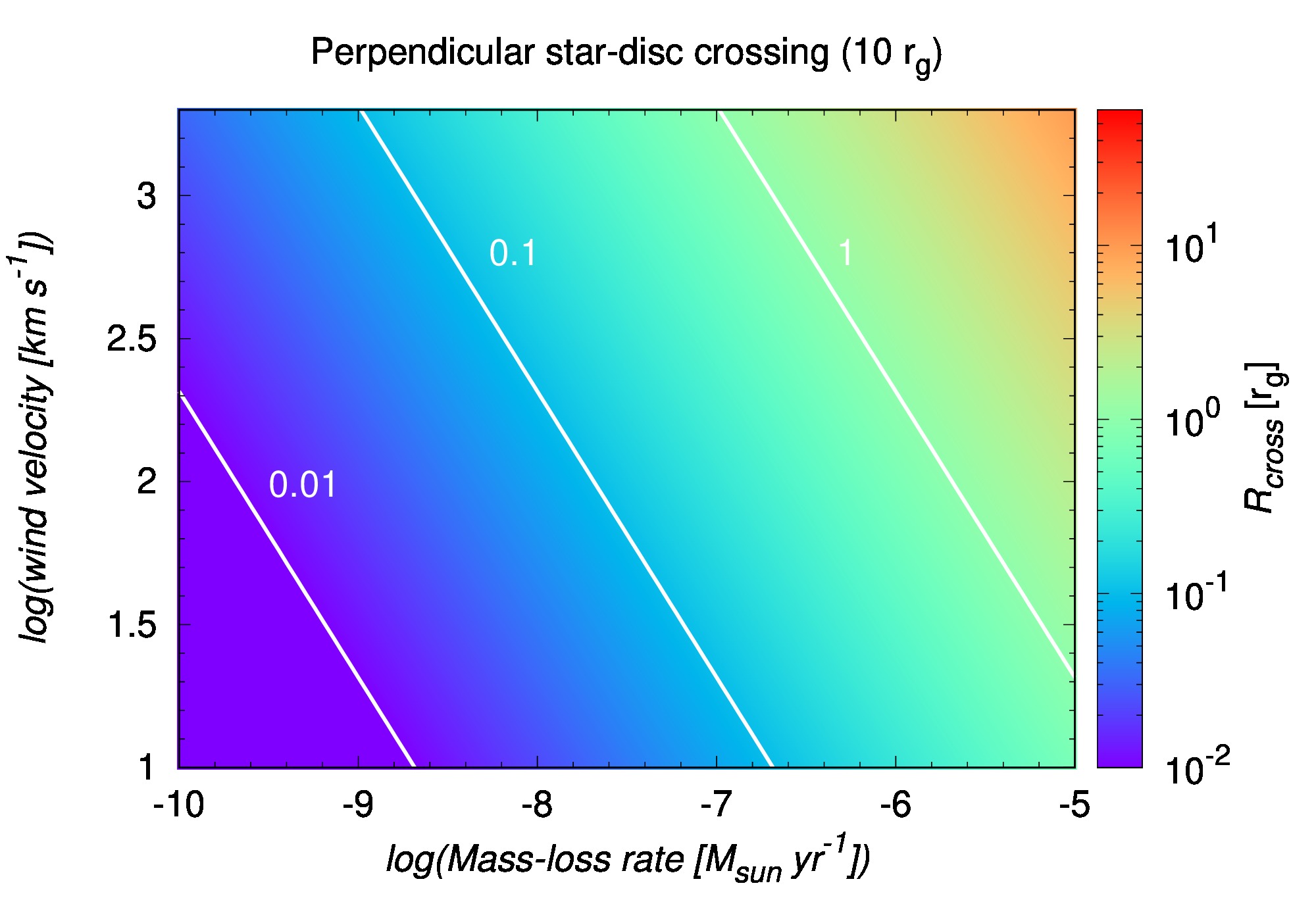}
    \caption{Characteristic cross-sectional radii of bow shocks of wind-blowing stars near a Sgr A$^*$-like accretion flow. {\bf Left panel:} A colour-coded cross-sectional radius of the wind-blowing star as a function of its mass-loss rate and of the terminal wind velocity. A star is assumed to move perpendicular to the accretion flow at the distance of $50\,r_{\rm g}$. White lines mark $R_{\rm cross}$ of 0.1, 1, 10, and 30 $r_{\rm g}$. Black dashed lines represent the values representative of the Galactic center star S2 in case it orbited the SMBH at 50 $r_{\rm g}$. The magnetization parameter is fixed to $\beta_{\rm m}=0.5$. {\bf Right panel:} The same as in the left panel, but for the perpendicular stellar passage at $10\,r_{\rm g}$. The white lines mark the parameter values for $R_{\rm cross}=0.01\,r_{\rm g}$, $0.1\,r_{\rm g}$, and $1\,r_{\rm g}$.}
    \label{fig_cross_sectional_radius}
\end{figure*}

Let us consider the momentum transition from the stellar wind to the ambient medium in front of the star as it passes through the accretion flow. The radius at which the stellar-wind kinetic pressure $P_{\rm sw} \approx \rho_{\rm w} v_{\rm w}^2$ on one side and the ram-pressure of the medium $P_{\rm ram} = \rho_{\rm a} v_{\rm rel}^2$ as well as the accretion flow thermal pressure $P_{\rm th} = \rho_{\rm a} c_{\rm s}^2$ on the other hand, are at equilibrium, is referred to as the stagnation radius $R_{\rm stag}$. Here $\rho_{\rm a}$ is the ambient gas density, $\rho_{\rm w}$ is the density of the wind at the given point, and $c_{\rm s}$ is the sound speed. At the distance $R$ from the star we have $\rho_{\rm w}(R) = \dot{m}_{\rm w}/(4\pi R^2 v_{\rm w})$. We further take into account the magnetic pressure of the ambient medium, for which we assume that it is a fraction of the thermal gas pressure, $P_{\rm mag}=\beta_{\rm m} P_{\rm th}$, where we assume the sub-equipartition value for $\beta_{\rm m}$ and we fix it to $\beta_{\rm m}=0.5$ \citep{2002A&A...383..854Y}. This is approximately consistent with the gas-to-magnetic field ratio in the accretion torus in our GRMHD simulations, see Fig.~\ref{fig:RB-beta} (where $\beta=1/\beta_{\rm m}$), while in the funnel region the magnetic field pressure dominates. The sub-equipartition value for $P_{\rm mag}$ is a necessary condition for the development of a strong shock as for $\beta_{\rm m}\gtrsim 1$ the fluid becomes mechanically incompressible \citep{2007MNRAS.382.1029K,2014ApJ...780..125V}. Finally, we obtain the equation for $R_{\rm stag}$ from the equilibrium conditions

\begin{align}
    P_{\rm sw} &= P_{\rm ram}+P_{\rm th}+P_{\rm mag}\,,\notag\\
    \frac{\dot{m}_{\rm w}v_{\rm w}}{4\pi R_{\rm stag}^2}&=\rho_{\rm a}[v_{\rm rel}^2+(1+\beta_{\rm m})c_{\rm s}^2]\,,\notag\\
    R_{\rm stag} &= \left[\frac{\dot{m}_{\rm w}v_{\rm w}}{4\pi \rho_{\rm a}[v_{\rm rel}^2+(1+\beta_{\rm m})c_{\rm s}^2]}\right]^{1/2}\,.\label{eq_stagnation_radius}
\end{align}

The shape of the thin bow shock was analytically derived by \citet{1996ApJ...459L..31W} as $R_{\rm bs}(\theta)=\frac{R_{\rm stag}}{\sin{\theta}}\sqrt{3(1-\frac{\theta}{\tan{\theta}})}$. For the perpendicular cross-section with respect to the relative stellar motion, we obtain the radius of $R_{\rm cross}\equiv R_{\rm bs}(\theta=\pi/2)=1.73R_{\rm stag}$, which we will consider as the effective radius of the interaction.

We are interested in the repetitive encounters of stars that have massive outflows and fast winds. They are supposed to be located outside the tidal disruption radius. As a prototype, we consider the B-type bright star S2 in the S cluster of the Galactic center, which has $m_{\star}\simeq 13.6\,M_{\odot}$, $R_{\star}=5.53\,R_{\odot}$ \citep{2017ApJ...847..120H}. The tidal disruption radius for this star orbiting Sgr~A* is,

\begin{equation}
    r_{\rm t}\sim R_{\star}\left(\frac{M}{m_{\star}}\right)^{1/3}\approx 43.4\,r_{\rm g}\,.
\end{equation}

We will look at the star-disc crossings at the scales of $r=50\,r_{\rm g}$ for a large range of mass-loss rates $\dot{m}_{\rm w}=10^{-10}-10^{-5}\,M_{\odot}{\rm yr^{-1}}$ and wind velocities, $v_{\rm w}=10-1000\,{\rm km\,s^{-1}}$. This essentially includes young massive stars with fast winds of a few 100 km/s as well as old stars with slow winds below 100 km/s.

For the density and the temperature of the hot ambient flow of the Sgr A$^*$ accretion disk, we apply the Radiatively Inefficient Accretion Flow (RIAF) profile with the flatter slope of $n_{\rm a}\propto r^{-1}$ and $T_{\rm a}\propto r^{-1}$ \citep{2013Sci...341..981W}, which includes both an inflow and an outflow. When we scale the profiles to the Bondi radius of $r_{\rm B}\approx 4''(T_{\rm a}/10^7\,{\rm K})^{-1}\sim 0.16\,{\rm pc}=837000\,r_{\rm g}$, using the values of the density and the temperature inferred from the X-ray spectroscopy \citep{2003ApJ...591..891B}, we get the following ambient profiles,

\begin{align}
    n_{\rm a} &=26  \left(\frac{r}{r_{\rm B}} \right)^{-1.0}\,{\rm cm^{-3}},\notag\\
    T_{\rm a} &=1.5\times 10^7  \left(\frac{r}{r_{\rm B}} \right)^{-1.0}\,{\rm K}.\label{eq_density_temperature}
\end{align}

In Fig.~\ref{fig_cross_sectional_radius}, we plot the colour-coded radii of the cross-sectional areas of stellar bow shocks as they interact with the hot flow. In this case, we consider the scenario when the stellar motion is perpendicular to the accretion flow and the star orbits at $50\,r_{\rm g}$ (left panel) and $10\,r_{\rm g}$ (right panel). Along the $x$-axis is the logarithm of the mass-loss rate, and along the $y$-axis is the logarithm of the wind velocity. In the left panel, we specifically mark the cross-sectional radius expected for the Galactic center star S2-like star -- $R_{\rm cross}=6.8\,r_{\rm g}$ -- for the inferred mass-loss rate of $\dot{m}_{\rm w}\lesssim 3\times 10^{-7}\,M_{\odot}\,{\rm yr^{-1}}$ and wind velocity of $v_{\rm w}=10^3\,{\rm km\,s^{-1}}$ \citep{2008ApJ...672L.119M} in case it orbited the SMBH at $50\,r_{\rm g}$.

It should be noted that this choice of $n_{\rm a}$ corresponds only to an example of a realistic RIAF scenario, specifically Sgr A$^*$. However, RIAFs are defined only as flows where a sizeable fraction of their binding energy stays in its thermal energy, $k_{\rm B} T_{\rm a} \gtrsim 0.1 GMm_{\rm p}/r$, where $m_{\rm p}$ is the proton mass, or $c_{\rm s}^2 \gtrsim 0.1 GM/r$. However, the ambient density $\rho_{\rm a} = m_{\rm p} n_{\rm a}$ in equation \eqref{eq_stagnation_radius} is set by a variety of other conditions in the neighborhood of the super-massive black hole. Hence, the cross-sectional radius $R_{\rm cross} \propto 1/\sqrt{n_{\rm a}}$ of a given star can vary by orders of magnitude in other galactic nuclei.


\subsection{Case of a young pulsar}\label{sec:pulsar}

A magnetized neutron star can form a sizeable cavity in the ambient accretion flow due to the pressure of its electromagnetic field \citep{2015AcPol..55..203Z}. Neutron stars can be understood as gravo-magnetic rotators that attract ambient plasma due to their strong gravitational field on one hand but can also effectively stop it from accreting due to electromagnetic pressure on the other hand. Neutron stars as rotating magnetic dipoles are characterized by a stationary electromagnetic field inside the light cylinder $R_{\rm l}=c/\Omega$, which transforms into a freely propagating electromagnetic wave outside it. Close to the magnetic axis, the electric component is aligned with the magnetic component and accelerates particles into relativistic velocities close to the speed of light. These escaping particles form collectively a pulsar wind, which can interact with the surrounding accretion flow and pass its impulse into it. The pressure of such a pulsar wind may be estimated as $P_{\rm m}=L_{\rm m}/(4 \pi r^2 c)$, where the electromagnetic power is given by the spin-down energy of a pulsar, $L_{\rm m}=-\dot{E}_{\rm sd}=-I\Omega\dot{\Omega}=4\pi^2 I \dot{P}/P^3$ with $I$ being the moment of inertia of a neutron star, $P$ is the rotational period of a pulsar, and $\dot{P}$ is a period derivative. The size of the contact discontinuity between the shocked pulsar wind and the ambient shock is scaled by the stagnation radius, where the mechanical pressure of the pulsar wind is equal to the sum of the ram pressure, the ambient thermal pressure as well as the magnetic pressure of the hot accretion flow. For the magnetic pressure, again we assume $P_{\rm mag}=\beta_{\rm m} P_{\rm gas}$ as in Subsection~\ref{sec:bowshock}. Considering the example of Sgr A$^*$, this yields the ambient magnetic field of $B_{\rm a}=69\,{\rm G}$ at 10$\,r_{\rm g}$ according to the density and the temperature profiles in Eq.~\ref{eq_density_temperature}. We note that this value is also consistent with the magnetic field strength in the range $10-100\,{\rm G}$ inferred from the flaring activity of Sgr~A* \citep{2012A&A...537A..52E}.

The stagnation radius for the pulsar wind can then be estimated as follows, 

\begin{equation}
    R_{\rm psr}=\left(\frac{\dot{E}_{\rm sd}}{4\pi c \rho_{\rm a}[v_{\rm rel}^2+(1+\beta_{\rm m})c_{\rm s}^2]} \right)^{1/2}\,.
    \label{eq_psr_stagnation}
\end{equation}
The radius of the cross-sectional area of the pulsar bow-shock can again be approximated as $R_{\rm cross}\approx 1.73R_{\rm psr}$, assuming that the contact discontinuity shape can again be described to the first approximation by an analytical solution of \citet{1996ApJ...459L..31W}, which is also confirmed by the X-ray, optical, and radio observations of several pulsar wind nebulae \citep{2014ApJ...784..154B}.

The innermost distance of the neutron star is not limited by the tidal disruption, since the critical radius for the tidal disruption event is smaller than the gravitational radius for SMBHs of mass at least $M\gtrsim [c^2R_{\rm NS}/(Gm_{\star}^{1/3})]^{3/2}\sim 14.9\,M_{\odot}$, where we used $R_{\rm NS}=10\,{\rm km}$ and $m_{\star}=1.4\,M_{\odot}$ as typical values for the radius and the mass of a neutron star respectively. Therefore, neutron stars can in principle orbit the SMBH close to the innermost stable circular orbit.

\begin{figure}
    \centering
    \includegraphics[width=\columnwidth]{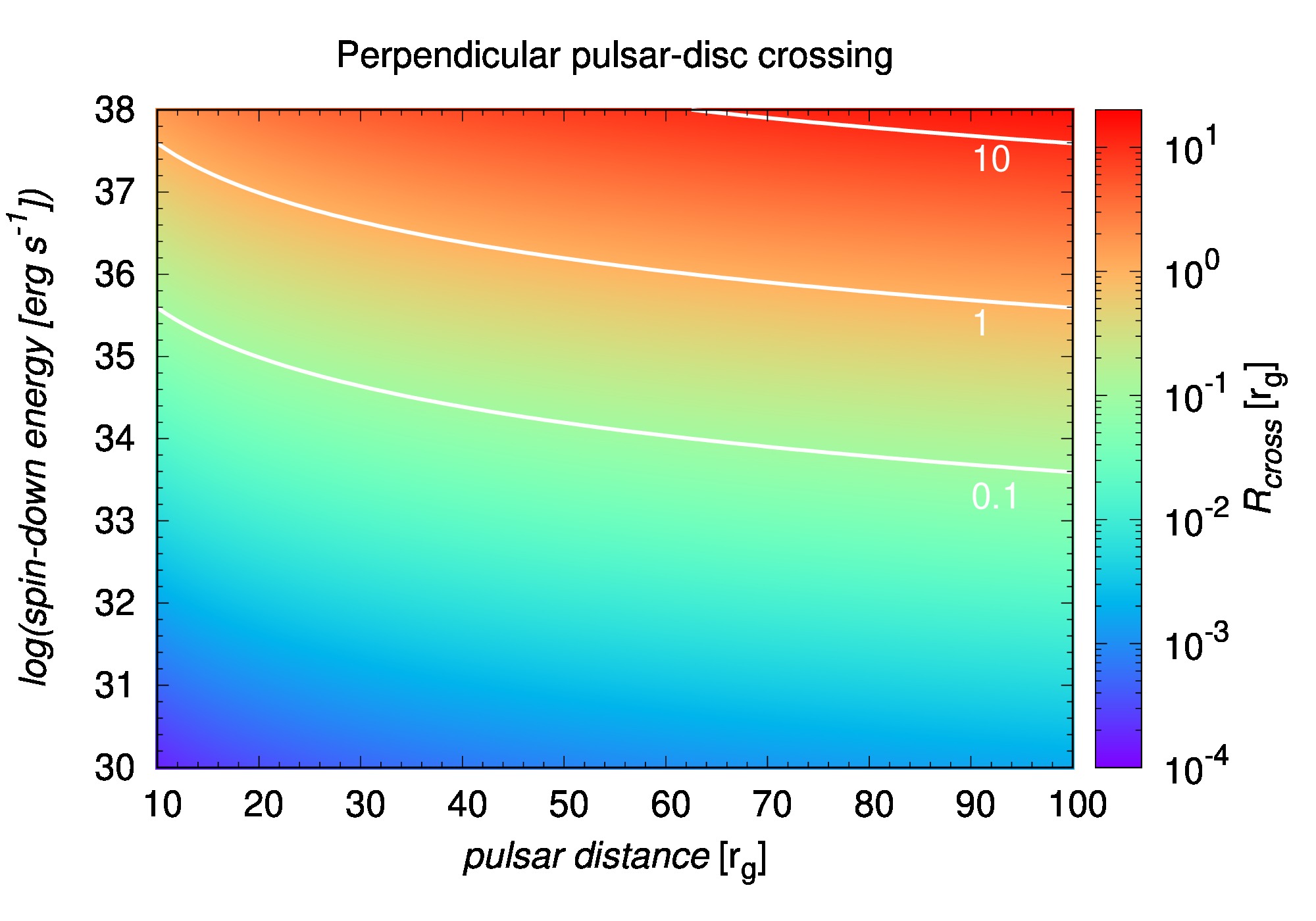}
    \caption{A colour-coded cross-sectional radius of the pulsar bow-shock as a function of the pulsar distance from a Sgr A$^*$-like SMBH (in gravitational radii) and of its spin-down energy (in ${\rm erg\,s^{-1}}$). The pulsar is considered to move perpendicular to the accretion flow. The white lines mark the parameter values for $R_{\rm cross}=0.1\,r_{\rm g}$, $1\,r_{\rm g}$, and $10\,r_{\rm g}$. The magnetization parameter is fixed to $\beta_{\rm m}=0.5$.}
    \label{fig_cross_sectional_radius_psr}
\end{figure}

In Fig.~\ref{fig_cross_sectional_radius_psr}, we plot the colour-coded cross-sectional radius of a magnetized neutron star with respect to the distance from the SMBH (in gravitational radii) and the neutron-star spin-down energy, or rather $-\dot{E}_{\rm sd}$ expressed in ${\rm erg\,s^{-1}}$. Again, we consider the perpendicular motion of a neutron star with respect to the accretion flow. We see that at the distance of $10\,r_{\rm g}$, the pulsar needs to be young and energetic with $-\dot{E}_{\rm sd}\approx 10^{38}\,{\rm erg\,s^{-1}}$, which of the same order as the Crab pulsar with $\dot{E}_{\rm sd}\sim 4.4\times 10^{38}\,{\rm erg\,s^{-1}}$ given its $P=0.033\,{\rm s}$ and $\dot{P}=10^{-12.4}\,{\rm s\,s^{-1}}$ \citep{2005AJ....129.1993M}, in order for the cross-sectional radius to be of the order of one gravitational radius. Again we remind the reader that $R_{\rm psr} \propto 1/\sqrt{n_{\rm a}}$ so a pulsar of given properties can create smaller or larger cavities in other scenarios.


\subsection{Case of a dark compact object} \label{sec:CO}

As a final scenario, we consider a compact object without magnetic fields or winds and with radiation that exerts only negligible pressure on the ambient medium. In this case we need to take into account mainly the purely gravitational interaction of the body with the disk. The body will accrete the disk matter in a Bondi-Hoyle-Lyttleton fashion \citep[see,e.g.,][]{2004NewAR..48..843E} and its effect on the flow was described by \citet{1999ApJ...513..252O}. The qualitative picture differs for supersonic and subsonic passages (high and low inclinations of the orbit). Supersonic passages correspond more closely to the Hoyle-Lyttleton scenario; they create a sonic surface in the shape of a trailing cone in the disk and accretion occurs through a wake that forms within the cone. 
Subsonic passages, on the other hand, correspond more closely to the Bondi accretion scenario. The disk matter is drawn towards the compact object and creates an oblate spheroidal sonic surface after the passage of which it is largely supersonically accreted. 

However, the meaning of these regions is different from the previously discussed cases, since the sonic surfaces are not surfaces at which the gas is synchronized with the motion of the compact object. Instead, one can model the impact of the gravitating body by considering the dynamical drag on the body and the fact that from the conservation of momentum the total gas momentum $P_{\rm gas}$ will evolve as \citep{1999ApJ...513..252O}

\begin{align}
    \frac{d P_{\rm gas}}{dt} = \frac{4 \pi (G m_{\ast})^2 \rho}{v_{\rm rel}^2} I (\mathcal{M})\,, \label{eq:COdpdt}
\end{align}
where $I(\mathcal{M})$ is a dimensionless factor of order one determined by boundary conditions on the surface of the compact object and the Mach number $\mathcal{M} = v_{\rm rel}/c_{\rm s}$. Specifically, at $\mathcal{M}=1$ there would be a singularity that is removed by the finite size of the compact body and non-equilibrium dynamics of the gas. We can then redistribute this gas momentum derivative into an extended but finite region to model the effect of the passing compact object. Let us consider momentum transfer through a \textit{synchronization sphere} such that all gas entering a sphere of radius $R_{\rm sync}$ obtains the velocity $v_{\ast}$. It is easy to show that the radius of this sphere in order for the synchronization to cause momentum transfer \eqref{eq:COdpdt} is 

\begin{align}
    R_{\rm sync} = \frac{G m_{\ast}}{v_{\rm rel}^{3/2} v_{\rm gas}^{1/2}} I (\mathcal{M})\,.
\end{align}
Note the similarity of this expression with that of the Bondi radius.
Then, considering that $v_{\rm gas} \sim \sqrt{GM/r}$ and $v_{\rm rel} \sim \sqrt{2(1-\cos i)GM/r}$ we finally obtain the order-of-magnitude estimate

\begin{align}
    R_{\rm sync} \sim 0.1 r_{\rm g} \left(\frac{m_{\ast}}{10^3 M_{\odot}}\right) \left(\frac{10^6 M_{\odot}}{M}\right) \left(\frac{r}{100 r_{\rm g}}\right)   \frac{}{} \label{eq:Rsync}
\end{align}
In other words, a black hole of mass $10^3 M_{\odot}$ passing through the disk at hundred gravitational radii near a black hole of mass comparable to Sgr A$^*$ can be modeled by a synchronization radius $\sim 0.1 GM/c^2$.


\section{Numerical framework} \label{sec:num}


\subsection{GRMHD evolution of plasma}\label{sec:GRMHD}
In our study we focus on the behaviour of the plasma around supermassive black holes. We perform global GRMHD 2D and 3D simulations of the flow using the publicly available code {\tt HARMPI} \citep{2015MNRAS.454.1848R,2007MNRAS.379..469T}, which is based on the original HARM code \citep{0004-637X-589-1-444,2006ApJ...641..626N} and which we have equipped with our own modifications. The code is 3D and parallelized using the domain decomposition with the Message Passing Interface.

The code solves the equations of ideal magneto-hydrodynamics under the assumption of magnetic field lines frozen in the plasma (in other words zero resistivity or infite conductivity of the material) on the curved background spacetime, which is described by the Kerr metric. In the present study, we use the fiducial value of the Kerr spin parameter $a=0.5$. The code uses a conservative, shock-capturing scheme with a staggered magnetic-field representation and adaptive time steps $\Delta t$, which are found on the basis of the shortest time needed for a wave to travel through a grid cell along the grid. 

The equations are solved in the modified Kerr-Schild coordinates in a way that the grid has logarithmic spacing in $r$ direction and non-uniform spacing in $\theta$ direction to focus the grid resolution towards the equatorial plane, where the majority of the accretion flow resides. Thanks to that, no grid refinement is needed for our purposes.
The grid stretches below the horizon in such a way that at least 5 cells are located below the horizon. 
To avoid the superfluous slow-down of the computation in 3D due to the small cell size near the rotational axis, the grid is deformed so that the innermost polar cell is cylindrical \citep{Sasha-smart-grid}. 

The gas is described by the polytropic equation of state $p = K \rho^\gamma$. The electrons in the accretion flow are relativistic, whereas the protons non-relativistic, therefore we choose the value $\gamma=13/9$, which lies in between the non-relativistic and relativistic limit, which is $\gamma = 5/3$ and $\gamma = 4/3$, respectively.

Our study aims at the low luminous galactic nuclei, such as Sgr A* in our own Galaxy, where the accretion rate onto the supermassive black hole is well below the Eddington limit (that is by eight to nine orders of magnitude). In this regime, the flow is known to be radiatively ineficient and cannot cool efficiently by the outgoing radiation. Therefore, we do not take into account the radiative transfer in the simulations. 

HARM naturally works in $G=c=1$ units so that velocities become dimensionless and times and lengths can be measured in terms of $M$ (gravitational radii $r_g = GM/c^2$ and gravitational times $GM/c^3$). All plots in this Section are given in these units.

The more detailed discussion about the chosen numerical method is given in Appendix~\ref{App_numerics}, while we test the capabilities of the code on simple test case in Appendix~\ref{App_test}.


\subsection{Motion of the star}\label{sec:star}
The interaction between the accreting plasma and the moving star, which can accrete gas or which can be equipped with magnetic field or strong outflows, is a complicated problem. In this study, we are interested in the dynamical effects of the passing star on the gas, hence we do not evolve the internal structure of the star itself, nor do we take into account the possible changes of the stellar trajectory due to the star-disc interactions.
Therefore, we assume that the star is a test solid body moving along the geodesic orbit. 

The geodesics of the star is found using Boyer-Lindquist coordinates, where the coordinate time $t_{\rm}$ is taken as the parametrization of the curve. We solve the trajectory numerically using the classical explicit Runge-Kutta method RK4 with the set of equations

\begin{equation}
    \frac{{\rm d^2} x^\mu}{{\rm d} t^2} = -\Gamma^\mu_{\alpha \beta} \frac{{\rm d} x^\alpha}{{\rm d}t} \frac{{\rm d} x^\beta}{{\rm d} t} + \Gamma^0_{\alpha \beta} \frac{{\rm d} x^\alpha}{{\rm d}t} \frac{{\rm d} x^\beta}{{\rm d} t}\frac{{\rm d} x^\mu}{{\rm d} t},
    \label{eqn:geodesic}
\end{equation}
where $x^\mu$ is the position of the star expressed in Boyer-Lindquist coordinates, $t\equiv x^0$, and $\Gamma^\mu_{\alpha \beta}$ are the Christoffel symbols for the Boyer-Lindquist metric. 

With this parametrization and thanks to the fact that coordinate time $t$ in Boyer-Lindquist and Kerr-Schild coordinates coincide, we evolve the star simultaneously with the gas in the coordinate time in the sense that we use the time step $\Delta t$ found by the adaptive time step feature of the GRMHD solver to find a new position of the star in the explicit RK4 scheme. Internally, the positions $x^\mu$ and the coordinate velocities $v^\mu = \frac{{\rm d} x^\mu}{{\rm d} t}$ of the star in the Boyer-Lindquist coordinates are stored, however, we find the four-velocity $u^\mu$ of the star in each step according to 
\begin{eqnarray}
    \left( u^t \right)^2 \equiv \left( \frac{{\rm d} t}{{\rm d} \tau} \right)^2 & = & - \left(g_{\alpha \beta} \frac{{\rm d} x^\alpha}{{\rm d}t} \frac{{\rm d} x^\beta}{{\rm d} t} \right)^{-1}, \label{eqn:ut} \\
    u^{\alpha} & = & u^t  \frac{{\rm d} x^\alpha}{{\rm d}t}.
\end{eqnarray}

We check the accuracy of the orbit integration by monitoring the constancy of the values of energy per unit mass $\varepsilon \equiv -u_t$ and azimuthal angular momentum per unit mass $l_z \equiv u_\phi$, which are the constant of geodesic motion in Kerr spacetime; the norm $u^\mu u_\mu = -1$ is always kept at machine precision due to the usage of the relation (\ref{eqn:ut}).
We verified that the HARM time step $\Delta t \in (10^{-3},10^{-2})\M$ in our simulations is small enough to obtain the orbits with sufficient precision, that is, the relative fluctuation of $u_t$ and $u_\phi$ were on the order $\lesssim 10^{-10}$ over the time span of the simulations. Additionally, we have validated the code using the {\em KerrGeodesics} Mathematica package from the BHPToolkit (\href{http://bhptoolkit.org/}{bhptoolkit.org}); we found agreement to at least a few parts per hundred thousand in the turning points of the orbits and the frequencies of motion.


\subsection{Perturbation of the flow}\label{sec:perturb}
The dynamical effect of the star on the surrounding medium is mimicked in the following way. In each time step, we find which grid cells (i.e. the center of the cell) have distance to the star position $d$ smaller than the interaction radius of the star $\mathcal{R}$, measured with respect to the Boyer-Lindquist metric:
\begin{eqnarray}
    d &=& \Big( g^{\rm BL}_{\alpha \beta}\Delta x^{\alpha}\Delta x^{\beta}\Big)^{1/2},\\
    \Delta x^\alpha &=& x^{\alpha}_{\rm star} - x^{\alpha}_{\rm fluid}  \label{eqn:rozdil}, \\
    \Delta t_{\rm BL} &\equiv& 0.
\end{eqnarray}
In these grid cells, we overwrite the velocity of the gas and set it equal to the velocity of the star while keeping the other primitive quantities of the gas intact. In this way, the number of cells affected by the star differs at different positions in the grid due to non-uniform grid spacing and, consequently, the shape of the star can be considered as only roughly spherical.

 This approach approximates the fact that the moving star is pushing the gas in front of itself, but it may also be slowing it down wherever the star is slower than the flow. The radius of the star $\mathcal{R}$ does not need to correspond directly to the physical dimension of the star, it rather describes the size of the sphere of influence of the star on the surrounding medium. 
 Estimates for $\mathcal{R}$ are given as $R_{\rm cross}$ in Subsections \ref{sec:bowshock} and \ref{sec:pulsar}, and as $R_{\rm sync}$ in Subsection \ref{sec:CO}.

\begin{figure}
    \centering
    \includegraphics[width=\columnwidth]{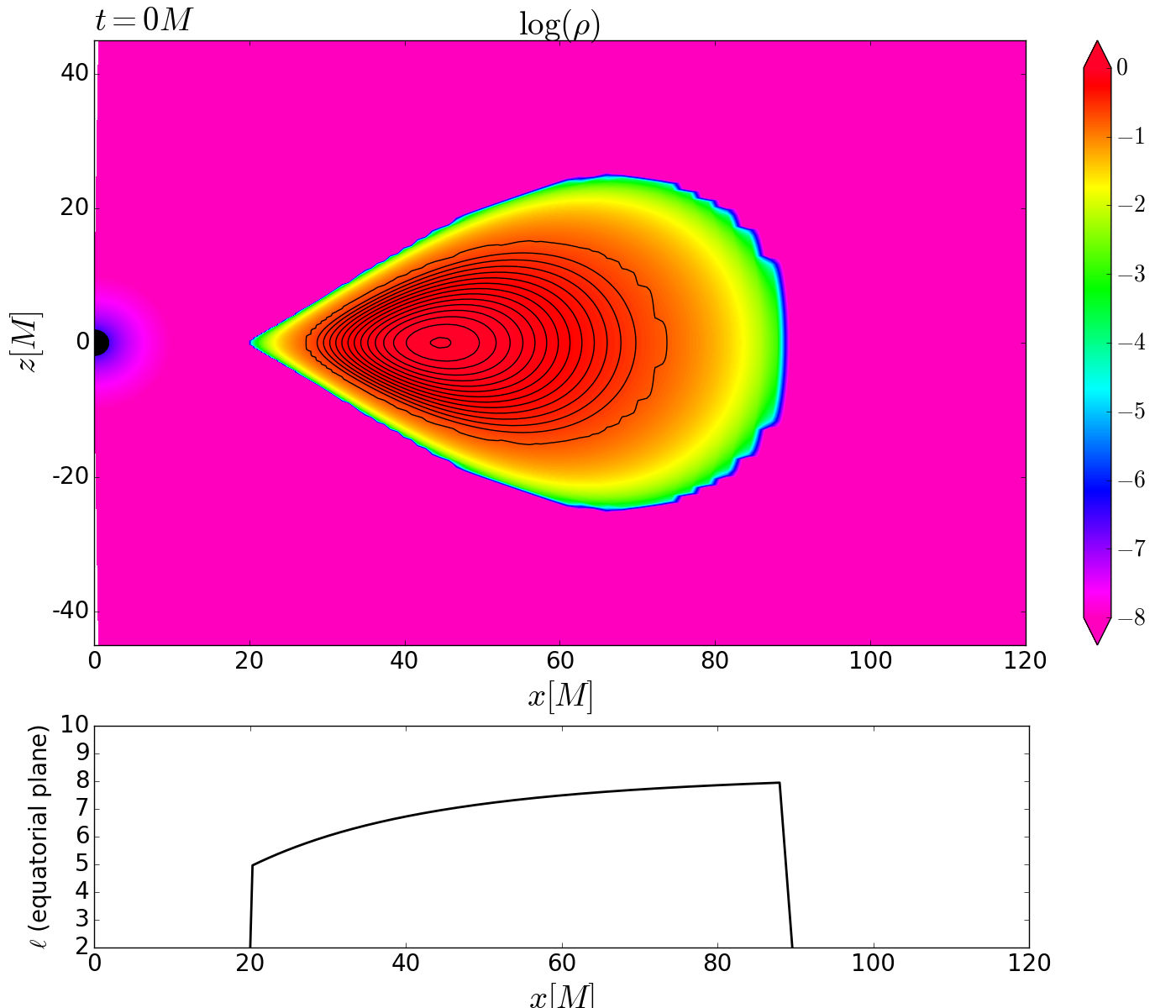}
    
    \includegraphics[width=\columnwidth]{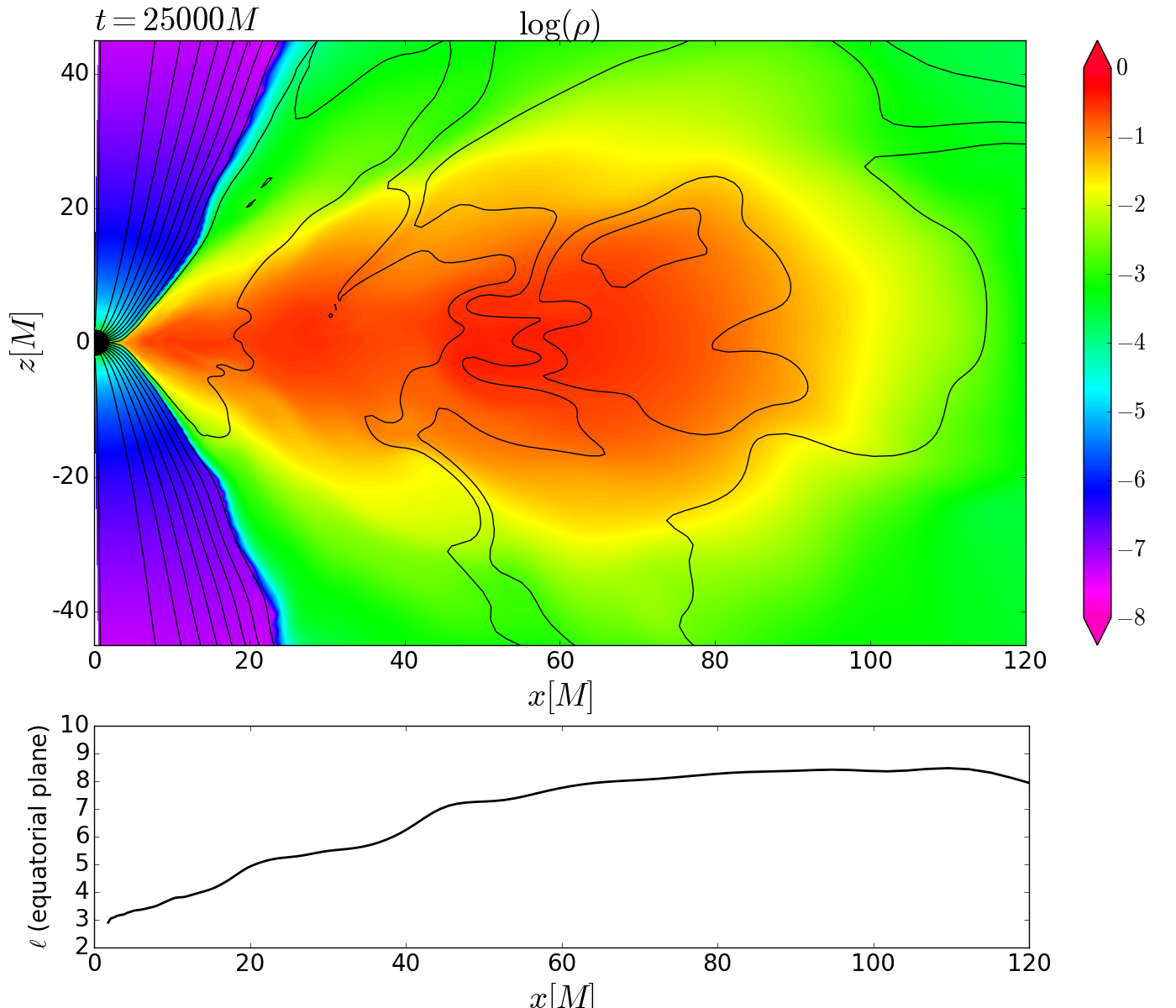}
    
    \caption{The map of density $\rho$ in the logarithmic scale and the equatorial profile of angular momentum $\ell=-u_\phi/u_t$.  {\bf Top:} Initial conditions for the non-perturbed Run~NP1. {\bf Bottom:} Evolved state of the non-perturbed torus at $t=25000\,M$, which is used as the initial condition for the perturbed runs.}
    \label{fig:NP-initial}
\end{figure}


\subsection{Initial conditions} \label{sec:init}

In order to study the influence of the passing star on the accreting medium, we have to start with a quasi-stationary unperturbed state and compare the unperturbed evolution with the perturbed ones. 

The quasi-stationary state of the accretion flow is obtained when an initial thick magnetized torus is left to spontaneously evolve for a long enough time. In the literature there are several examples of exact solutions describing the torus in equilibrium, starting with the well-known non-magnetized tori with constant angular momentum \citep{1978A&A....63..209K,1978A&A....63..221A,1976ApJ...207..962F} and later generalized to configurations with toroidal magnetic fields \citep{2006MNRAS.368..993K}. Recently \citet{Witzany_Jefremov-tori} generalized these solutions further into a closed-form two-parametric family of solutions with non-constant profiles of angular momentum with various possibilities of rotation curves
and geometric shapes. 
We chose to initiate the simulation using the tori from the Witzany-Jefremov family, which are large enough, so that they serve as a suitable reservoir of the accreting matter during the time span of the whole simulation, and which have an increasing profile of angular momentum. In particular, here we start the simulation with the torus described by $\kappa = 7.61,l_0 = 8.46098\M$, which stretches from $r_{\rm min}=20\M$ to $r_{\rm max}=90\M$ (for the meaning of the parameters, see \citet{Witzany_Jefremov-tori}).

The magnetic field lines in the equilibrium solutions are purely toroidal. However, the initiation of the magneto-rotational instability and thus of the accretion process requires at least a small poloidal field \citep{1991ApJ...376..214B}. Hence, we omit the toroidal magnetic field in the initial conditions and equip the torus with magnetic field lines that follow the isocontours of density. We choose the gas-to-magnetic pressure ratio as equal to $\beta=p_g/p_m = 100$ so that the magnetic field acts only as a small perturbation to the equilibrium. However, once the simulation is started, the magneto-rotational instability quickly enhances the magnetic fields and turbulent accretion commences.

The example of the non-perturbed Run~NP1 is given in Fig~\ref{fig:NP-initial}. In the upper two panels, the initial state of the torus is shown. The density in the logarithmic scale is overlaid with the magnetic field lines. The radial extent of the torus is quite large, so that the torus contains enough matter to supply the accretion in the inner region throughout the time span of the simulation. Below we show the increasing profile of angular momentum along the equatorial plane. 
The grid spans from below the horizon up to $R_{\rm out} = 2\cdot10^4\,M$ and the fiducial resolution is $n_r=252, n_\theta=192$.

\begin{figure}
\includegraphics[width=\columnwidth]{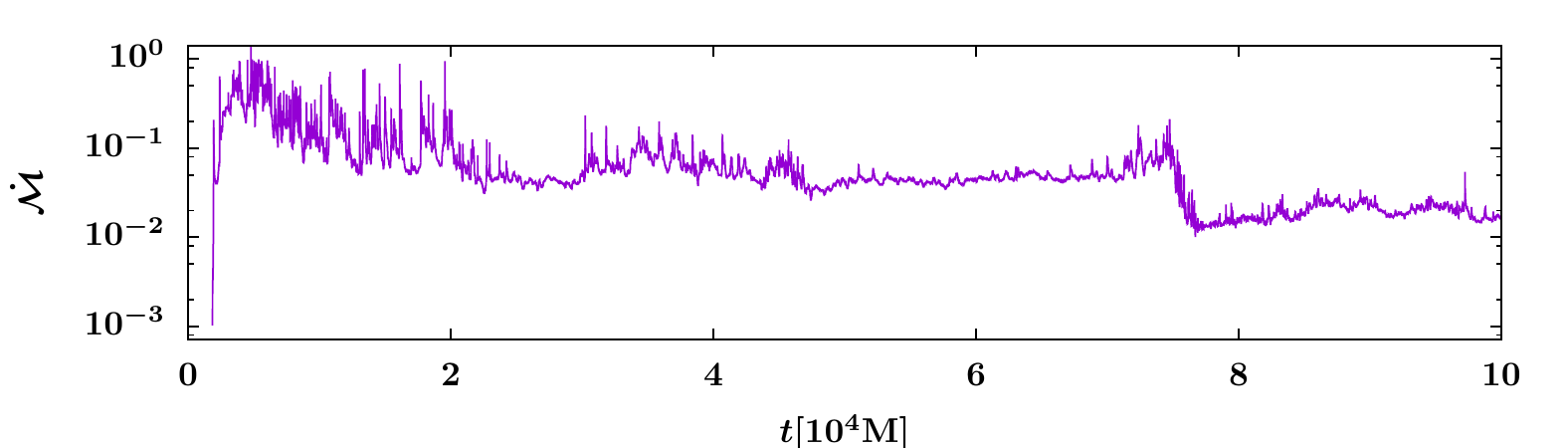}
\caption{Accretion rate $\dot{\mathcal{M}}$ of Run NP1 in code units.}
\label{fig:Mdot-NP1}
\end{figure}

The accretion rate profile of Run~NP1 is shown in Fig.~\ref{fig:Mdot-NP1}. MRI develops on the order of several hundreds of $M$ and matter from the inner part of the torus starts to accrete into the black hole at $t\sim2\cdot10^3 \, M$. Between $t\sim2\cdot10^3 \, M$ and $t\sim2\cdot10^4 \, M$, the accretion proceeds in quite a violent way, which is reflected in the accretion rate in form of high peaks and variable mean value. Roughly at $t\sim2\cdot10^4 \, M$ the quasi-stationary accretion state is achieved, which means that there are no abrupt peaks and the mean value stabilises. 

Therefore, we choose the state at $t_{\rm start} = 2.5\cdot10^4 \, M$ of the Run~NP1, which is shown in Fig.~\ref{fig:NP-initial} in the bottom set of plots, as the starting point for our simulations with the orbiting stars. In the 3D case, we copy the 2D slice into each of the $\phi$-direction slices and perturb the density $\rho$ by a random factor $X^{\rm rand}\in[0,1]$ such that $\rho^{\rm 3D}(r,\theta,\phi) =\rho^{\rm 2D}(r,\theta)(1 + 10^{-2}(X^{\rm rand}-0.5))$, which ensures the development of the flow in the azimuthal direction. In this way, we speed up the simulations (in particular the 3D runs) significantly compared to the situation, where all the runs would be initiated with the non-evolved torus at $t=0$.

\begin{table*}
    \centering
    \begin{tabular}{ccccccccccc}
    \hline
    \hline
         Run & $u_t$ & $u_\phi$ & $t_{\rm end} [\,M]$ & $r [\,M]$ & $\mathcal{R} [\,M]$ & $z_{\rm max}[\,M]$ & $i[\degree]$ & Type & $\mathcal{M}_{\rm in}$ ($t>3\cdot10^4\,M$) & $\mathcal{M}_{\rm out}$ ($t>3\cdot10^4\,M$) \\
    \hline     
         A & -0.9557 & 0.479 & $5\cdot10^4$ & 10 & 1.0 & 9.9 & 82.6 & I & 327.6 (4.1) & 355.2 (78.9)\\ 
         B & -0.9761 & 3.295 & $5\cdot10^4$ & 15 -- 25 & 1.0 & 17.8 & 45.3& E & 370.1 (97.5) & 22.0 (6.1) \\
         C & -0.9871 & 5.955 & $1\cdot10^5$ & 26 -- 50 & 1.0 & 10.5 & 12.2 & E & 2033.1 (1608.2) & 39.1 (29.4)  \\
         D & -0.9901 & 0.237 & $1\cdot10^5$ & 50 & 1.0 & 50.0 & 88.1 & I & 5500.9 (3096.2) & 90.1 (72.3) \\
         E & -0.9902 & 3.082 & $1\cdot10^5$ & 50 & 1.0 & 45.7 & 65.0 & E & 4103.2 (3329.4) & 54.9 (41.6)  \\
         F & -0.9902 & 3.082 & $1\cdot10^5$ & 50 & 10.0 & 45.7 & 65.0 & E & 1592.4 (510.0) & 73.5 (40.9) \\
         G & -0.9557 & 0.479 & $5\cdot10^4$ & 10 & 0.1 & 9.9 & 82.6 & I & 1631.0 (447.7) & 75.5 (39.4)\\ 
         H & -0.9539 & 3.352 & $5\cdot10^4$ & 10 & 1.0 & 3.6 & 21.4 & E & 207.8 (64.6) & 19.4 (4.2)\\ 
         I(3D)& -0.9557 & 0.479 & $3\cdot10^4$ & 10 & 1.0 & 9.9 & 82.6 & I & 1157.1 & 22.1\\ 
    \hline      
    \end{tabular}
    \caption{Summary of the star orbits in the runs. The first two columns show the constants of geodesic motion $u_t$ and $u_\phi$, next is the end time of the simulation $t_{\rm end}$, the radial range of the orbit, the radius of the star $\mathcal{R}$, maximal vertical position $z_{\rm max}$, the inclination angle $i$ and the type of the orbit (I for highly inclined orbits, E for orbits embedded in the torus). Last two columns show the total accreted amount of mass $\mathcal{M}_{\rm in}$ and the total fast-outflowing mass $\mathcal{M}_{\rm out}$, in the brackets the value is the total amount computed for $t>3.5\cdot10^4\,M$.}
    \label{Table:runs}
\end{table*}


\section{Results}\label{sec:results}


Here we present the results of the runs with the transiting star. Most of our runs are performed in 2D; for comparison the full 3D simulations of one non-perturbed run and one perturbed run are provided. The orbital parameters of the star for different runs are summarised in Table~\ref{Table:runs}. 


\subsection{2D runs}\label{sec:2D}
\begin{figure}
\includegraphics[width=0.47\textwidth]{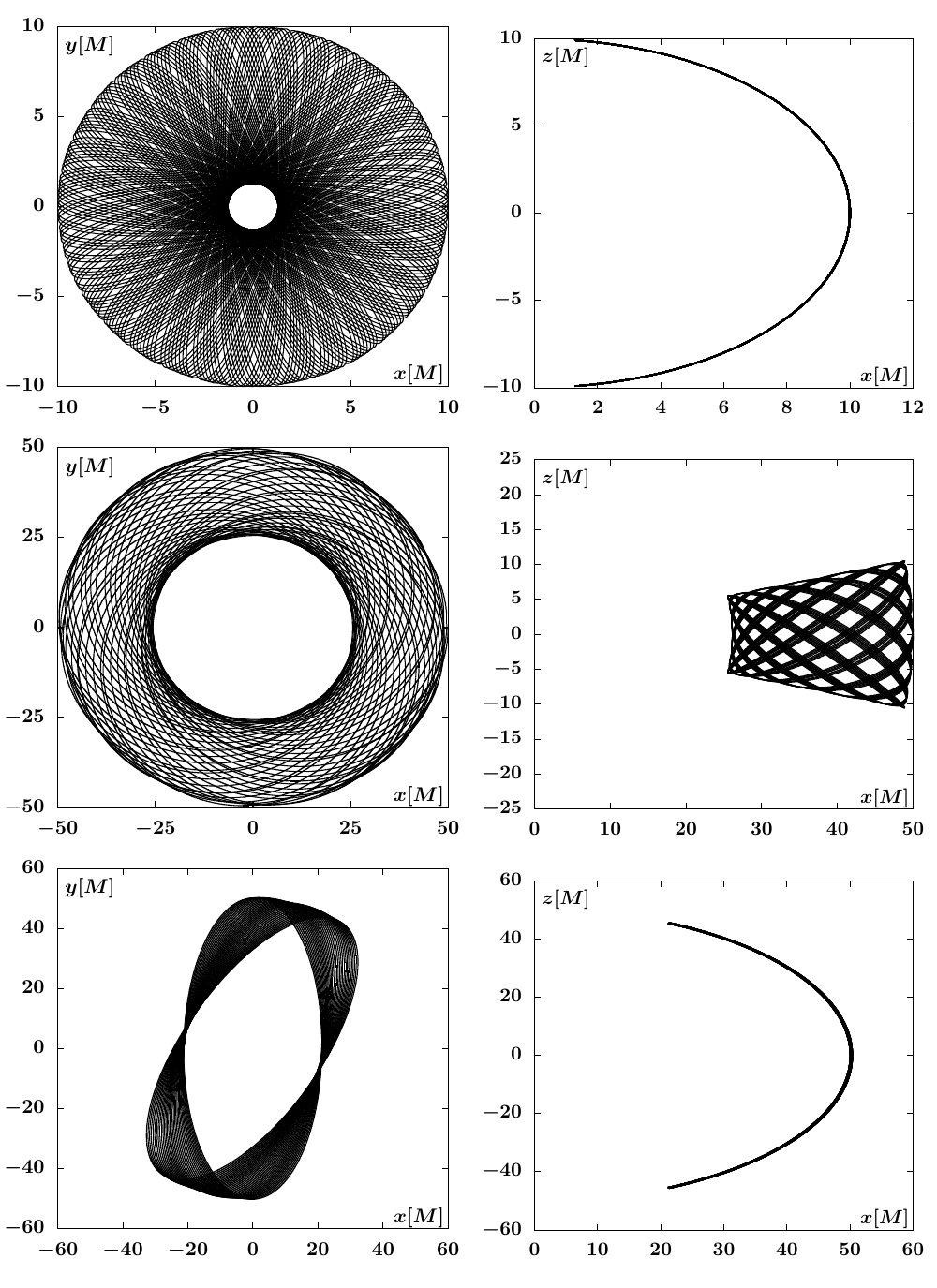}
\caption{Several examples of projections of the orbit of the star onto the equatorial plane (left column). Precession is caused by the Lense-Thirring effect near the rotating black hole. The trajectory collapsed onto a single 2D poloidal slice (right column). Orbit of Run~A in the first row, orbit of Run~C in the second row and orbit of Run~E in the last row. See the text for further details.}
\label{fig:orbits}
\end{figure}

We have studied several exemplary cases, which differ from each other by the orbital parameters of the star. In general, we can divide the star trajectories into two groups; \emph{the inclined orbits} which penetrate into the funnel region and \emph{the orbits embedded inside the torus} which move closer to the equatorial plane. 
 In most cases, the radius of the star is set to $\mathcal{R}=1\,M$ and the radial position of the star differs between $10\,M$ and $50\,M$. 
 
 As discussed in Section \ref{sec:bowshock}, the maximal physically plausible radius of the star at the distance of about $50\,M$ from Sgr A$^*$ is roughly $\sim 10\,M$, therefore we provide one run with $\mathcal{R}=10\,M$ (Run~F). On the other hand the smallest star with $\mathcal{R}=0.1\,M$ was studied in Run~G, which approximately corresponds to the \textit{physical} radius of the solar type star around Sgr A* ($\sim 1/9\,M$).

The fiducial resolution is given by the resolution of the non-perturbed Run~NP1 $n_r=252, n_\theta=192$ with logarithmic spacing in $r$-direction (superlogarithmic above a break radius $r_{\rm br}=200\,M$). For radii close to $r=10\,M$ the grid spacing is roughly $\Delta r \sim 0.2\,M$, while for $r=50\,M$ the grid spacing increases to $\Delta r \sim 1\,M$. This of course sets the lower limit on the possible size of the star in our simulation,  which is at the point where the star is so small that it occupies only a handful of the grid cells. This lower limit is met in Run D, E and G. On the other hand, in Runs A and F a few hundreds of cells are perturbed in each time step.

In the 2D runs, the full 3D trajectory of the star is computed according to eq.~(\ref{eqn:geodesic}), however, the $\phi$ coordinate of the star is "forgotten" in the sense that in equation~(\ref{eqn:rozdil}) we put $\Delta x^\phi=0$ (or in other words, the $\phi$ coordinate is reset to $\phi = \pi/2$). That corresponds to merging all $\phi$-slices into one 2D slice and averaging in the azimuthal direction. Therefore, we can expect that the effect of the star on the accretion flow will be enhanced by this approximation.

We vary the initial conditions of the star, which are given by the values of $x_{0}^{\mu}$ and $u_0^\mu = (1,u_0^r,u_0^\theta,u_0^\phi)$ in Boyer-Lindquist coordinates. 
The relation between Boyer-Lindquist coordinates and Kerr-Schild coordinates $x,y,z$ reads
\begin{eqnarray}
    x &=& \sqrt{(r^2 + a^2)}\sin{\theta}\sin{\phi} \,, \\
    y &=& \sqrt{(r^2 + a^2)}\sin{\theta}\cos{\phi} \,,\\
    z &=& r\cos{\theta} \,.
\end{eqnarray}
In Fig.~\ref{fig:orbits} we show projections of some of the star trajectories onto the equatorial plane (obtained by plotting $x(t),y(t)$), and the motion of the star in the 2D simulations' grid (obtained by plotting $\sqrt{x(t)^2+y(t)^2},z(t)$).

In some plots we provide snapshots from the simulations that contain the ``fast-outflowing rate density'' $\dot{m}_{\rm out}$, which is computed as
\begin{eqnarray}
    \dot{m}_{\rm out} &=& \rho \, u^r \sqrt{-g} \, {\rm d}\theta \, {\rm d}\phi \qquad {\rm for \, } \Gamma > 1.155 \\
    \dot{m}_{\rm out} &=& 0 \hspace{2.85cm} {\rm for \, } \Gamma \leq 1.155 ,,
\end{eqnarray}
where $\Gamma=\sqrt{-g^{00}}u^t$ is the gamma-factor of the gas.
We define the outflowing rate density in this way to separate the fast moving matter inside the funnel region from the slowly moving, but much denser gas in the torus. The value $\Gamma = 1.155$ corresponds to the velocity equal to half of the velocity of light and only wind escaping along the funnel-torus boundary from the torus and the expelled blobs exceed this velocity.

For comparison of the amount of accreting versus outflowing matter, we define the outflowing rate $\dot{\mathcal{M}}_{\rm out}$, which is a surface integral of $\dot{m}$ over two spherical sectors along the funnel with $0 < \theta < 35 ^{\circ}$ and $145^{\circ} < \theta < 180 ^{\circ}$ at a chosen radius $r_{\rm diag}=100M$
\begin{eqnarray}
    \dot{\mathcal{M}}_{\rm out}  &=& \int\limits_{\theta = 0}^{\theta = 0.61} \int\limits_{\phi = 0}^{\phi = 2\pi} \dot{m}_{\rm out}(r=r_{\rm diag})  + \nonumber \\
    & & + \int\limits_{\theta = 2.53}^{\theta = \pi} \int\limits_{\phi = 0}^{\phi = 2\pi} \dot{m}_{\rm out}(r=r_{\rm diag}). \label{eqn:Mdot_out}
\end{eqnarray}

In Table~\ref{Table:runs} we provide the total accreted matter $\mathcal{M}_{\rm in}$ during the time span of the simulation given in code units. In the brackets the amount of gas accreted for $t>3.5\cdot10^4M$ is given. Similarly, we provide the amount of outflowing matter $\mathcal{M}_{\rm out}$, given as the time integral of eq.~(\ref{eqn:Mdot_out}).

\begin{figure}
\includegraphics[width=\columnwidth]{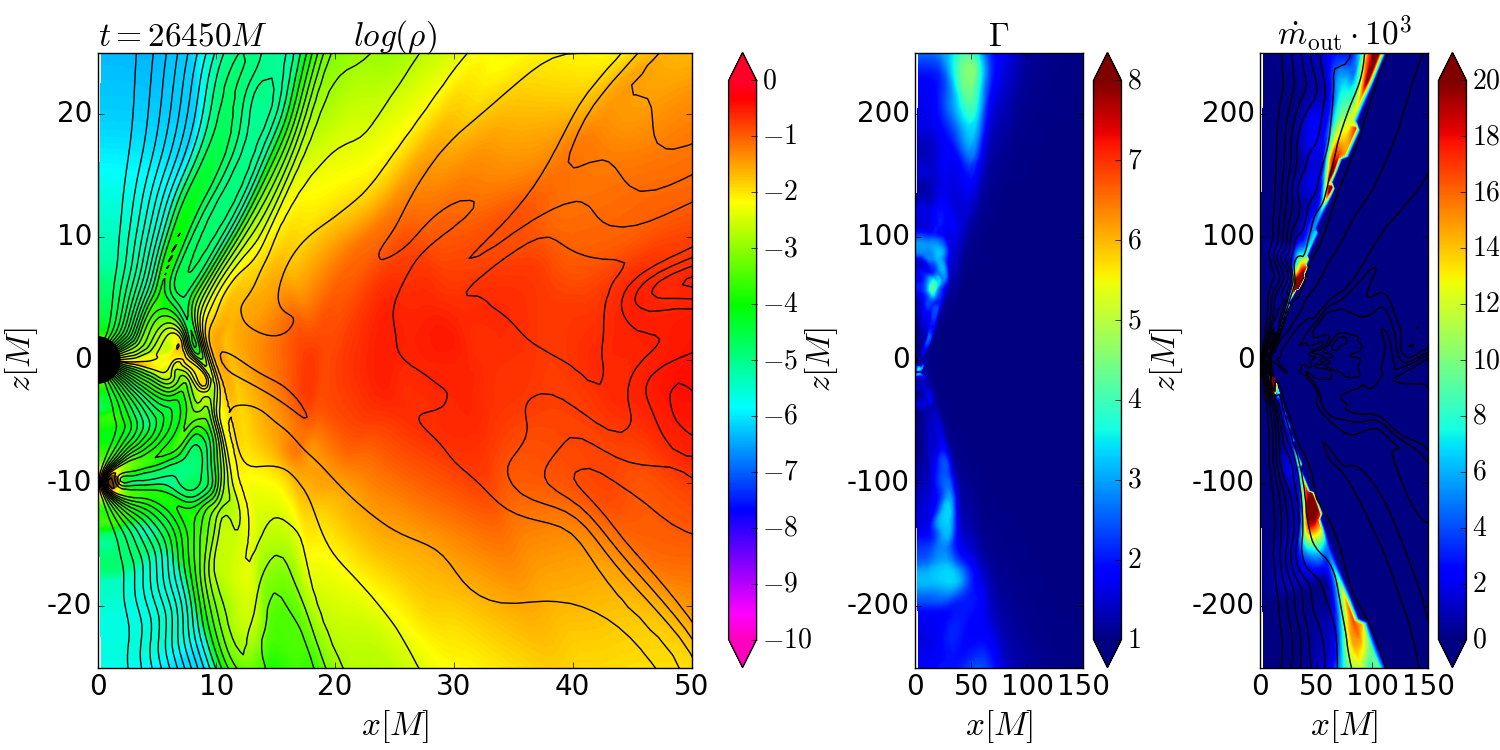}
\includegraphics[width=\columnwidth]{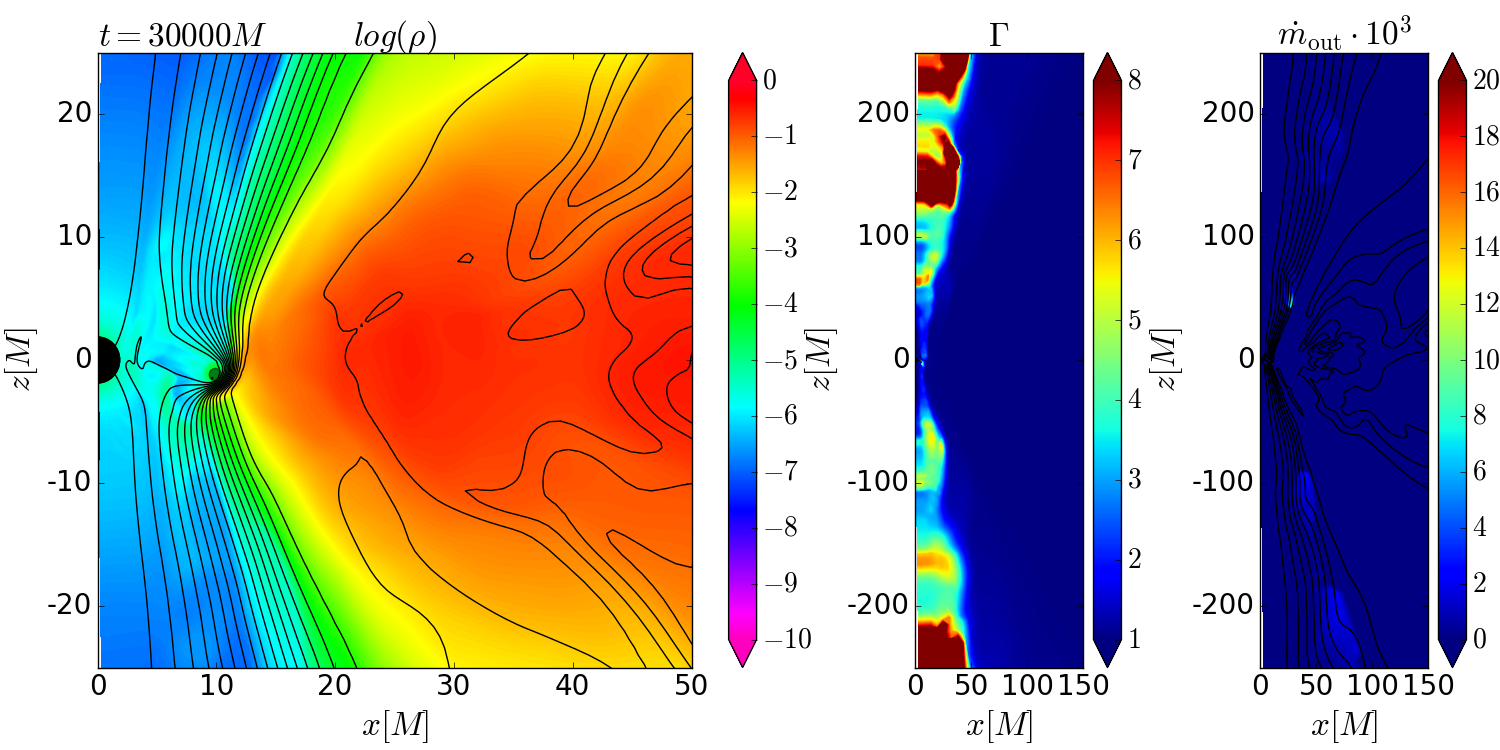}
\includegraphics[width=\columnwidth]{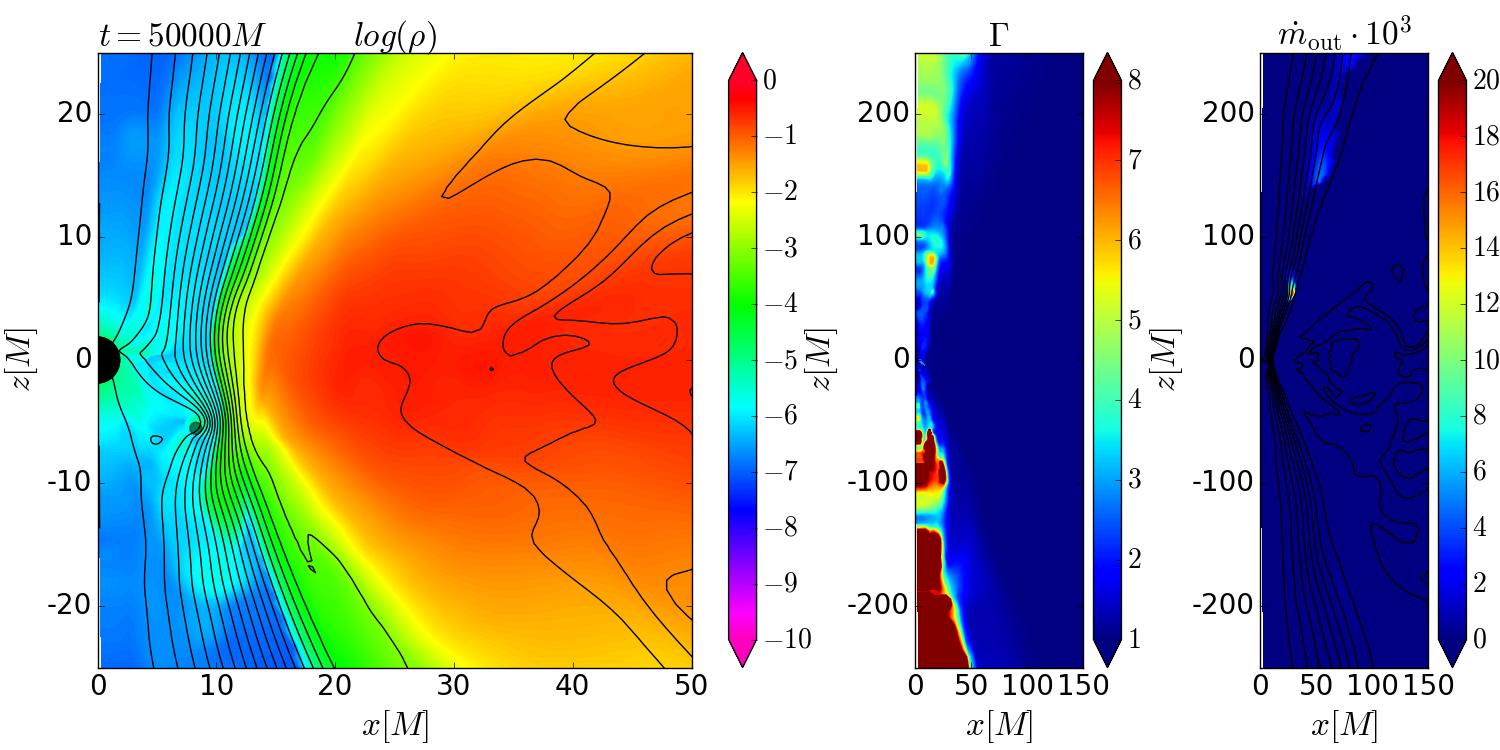}
\caption{Evolution of Run~A: snapshots at three different time instances showing the map of density in logarithmic scale, the Lorentz factor $\Gamma$ and the outflowing rate $\dot{m}_\textrm{out}$. Note the difference in the coordinate ranges of the plots (only a part of the entire simulation domain is shown). The animation of Run~A is available at \url{https://youtu.be/eERYioirgQc}. The video first resolves the initial phase with a short time-step, while the second part captures the long-time evolution at course time resolution (let us remind the reader that the orbital motion of the star is projected onto $xz$-poloidal plane in the animation).}
\label{fig:RA-slices}
\end{figure}


\subsubsection{Run A}
\begin{figure}
\includegraphics[width=\columnwidth]{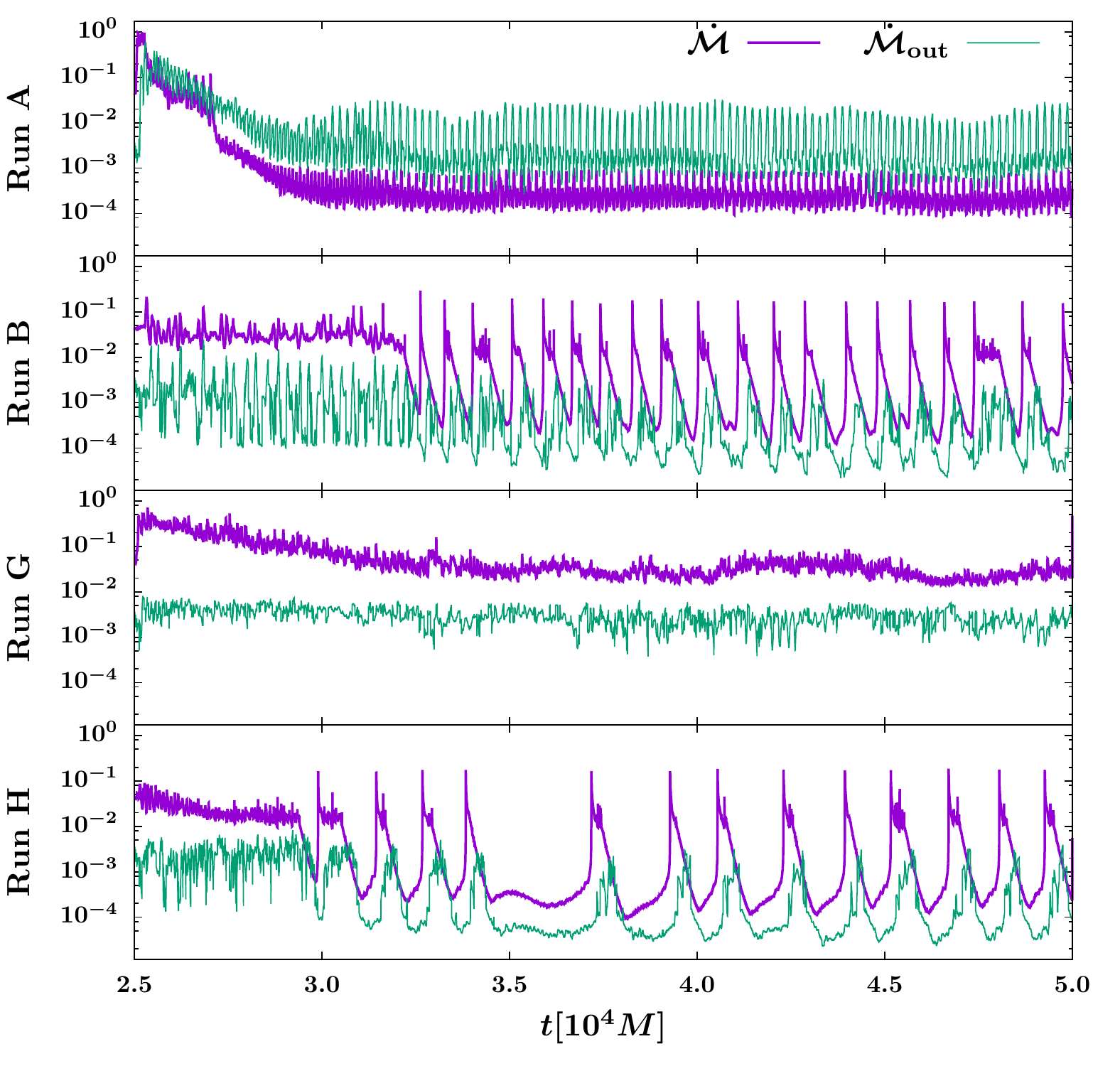}

\includegraphics[width=\columnwidth]{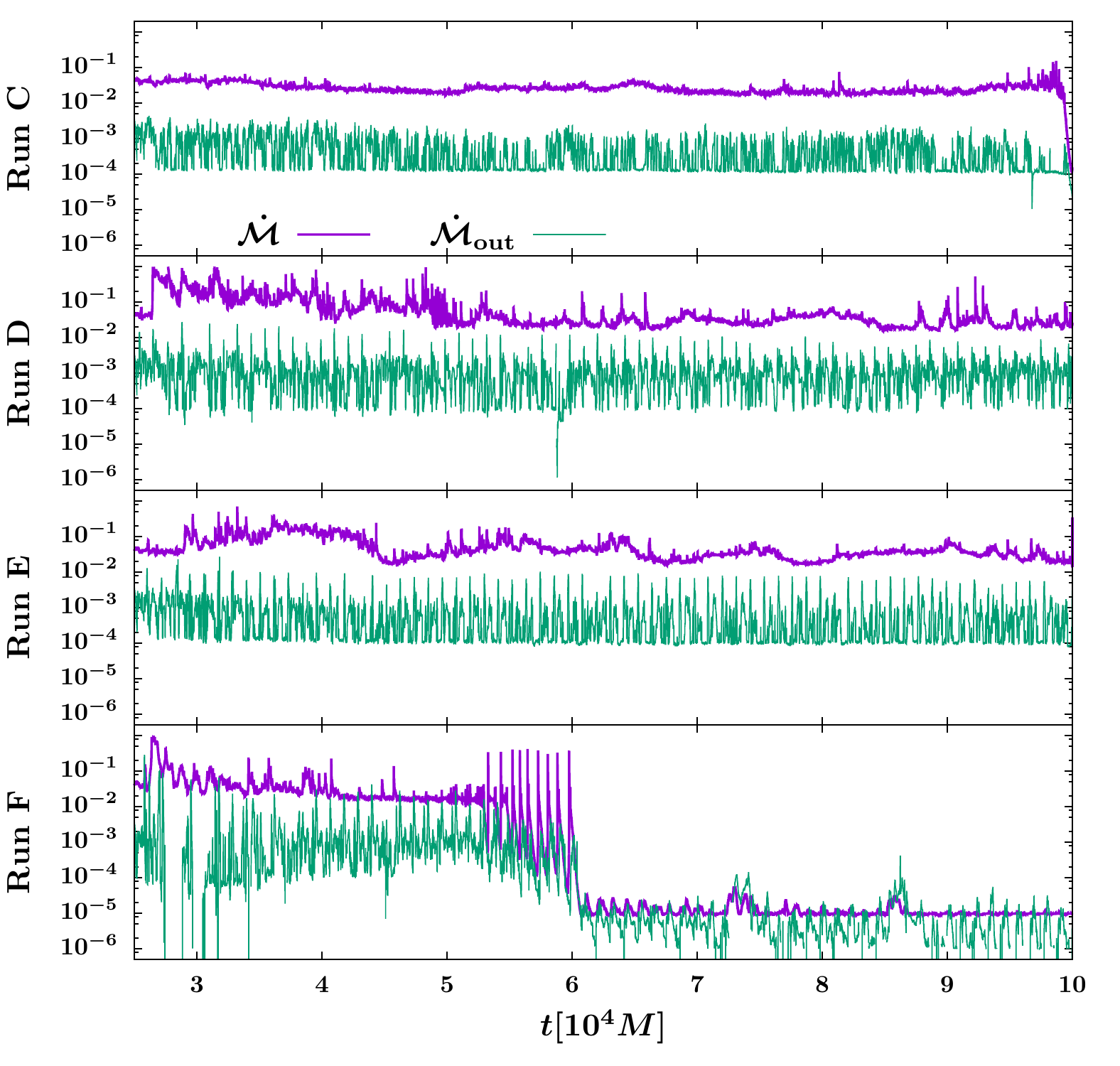}
\caption{Accretion rate $\dot{\mathcal{M}}$ and outflowing rate $\dot{\mathcal{M}}_{\rm out}$ of all 2D runs. The upper four panels show the shorter runs with  $t_{\rm end}= 5\cdot10^4 M$, while the bottom four panels show the longer runs with $t_{\rm end} = 1\cdot10^5 M$. }
\label{fig:Mdot-2D}
\end{figure}

In Run A the star moves on an almost spherical highly inclined orbit with low azimuthal angular momentum ($u_\phi = 0.4793$), so that it approaches close to the rotational axis. Each transit through the torus expels a blob of matter into the funnel region, where a large-scale magnetic field ordered along the rotational axis has developed already during NP1. In Fig.~\ref{fig:RA-slices} in the upper panel one of the first transits of the star is shown, which launches density waves propagating outwards in the torus. The density in the innermost region decreases, while the magnetic field lines are being entangled by the star. The accretion rate, which is plotted in Fig.~\ref{fig:Mdot-2D}, gradually drops down until $t=3\cdot 10^4\,M$ when it stabilises at a level more than three orders of magnitude lower than in NP1. On the corresponding slice in the middle panel of Fig.~\ref{fig:RA-slices} it can be seen that the star cuts down the inner part of the torus completely and the torus is only able to repeatedly split small blobs of gas into the black hole on the orbital time scale of the star. This reflects in the quasiperiodic peaks in the accretion rate. The last panel shows the state of the simulation at the end of the run, which is very similar to the middle panel. Because the torus is still large with a lot of matter and the evolution settles into a quasi-stationary state, we can expect that such behaviour will last on much longer time scales than the span of our simulation.

In the right plot in each panel the outgoing blobs of gas are seen on the plot of $\dot{m}_{\rm out}$ and they are present during the whole time span of the simulation. The blobs are moving along the boundary between the empty funnel and the torus with mildly relativistic speed and those going upwards are more pronounced than those moving downwards with progressing time. The animation of the whole run~A is available at \url{https://youtu.be/eERYioirgQc}.

The temporal profile of $\dot{\mathcal{M}}_{\rm out}$ is shown in Fig.~\ref{fig:Mdot-2D}, where the peaks correspond to the outflowing blobs. It can be seen that after the flow settles into a stationary state, the amount of the expelled matter is much larger than the amount of accreted matter, with the outflowing rate being more than one order of magnitude higher than the accretion rate. In the time period $t\in(3.5,5)\cdot10^4M$ almost 20 times more gas was expelled from the inner region than it was accreted onto the black hole (see Table~\ref{Table:runs}).

\begin{figure}
\includegraphics[width=0.49\columnwidth]{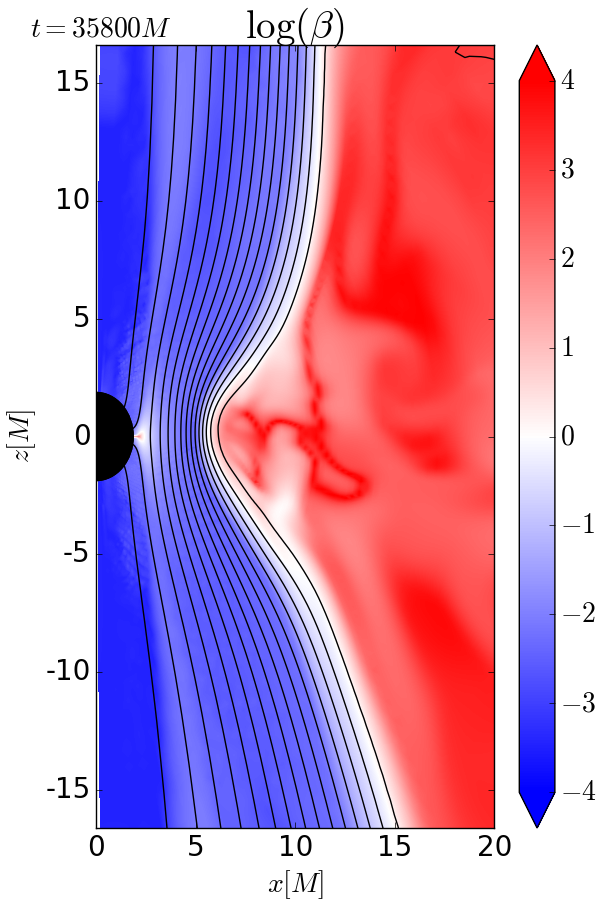}
\includegraphics[width=0.49\columnwidth]{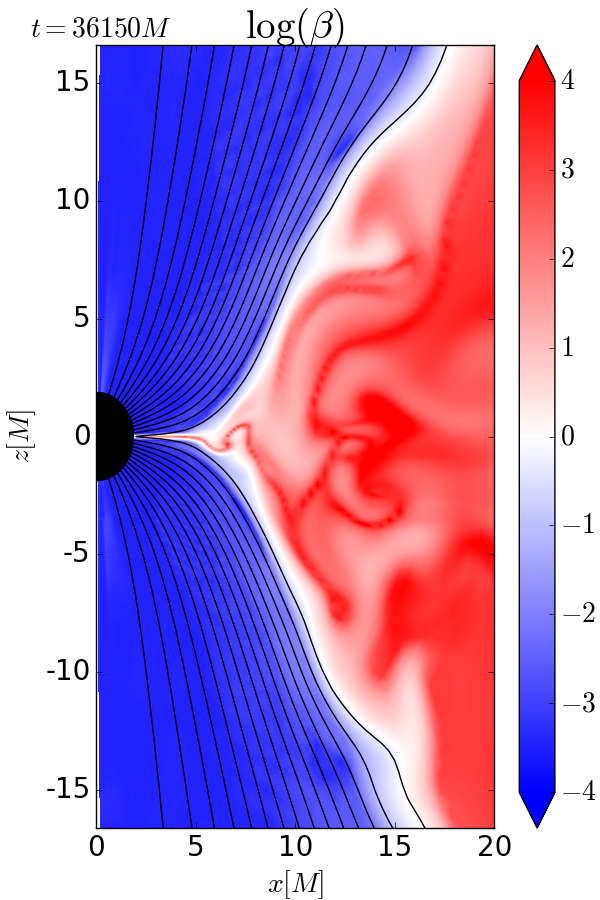}
\caption{The map of the gas-to-magnetic pressure ratio  $\beta$ of Run B in logarithmic scale at two time instances during the episodic accretion state. The white color corresponds to the equipartition region. {\bf Left:} The slice shows the MAD state, where the accretion is ceased by the large-scale magnetic field parallel with the axis. {\bf Right:} After few hundreds $M$ the accretion proceeds through narrow neck along the equatorial plane, the field lines are dragged with the flow into the black hole and they squeeze the flow into a thin layer. The evolution of density, Lorentz factor and fast-outflowing rate of the whole run B is available at \url{https://youtu.be/7Fz4g0lfpP4}.}
\label{fig:RB-beta}
\end{figure}


\subsubsection{Run B}
Run B follows a star embedded in the torus  that is moving in the radial range $r\in(15,25)\,M$. As the star moves through the gas, a bow shock appears along its path and expands into the torus as ingoing and outgoing density waves. These waves partially prevent the matter from falling into the inner region of the accretion flow. The density in the innermost region decreases as the accretion proceeds, the inner part of the torus for $r<10\,M$ is squeezed into a thin layer, which is enclosed by the large-scale magnetic field of the funnel. With progressing time equipartition between the magnetic energy density and the thermal energy density is achieved at the torus surface, see Fig.~\ref{fig:RB-beta}, where $\beta$ is shown in logarithmic scale and the white color corresponds to the equipartition region. The magnetic field close to the black hole is then strong enough to suppress the accretion. At about $t\sim3.2\cdot10^4\,M$ the torus detaches from the black hole, as the outermost magnetic field lines in the funnel reconnect. For a while large-scale magnetic field lines stretching from the bottom to the top of the simulation domain parallel to the axis exist. In other words, vertical magnetic-fields lines that do not enter the black hole emerge, see left panel of Fig.~\ref{fig:RB-beta}. After several hundreds of gravitational times the incoming gas pushes through the magnetic field, pulling the field lines back into the black hole and a blob of matter accretes onto the black hole (right panel of Fig.~\ref{fig:RB-beta}). The accretion then proceeds via the local interchanges of gas followed by repeated reconnection of the magnetic field lines. This process repeats on a time scale slightly larger than the orbital period of the star and leads to quasi-periodic peaks and drops in the accretion rate by three orders of magnitude. The evolution of density, Lorentz factor and fast-outflowing rate of the whole run B is available in the movie at \url{https://youtu.be/7Fz4g0lfpP4}.

\begin{figure}
\includegraphics[width=\columnwidth]{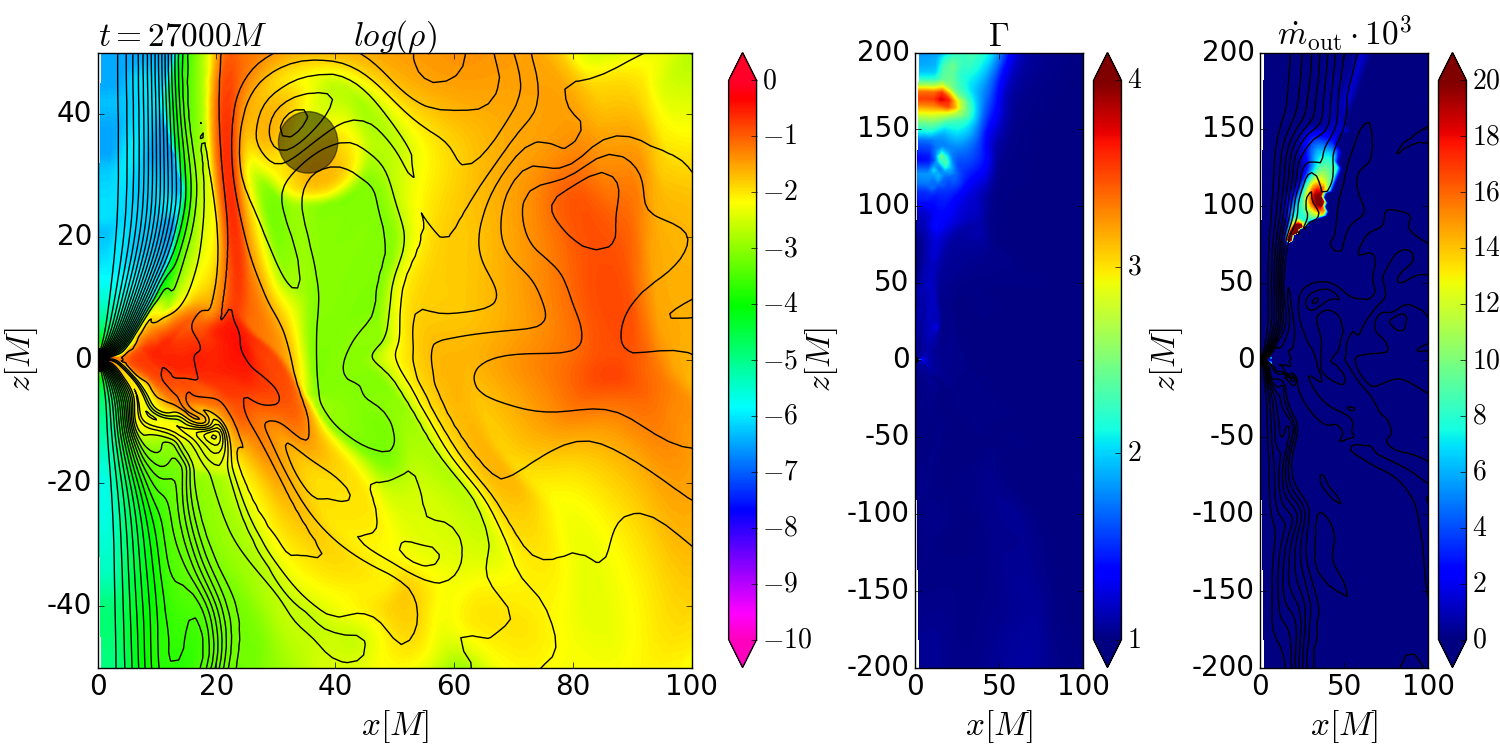}
\caption{Evolution of Run~F (radius of the star $\mathcal{R}=10\,M$). slice at $t=27000\,M$ showing the map of density in logarithmic scale, the Lorentz factor $\Gamma$ and the outflowing rate $\dot{m}_\textrm{out}$. }
\label{fig:RF-slices}
\end{figure}


\subsubsection{Runs C,D,E}
Run C presents the star on an orbit farther away from the black hole, ($r\in(25,50)\,M$), which is embedded inside the torus (it gets maximally up to $z=10\,M$ above the equatorial plane). The evolution proceeds in a similar way as in Run~B, only with longer time scales. The star again empties the part of the torus where it moves and detaches the inner and outer part of the torus. The inner part accretes and comes to a state of episodic accretion accompanied by abrupt drop of accretion rate just at the end of the simulation run.

Runs D and E show star on almost circular (precessing) orbits\footnote{An orbit that stays at a constant radius while having a strong inclination in Kerr space-time will have a precessing orbital plane. In return, it will trace out a section of a sphere of radius $r$. It would thus often be called a \textit{spherical orbit}.} with $r\sim50\,M$, the former is highly inclined going deep into the funnel, the latter coming only to the boundary between the funnel and the torus. 


\subsubsection{Run F}
Run F evolves a large star (radius $\mathcal{R}=10\M$) on the same orbit as in Run~E. It shows that for the perturbers with the largest possible effective radius (see Sections~\ref{sec:bowshock} and \ref{sec:pulsar} for relevant estimates) the effect on the accretion flow can be devastating. In Fig.~\ref{fig:RF-slices} the slice of the simulation at $t=2.7\cdot10^4\,M$ shows the moment when the star passed once down and up through the torus. It can be seen that even one such passage destroys the torus in the outer part and pushes a lot of material into the polar region. Later the motion of the star effectively cuts the torus into two parts and prevents the matter from the outer part to flow closer to the center. The outer part is dispersed by the star into large radii and has low density. The inner part of the torus feeds the accretion up to the time $t\sim5.5\cdot 10^4\,M$. At that moment the density in the inner torus decreases, the torus comes into the episodic accretion state which lasts for 9 cycles, after which the gas pressure in the innermost region is not sufficient to push through the magnetic field into the black hole. The torus detaches completely from the black hole from which it is separated by a large-scale magnetic field parallel to the rotational axis. The accretion rate thus drops down abruptly by four orders of magnitude and stabilizes at the level around $10^{-5}$.

\begin{figure}
\includegraphics[width=\columnwidth]{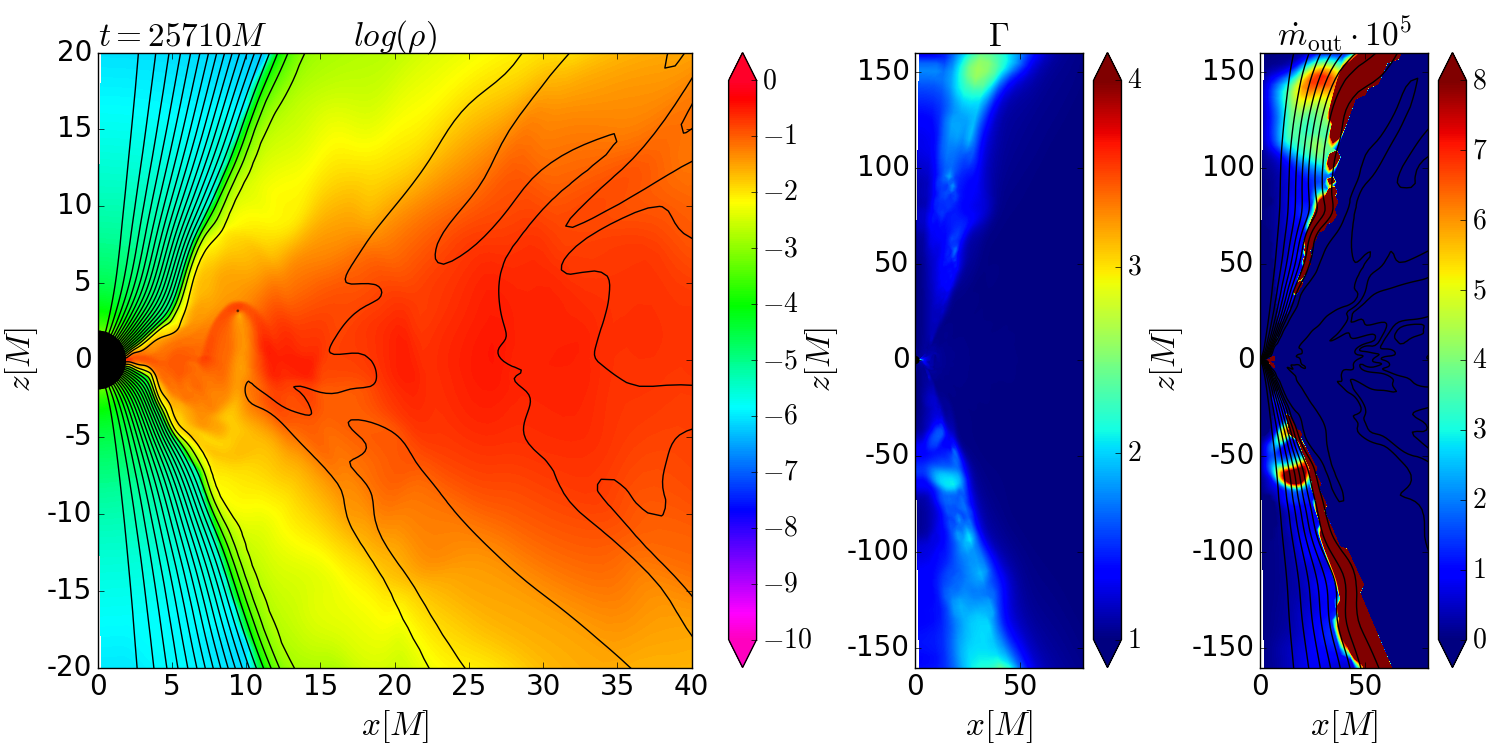}
\includegraphics[width=\columnwidth]{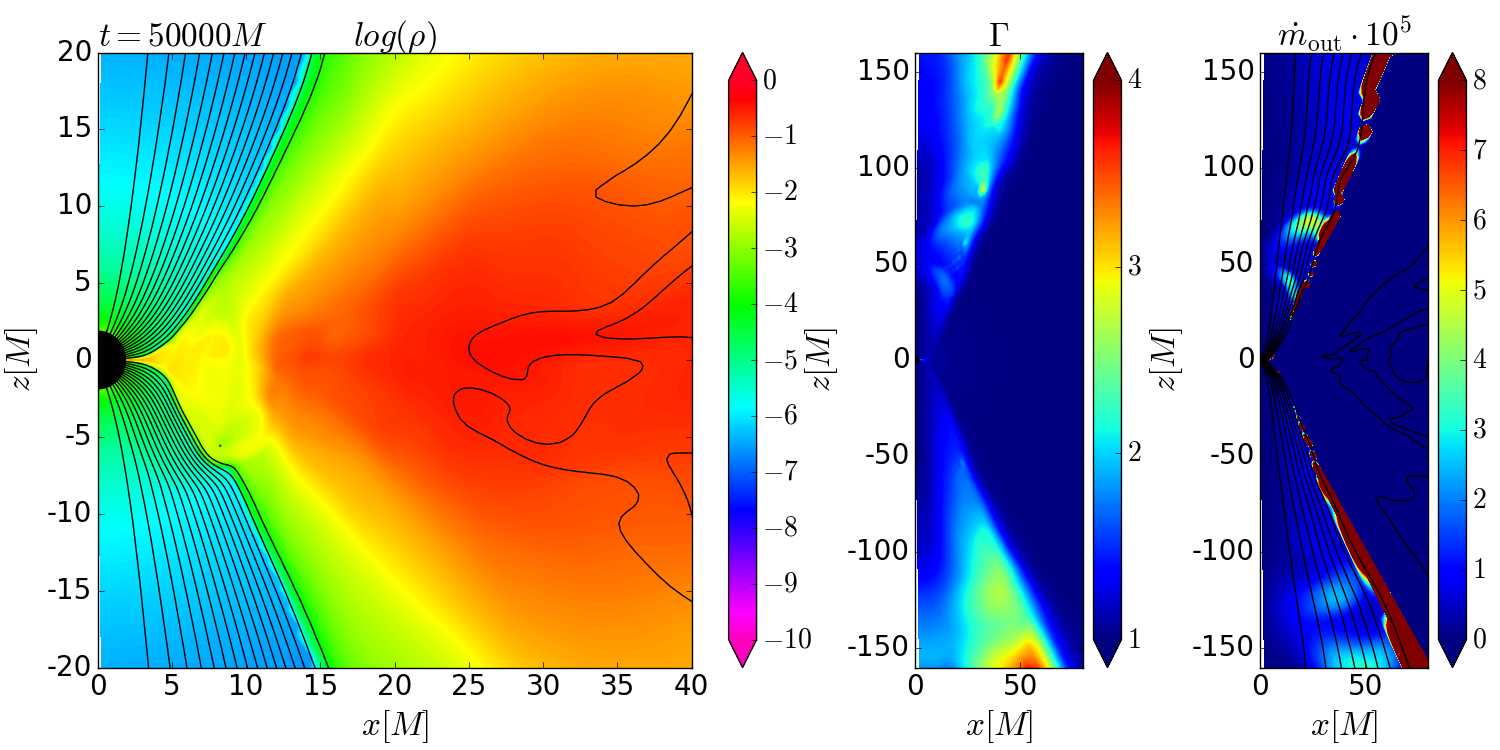}
\caption{Evolution of Run~G (radius of the star $\mathcal{R}=0.1\,M$). Slices at two different time instances show the map of density in logarithmic scale, the Lorentz factor $\Gamma$ and the outflowing rate $\dot{m}_\textrm{out}$. }
\label{fig:RG-slices}
\end{figure}


\subsubsection{Run G}
In contrast to run~F, run~G shows the case with the smallest radius of the star $\mathcal{R}=0.1\,M$, which has the same orbit as the star in Run~A. For Sgr A*, this setup corresponds to a Solar-type star moving very close to the black hole at $r\sim10\,M$. In fact, for star with $M_{\rm star}=1M_\odot, r_{\rm star}=0.9r_\odot$, the tidal radius is $r\sim16.1\,M$, so the chosen setup puts the star even below its tidal radius. We place the star on such a tight orbit, because the resolution of the grid is higher closer to the black hole and the small radius of the star is on the lower limit allowed by our resolution -- only 1~--~3 grid cells are occupied by the star volume at this radius. At some time instances, the star even does not  encompass any grid cell. However, despite that the motion of the star has a significant effect on the accretion flow. In Fig.~\ref{fig:RG-slices} the slice shortly after the launch of the star shows the bow shock generated by the star and the expanding density waves in the torus. The outflowing blobs are smaller in this case, but they are expelled until the end of the simulation as shown in the second slice in Fig.~\ref{fig:RG-slices}. The accretion rate increases at the beginning of the simulation and during $t\sim5000\, M$ it returns to levels slightly lower than in NP1. 


\subsubsection{Run H and resolution effects on reconnection} \label{sssec:resolution}
Finally, Run~H represents the case of an embedded star on a nearly-circular orbit near to the black hole with $\mathcal{R}=1M$. In this case, a similar state of episodic accretion as in Run~B is achieved quite shortly after the start of the simulation. Because the episodic accretion is accompanied by interchanges of the gas and the repeated reconnections of magnetic field lines, the resistivity of the material seems to play some role in the process. However, our code deals with the ideal MHD equations, hence it would seem that only their discretization on the finite grid can provide the numerical resistivity needed for this effect to work.
Therefore, we should explore to which degree the result will be affected by both the spatial and temporal discretization.

\begin{figure}
\includegraphics[width=\columnwidth]{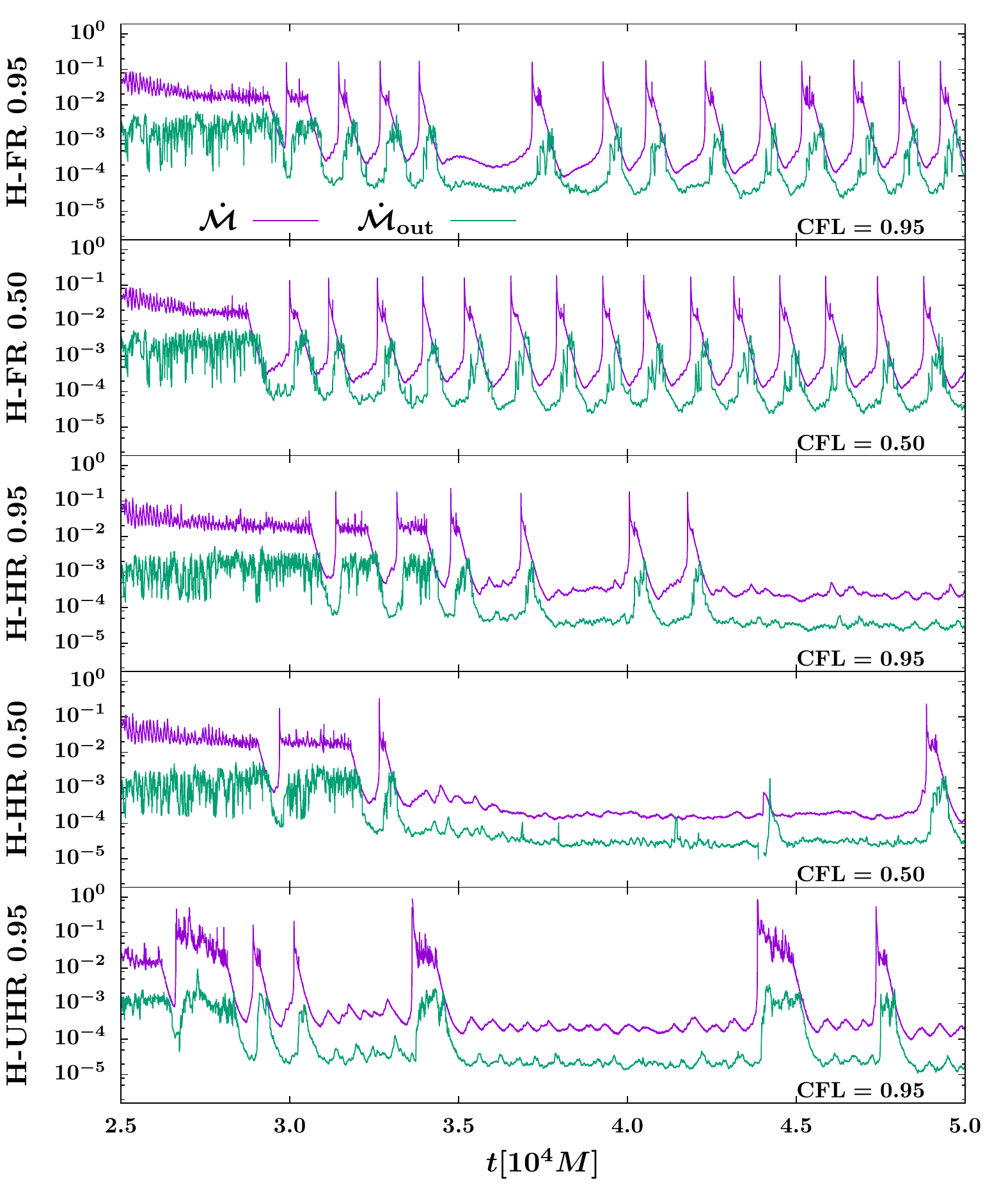}

\caption{Accretion rates of run H computed with different resolution and Courant number $c_{\rm CFL}$. The amount of accreted versus expelled matter is FR + $c_{\rm CFL}=0.95$: $\mathcal{M}_{\rm in}=207.8,\mathcal{M}_{\rm out}=19.4$; FR + $c_{\rm CFL}=0.5$: $\mathcal{M}_{\rm in}=199.6,\mathcal{M}_{\rm out}=17.9$; HR + $c_{\rm CFL}=0.95$: $\mathcal{M}_{\rm in}=227.3,\mathcal{M}_{\rm out}=14.1$; HR + $c_{\rm CFL}=0.5$: $\mathcal{M}_{\rm in}=185.8,\mathcal{M}_{\rm out}=11.2$; UHR + $c_{\rm CFL}=0.95$: $\mathcal{M}_{\rm in}=279.6,\mathcal{M}_{\rm out}=8.2$ in code units. }
\label{fig:Mdot-runH}
\end{figure}

To study this effect, we recomputed the same scenario with three different resolutions, the fiducial resolution (FR) of $n_r=252, n_\theta=192$, the high resolution (HR) set-up of $n_r=384, n_\theta=288$ and the ultra-high resolution (UHR) with $n_r=576, n_\theta=512$. We also varied the value of Courant number $c_{CFL}$ from the fiducial value $c_{CFL}=0.95$ to $c_{CFL}=0.5$. 

Similarly like we did in \citet{RAGtime2020} we check the quality factor $Q_z$, which quantifies the ability of the given grid discretization to describe the MRI properly. At the radius of the density maximum we find the number of cells in $\theta$-direction which cover the scale height $H$ of the torus - the height, at which the density decreases to $\rho_{\rm max}/e$. Then in line with \citet{Hawley_2011} we find the estimate of $Q_z$ according to the relation

\begin{equation}
    Q_z \eqsim 0.6 \, N_\theta \left(\frac{100}{\beta}\right)^{1/2}\left(\frac{\langle v_{\mathrm{A}z}^2\rangle}{\langle v_{\mathrm{A}}^2\rangle }\right)^{1/2} ,
\end{equation}
where $v_\mathrm{A}\propto B/\sqrt{\rho}$ is the Alfv\'en speed and $v_{\mathrm{A}z}$ its $z$-component. The general  requirement for a satisfactory resolution of the vertical MRI modes is that $Q_z\gtrsim 10$. In our geometry we have $B_r \sim B_\theta$ and $B_\phi=0$ and thus $v_{\mathrm{A}z} \sim v_\mathrm{A}$. Therefore, with our initial choice $\beta=100$, the value of $Q_z\sim12$ for FR,  $Q_z\sim18$ for HR and $Q_z\sim30$ for UHR, so that the initial MRI is well captured by our grid. However, as seen in Figure~\ref{fig:RB-beta}, in some of the regions of the torus interior one develops $\beta \sim 10^4$. In these regions the production of turbulence by MRI may be somewhat underrepresented later in our simulations. 

The resulting temporal profile of accretion rate for each version of Run~H is shown in Figure~\ref{fig:Mdot-runH}. The panels of the figure are ordered from the worst resolution (top) to the best one (bottom).
While the mean values of the high as well as the low accretion state do not change significantly with resolution, the duration of dips between successive accretion events does vary considerably, generally yielding longer dips for a better resolution.

In Figure~\ref{fig:Mdot-runH}, we also show the temporal behaviour of $\dot{\mathcal{M}}_{\rm out}$ computed at $r_{\rm diag}=100M$. 
During the high-accretion state at the beginning of the computation, the outflowing blobs of gas along the funnel translate into large and fast peaks of $\dot{\mathcal{M}}_{\rm out}$. Later, it is seen that the episodic accretion expels plasmoids into the funnel region, while during the dips the outflow is ceased. The time shift between $\dot{\mathcal{M}}$ and $\dot{\mathcal{M}}_{\rm out}$ is caused by the spatial distance between the position of the star, the black hole horizon and $r_{\rm diag}$.
The total amount of the expelled matter decreases with the resolution from 9\% to 3\% of the accreted gas.


\subsubsection{Effect on the torus shape}
\begin{figure}
\includegraphics[width=\columnwidth]{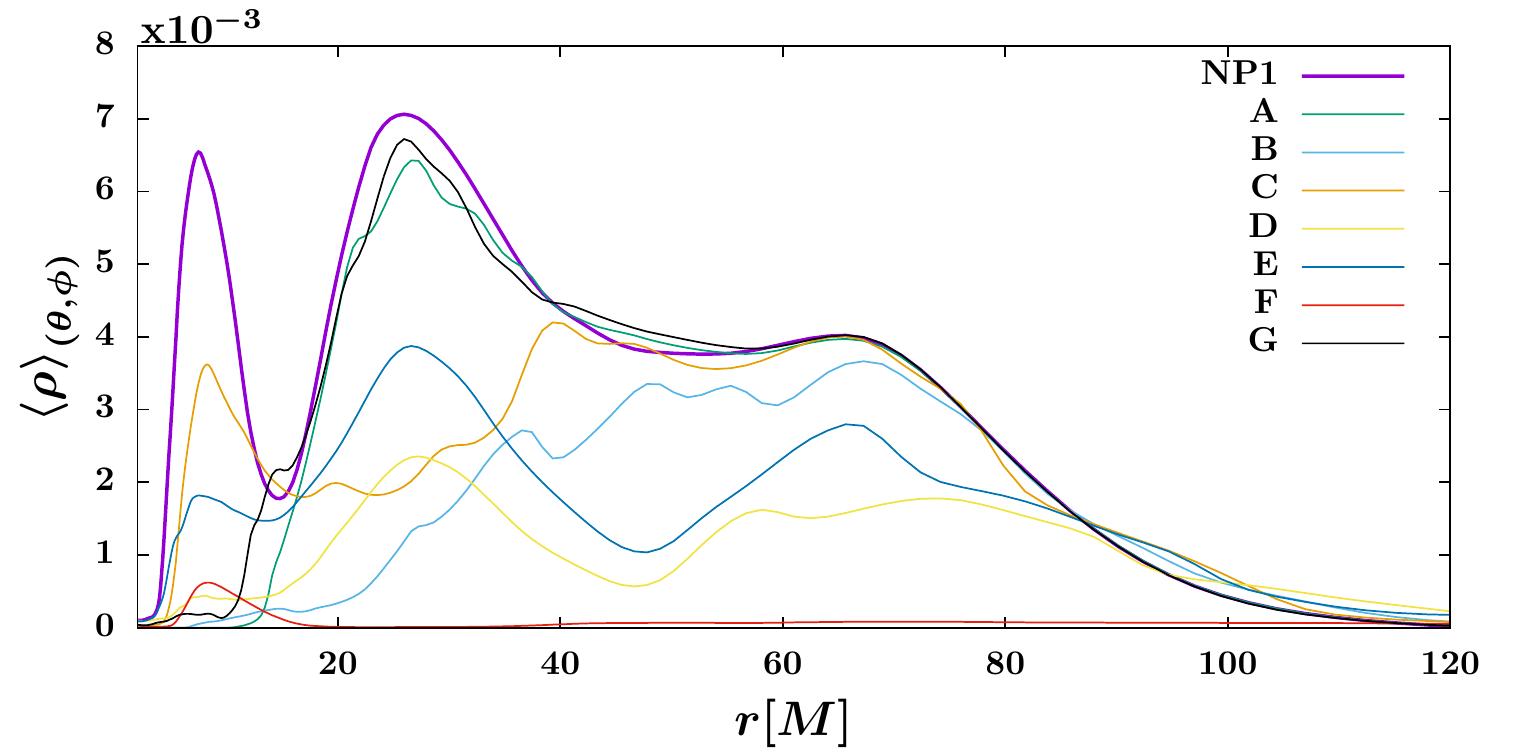}
\includegraphics[width=\columnwidth]{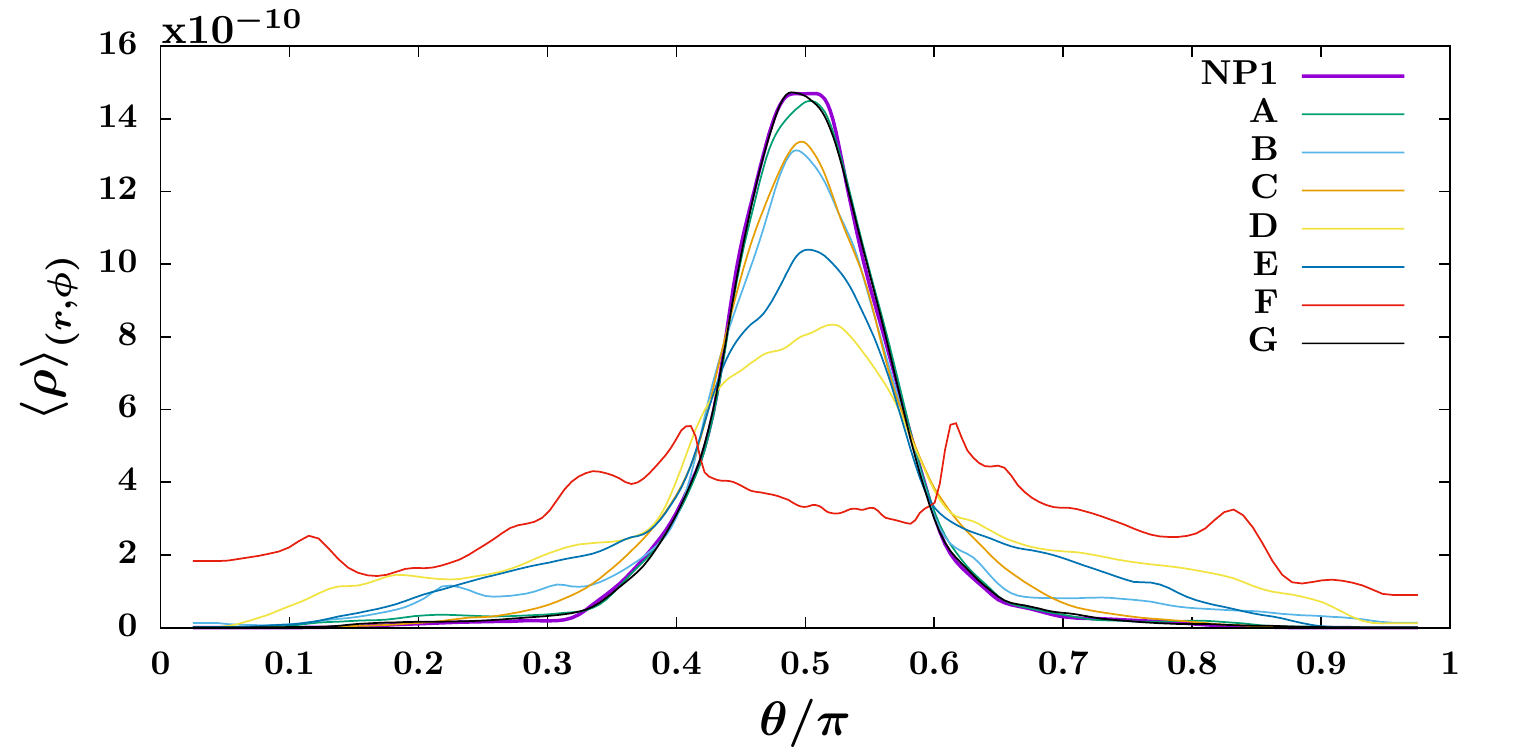}
\caption{The averaged radial and angular profile of density  at $t = 5\cdot10^4 \,M$ of 2D runs. }
\label{fig:rho-profile-2D}
\end{figure}

The effect of the star passing through the torus can be seen in the distribution of matter that is gradually established inside the torus. In Fig.~\ref{fig:rho-profile-2D} we show the averaged radial and angular profile of density computed according to the relations
\begin{eqnarray}
\left<\rho\right>_{(\theta,\phi)}(r) &=& \frac{\int^{2\pi}_0 \int^{\pi}_0 \rho \sqrt{-g} \,{\rm d}\theta {\rm d}\phi}{\int^{2\pi}_0 \int^{\pi}_0 \,\sqrt{-g}\, {\rm d}\theta {\rm d}\phi}, \\
\left<\rho\right>_{(r,\phi)}(\theta) &=& \frac{\int^{2\pi}_0 \int^{R_{\rm out}}_0 \rho \sqrt{-g}\, {\rm d}r {\rm d}\phi}{\int^{2\pi}_0 \int^{R_{\rm out}}_0 \sqrt{-g} \,{\rm d}r {\rm d}\phi},
\end{eqnarray}
where the averages are computed for a snapshot at a single value of $t$.
In each run, the presence of the star clearly imprints into the density profile. In general, the density is reduced mainly in the radial range of the star motion. In case of the star moving very close to the black hole (Runs~A and G), the density in the inner part is low, while the outer part of the torus is left almost intact. For Run~A, the very low value of $\left<\rho\right>_{(\theta,\phi)}$, which is $10^{2} - 10^{5}$ times smaller than in NP1 up to $r\sim 10\,M$, corresponds to a significant decrease in the accretion rate. In Run~G, where only the radius of the star is smaller, the density is $10^{1} - 10^{2}$ times lower than for NP1 and the accretion rate drops only by a factor of a few. The angular density profile of both runs do not differ from the Run~NP1 noticeably. 

In Runs~B and C, the decrease of density stretches to larger radii, as the star moves further away from the black hole. At the same time the angular profile shows a decrease of density along the equatorial plane and a slight increase at the intermediate angles, hence the torus puffs up.  

Runs~D and E, in which the star moves almost on a near-circular orbit with $r_{\rm star}\sim50\,M$, shows a significant dip in the density profile corresponding to the star position. However, the density profile is decreased along almost the whole radial extent of the torus, only the outermost parts are similar to NP1. In the angular direction, the tori are more dissolved and some gas can be found even in the highest inclinations in the funnel region.

The destruction of the torus in Run~F is documented by very low values of averaged radial density, while on the angular profile no peak in the equatorial plane is found, hence the gas is dispersed in all directions. 


\subsection{3D runs}\label{sec:3D}
\begin{figure}
\includegraphics[width=\columnwidth]{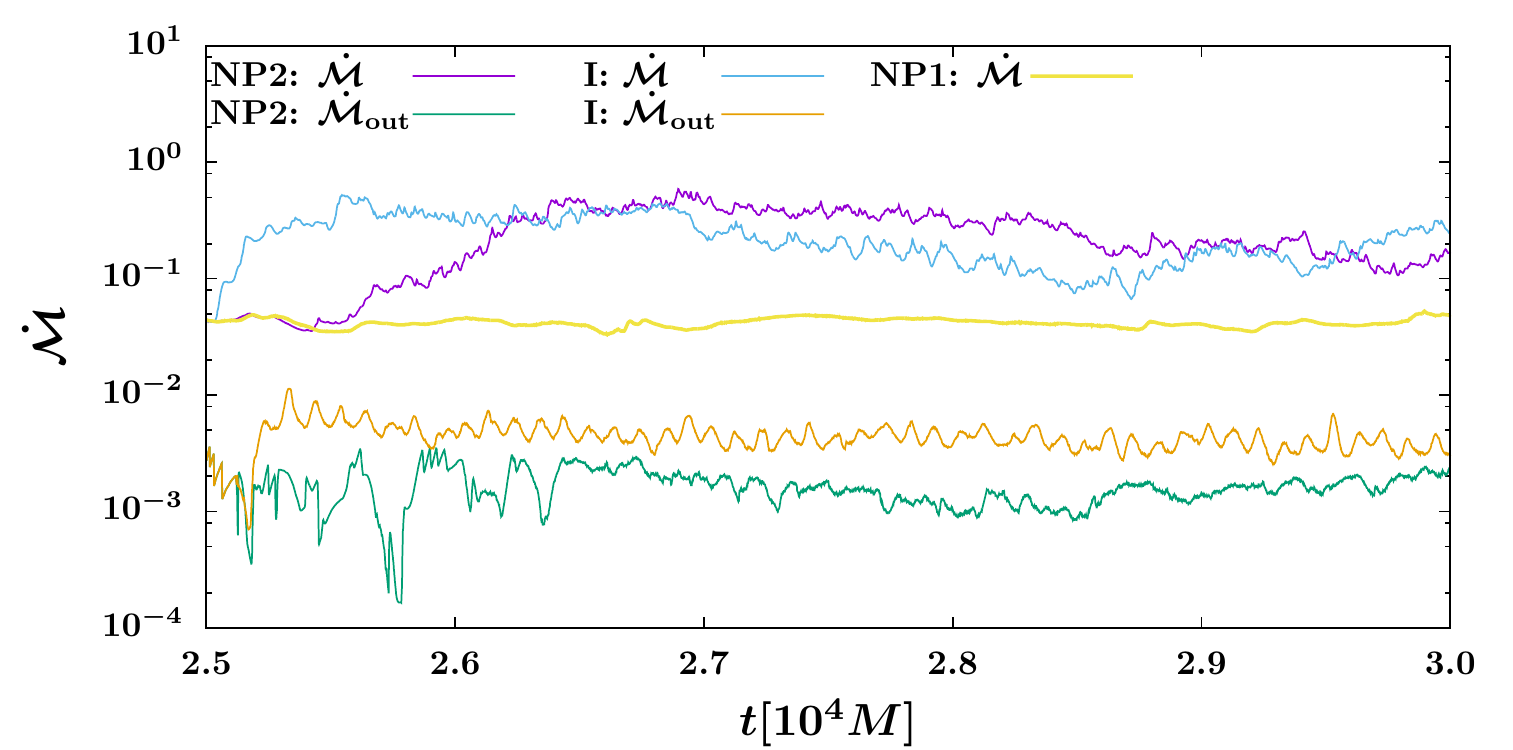}
\caption{The temporal profile of accretion rate $\dot{\mathcal{M}}$ and outflowing rate $\dot{\mathcal{M}}_\textrm{out}$ for the non-perturbed 3D Run~NP2 and perturbed Run~I. The accretion rate of the non-perturbed 2D Run~NP1 is given for comparison. The outflowing rate in Run~I is about 2-3 times higher than in NP2 and shows clear variability caused by the matter expelled in the funnel by the star.}
\label{fig:Mdot-3D}
\end{figure}
3D runs are much more computationally demanding, therefore, our 3D runs were computed for significantly lower total times, and we carried out only a single run with a perturbing star. The trajectory of the star in Run~I is the same as in Run~A, which was computed in 2D. Additionally, when we compare the evolution time that was needed in the 2D runs to achieve the new quasi-stationary accretion state, we conclude that we probably do not reach the stationary state in during our 3D evolutions. However, we can observe several features emerging in the simulations that support the results obtained in longer 2D runs, as well as some differences such as in the amounts of influenced and expelled gas. 

To be able to better compare the perturbed and unperturbed evolution of the flow we ran the 3D non-perturbed Run~NP2, which was also initializated by the state of NP1 at $t_{\rm start} = 25 000 \, M$. The total integration time is $t_{\rm run}=5000\, M$. The resolution is $n_r=252, n_\theta=192, n_\phi = 96$.

The comparison of the accretion rates $\dot{\mathcal{M}}$ of the non-perturbed Runs~NP1 and NP2 and the perturbed Run~I is given in Fig.~\ref{fig:Mdot-3D} together with the outflowing rate $\dot{\mathcal{M}}_\textrm{out}$ of NP2 and I. 

It is interesting to note that the accretion rate enhanced significantly at the beginning of the run even without the perturbing star. This is likely associated with the establishment of the turbulent structure in the azimuthal direction $\phi$. Later, however, the value decreases and seems to approach the values from the 2D simulation. The presence of the star speeds up the rise of the accretion rate, which reaches its maximal value on a shorter time scale than NP2, but the maximal level is comparable in both cases. From the short-term simulations, we are not able to conclude how the settled state looks like. However, the presence of the star is imprinted into the timing properties of the accretion rate, which we will discuss in the following Section. 

The outflowing rate in Run~NP2 is approximately 200-100 times smaller than the accretion rate. The outflowing rate in Run~I is cca 40 times lower than the accretion rate and about 2-3 times higher than in NP2 and it shows clear peaks corresponding to the outflowing blobs expelled by the star. 

The trajectory of the star in Run~I is the same as in Run~A, which was computed in 2D. In figure~\ref{fig:Mdot-2D} we can see that at $t=3\cdot10^4\,M$, Run~A has just reached the settled state, in which the outflowing rate is about 10 times higher than the accretion rate. It is thus seen that the effect of the star passage is enhanced in the 2D simulations by about 2-3 orders of magnitudes, which corresponds to the fact that a real star perturbs just a small portion of the torus in the $\phi$-direction, not the full 2$\pi$ angle as it is implicitly assumed in the 2D approximation.


\subsection{Periodicity and Power Spectral Density analysis}

\label{subsec_periodicity_psd}

\begin{table*}
    \centering
    \begin{tabular}{ccccc}
    \hline
    \hline
    Run  & periodogram peak $\dot{\mathcal{M}}$ $[M^{-1}]$ &HFPSD slope $\dot{\mathcal{M}}$&  periodogram peak $\dot{\mathcal{M}}_{\rm out}$  &HFPSD slope $\dot{\mathcal{M}}_{\rm out}$ \\
    \hline
    A     & $10^{-2}$\,$(100)$ (narrow peak) & $-1.39 \pm 0.17$ & $5\times 10^{-3}$\,$(200)$ (narrow peak)   & $-2.06 \pm 0.23$ \\
    B     & $1.04 \times 10^{-3}$\,$(962)$ (broad peak) & $-1.28 \pm 0.02$ & $1.03 \times 10^{-3}$\,$(971)$ (multiple peaks)  & $-2.26 \pm 0.04$  \\
    C     & $2.4\times 10^{-4}$\,$(4167)$ (broad peak) & $-1.73 \pm 0.01$ & $1.42 \times 10^{-4}$\,$(7042)$ (multiple peaks) & $-1.76 \pm 0.01$  \\
    D     & ---  (no clear peak) & $-1.47 \pm 0.01$& $9\times 10^{-4}$\,$(1111)$ (narrow peak)   & $-2.05 \pm 0.01$\\
    E     &  --- (no clear peak) & $-2.66 \pm 0.03$& $1.79\times 10^{-3}$\,$(560)$ (narrow peak)   & $-0.90 \pm 0.05$  \\
    F     &  $8.9\times 10^{-4}$\,$(1124)$ (narrow peak) & $-3.82 \pm 0.10$& $8.94 \times 10^{-4}$\,$(1119)$ (narrow peak) & $-2.53 \pm 0.05$\\
    G     & $10^{-2}$\,$(100)$ (narrow peak) & $-3.87 \pm 0.08$& $10^{-2}$\,$(100)$ (narrow peak) & $-2.26 \pm 0.03$\\
    H     & $7.24 \times 10^{-4}$\,$(1381)$ (broad peak) &  $-1.33 \pm 0.01$& $7.84 \times 10^{-4}$\,$(1276)$ (broad peak)  & $-2.16 \pm 0.03$\\
    I(3D) & $9.01\times 10^{-4}$\,$(1110)$ (broad peak) & $-2.41 \pm 0.09$& $10^{-2}$\,$(100)$ (narrow peak)  & $-2.64 \pm 0.09$\\
    NP2(3D) &  ---  (no peak)  &  $-1.83 \pm 0.01$&  --- (no peak)    & $-2.23 \pm 0.08$  \\
    \hline
    \end{tabular}
    \caption{Summary of the highest frequency peaks (with the corresponding period in parenthesis) for individual runs. We characterize the peak according to the width. If there is no well-defined peak in the periodogram, we do not include it. The second and the third columns list periodogram peaks and the slopes of high-frequency power spectral densities (HF PSD, for $\nu>10^{-2}\,M^{-1}$) for the inflow rate. The fourth and the fifth columns contain the same information (periodogram peaks and the slopes of HF PSD) for the outflow rate.}
    \label{tab_periodicity}
\end{table*}

\begin{table}
    \centering
    \begin{tabular}{cccc}
    \hline
    \hline
    Run & $f_r [10^{-3} M^{-1}]$ & $f_\theta [10^{-3} M^{-1}]$ & $f_\phi [10^{-3} M^{-1}]$\\
    \hline
    A     &  $3.26$ & $5.00$ & $5.16$ \\
    B     &  $1.50$ & $1.76$ & $1.78$ \\
    C     &  $0.627$ & $0.680$ & $0.683$ \\
    D     &  $0.423$ & $0.451$ & $0.451$ \\
    E     &  $0.421$ & $0.447$ & $0.449$ \\
    F     &  $0.421$ & $0.447$ & $0.449$ \\
    G     &  $3.26$ & $5.00$ & $5.16$ \\
    H     &  $3.55$ & $4.83$ & $4.97$ \\
    I(3D) &  $3.26$ & $5.00$ & $5.16$ \\
    \hline
    \end{tabular}
    \caption{Characteristic stellar orbital frequencies $f_{r}$, $f_{\theta}$, and $f_{\phi}$ are given for each run and they are expressed in geometrized units. The azimuthal frequency is important only for the 3D case (denoted I), while in 2D computations (A--H) the $\phi$-dimension has been disregarded by projecting the trajectories onto the poloidal plane. 
    }
    \label{tab_star_periodicity}
\end{table}

The passage of a star through the accretion flow during an orbital period can potentially lead to periodic or the quasi-periodic signals that could be detected via the periodicity analysis of light curves in the X-ray, optical/UV, and radio domains.
\citet{1999PASJ...51..571S} proposed that, under suitable circumstances, stellar transits can produce the desired modulation of the inner regions of an AGN accretion torus. If the central black hole rotates and the companion star is at a low orbit, the Lense-Thirring orbital precession will show up in the signal and parameters of the central black hole can be thus determined. Further, \citet{2010MNRAS.402.1614D} examined simulated light curves of flares generated by these inelastic collisions between a star bound to a supermassive black hole and the accretion flow. The authors confirmed that the behaviour of the quasi-periodicity is influenced by the mass and spin of the black hole together with the orbital elements of the stellar orbit. Among the most promising candidates is the well-known quasar OJ 287 \citep{1988ApJ...325..628S,2016ApJ...819L..37V,2018MNRAS.478.3199B} and the bright Seyfert I galaxy RE J1034+396 \citep{2008Natur.455..369G,2010A&A...524A..26C,2020MNRAS.495.3538J,2021MNRAS.500.2475J}. Likewise, a similar mechanism including orbiting plasmoids or hot spots was proposed to explain the modulation detected in some bright flares from SgrA* supermassive black hole \citep{2017MNRAS.472.4422K}. In particular, some near-infrared flares of Sgr~A* appear to exhibit quasiperiodic features with a period of $17-20$ min that could be interpreted by an infalling, sheared hot spot \citep{2006A&A...460...15M,2003Natur.425..934G}, although the significance of $\sim 20$ min peak in the periodogram has been questioned and found insignificant \citep{2008ApJ...688L..17M,2009ApJ...691.1021D}. However, apart from periodicities, stars embedded in the accretion flow can affect and ``shepherd'' the accretion state, in particular by inducing the MAD state, as we have shown e.g. for Run~H.

To examine the mentioned effect, we consider the computed accretion rate $\dot{\mathcal{M}}$ as well as the outflow rate $\dot{\mathcal{M}}_{\rm out}$ as proxies for the light curves. For all the runs listed in Table~\ref{Table:runs}, we analyse the periodic properties of $\dot{\mathcal{M}}$ and $\dot{\mathcal{M}}_{\rm out}$ using the multi-harmonic analysis of the variance \citep[MHAOV;][]{1996ApJ...460L.107S}. The MHAOV method uses the expansion into periodic orthogonal polynomials and applies the analysis of variance statistic to assess the fit quality. It is thus suitable also for nonsinusoidal pulsations (in comparison with discrete Fourier transform) and it is efficient in suppressing aliases that may arise due to a regular sampling of the signal. In the periodicity analysis, we focus on the frequence range between $f_{\rm min}=10^{-4}\,M^{-1}$ up to the Nyquist limit, which is usually $f_{\rm max}=0.5M^{-1}$ due to the $1M$ step size. The most prominent frequency peaks in the MHAOV periodograms are summarized in Table~\ref{tab_periodicity}.


\subsubsection{Periods in accretion rate}

 Let us now discuss the periodogram peaks of the mass-acretion time series $\dot{\mathcal{M}}(t)$. For Runs~A and G, we detect a clear narrow peak at $f=10^{-2}\,M^{-1}$, which is more prominent for Run A. These runs are characterized with the star orbiting at the close distance of $r=10\,M$, with Run G having just ten times smaller stagnation radius. Run~H, where the star orbits at $10\,M$ as for Run~A but is embedded within the torus, does not exhibit a clear peak at $f=10^{-2}\,M^{-1}$. Instead, there is a broader peak at lower frequencies corresponding to period $\sim 1400\,M$, which is related to episodic accretion events. Then for Runs B and C, we can detect broader peaks at longer timescales of the order of 1000$M$, which is due to larger orbital distances of the embedded star -- $r=15-25\,M$ and $r=25-50\,M$ for Runs B and C, respectively. 
 
 Runs D and E are not characterized by a clear peak in the periodogram, which can be related to a large orbital distance of the star ($r=50\,M$) and a relatively small stagnation radius of $1\,M$ in comparison with that. For Run F, where the star is embedded within the flow at the same orbital distance of $r=50\,M$, we can see a clear broader peak corresponding to period $\sim 1124\,M$, which is at twice the orbital frequency $f_\theta$. In comparison with Runs D and E, the presence of a clear peak is related to a ten-times larger stagnation radius of the star. Hence, both the proximity of a star and the length-scale of its stagnation radius play a role in inducing quasi-periodic features in the accretion rate. In other words, the ratio between the stagnation radius and the distance $\mathcal{R}/r$ is of an importance.

\begin{figure*}
    \centering
    \includegraphics[width=\columnwidth]{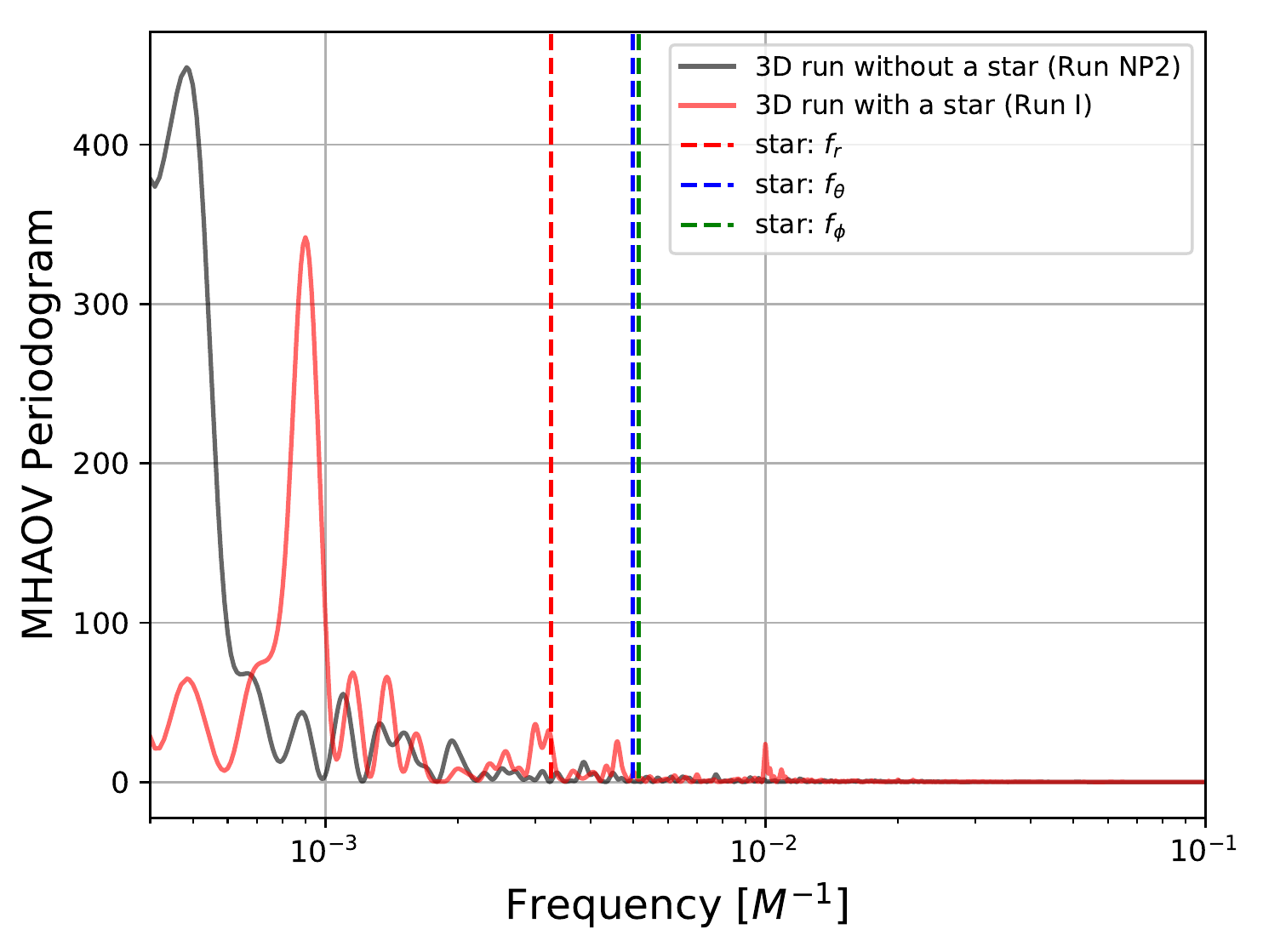}
    \includegraphics[width=\columnwidth]{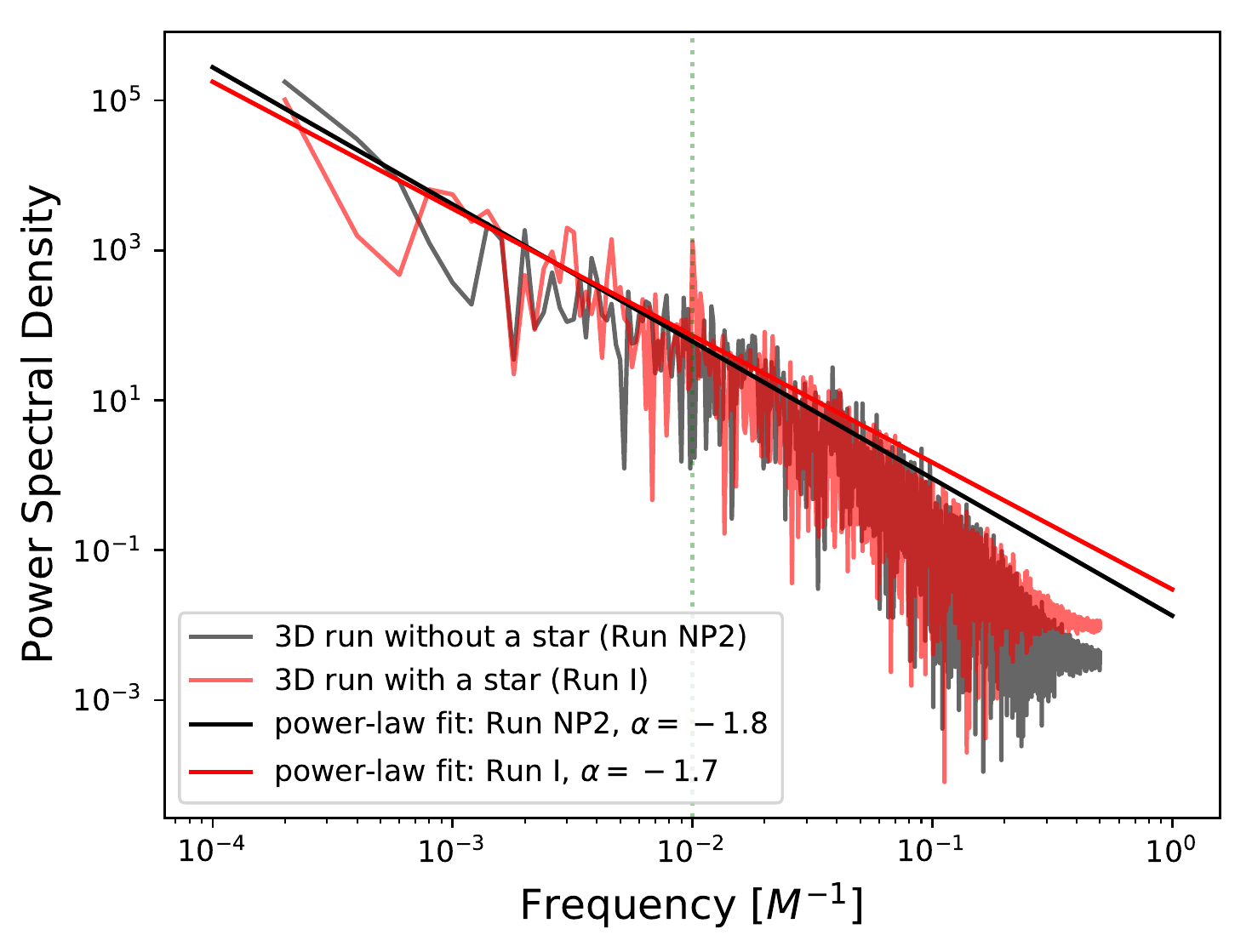}
    \caption{Comparison of the unperturbed 3D run (Run~NP2) with the 3D run perturbed by the star (Run~I). {\bf Left panel:} Comparison of the MHAOV periodogram between the 3D run without a star (NP2; black solid line) and the 3D run with a star (Run I; red solid line). The perturbed Run~I exhibits more pronounced periodogram peaks, especially at $10^{-2}\,M^{-1}$ and close to $10^{-3}\,M^{-1}$. For comparison, we also depict by vertical dashed lines characteristic stellar frequencies $f_r$, $f_{\theta}$, and $f_{\phi}$ according to Table~\ref{tab_star_periodicity}. The small peak at $10^{-2} M^{-1}$ corresponds to $2 f_\theta$. {\bf Right panel:} The power spectral density (PSD) for Run~NP2 (black solid line) and Run~I (red solid line). At intermediate frequencies, both PDS can be described by a simple power-law function with the slope of $\alpha=-1.8$ and $\alpha=-1.7$ for Run~NP2 and Run~I, respectively. The PDS for Run~I has a clear peak at $10^{-2}\,M^{-1}$, which corresponds to twice the orbital frequency of the star.}
    \label{fig_comparison_runI_NP}
\end{figure*}

The orbiting star can induce periodic features also in the 3D accretion flow, although with more complex properties than for 2D runs. In Fig.~\ref{fig_comparison_runI_NP}, we compare the 3D run without any star (black solid line) with the 3D run perturbed by a star (red solid line). Each of these runs lasts about 5000$M$. For Run I (with a star), we see a clear peak at $1110\,M$, which is not present for Run NP2 (without a star). In addition, for Run I we can also detect the peak at higher frequencies at $10^{-2}\,M^{-1}$, which is clearly present for the 2D analogue represented by Run A.

\subsubsection{Periods seen in outflow rates}

\begin{figure*}
    \centering
    \includegraphics[width=\columnwidth]{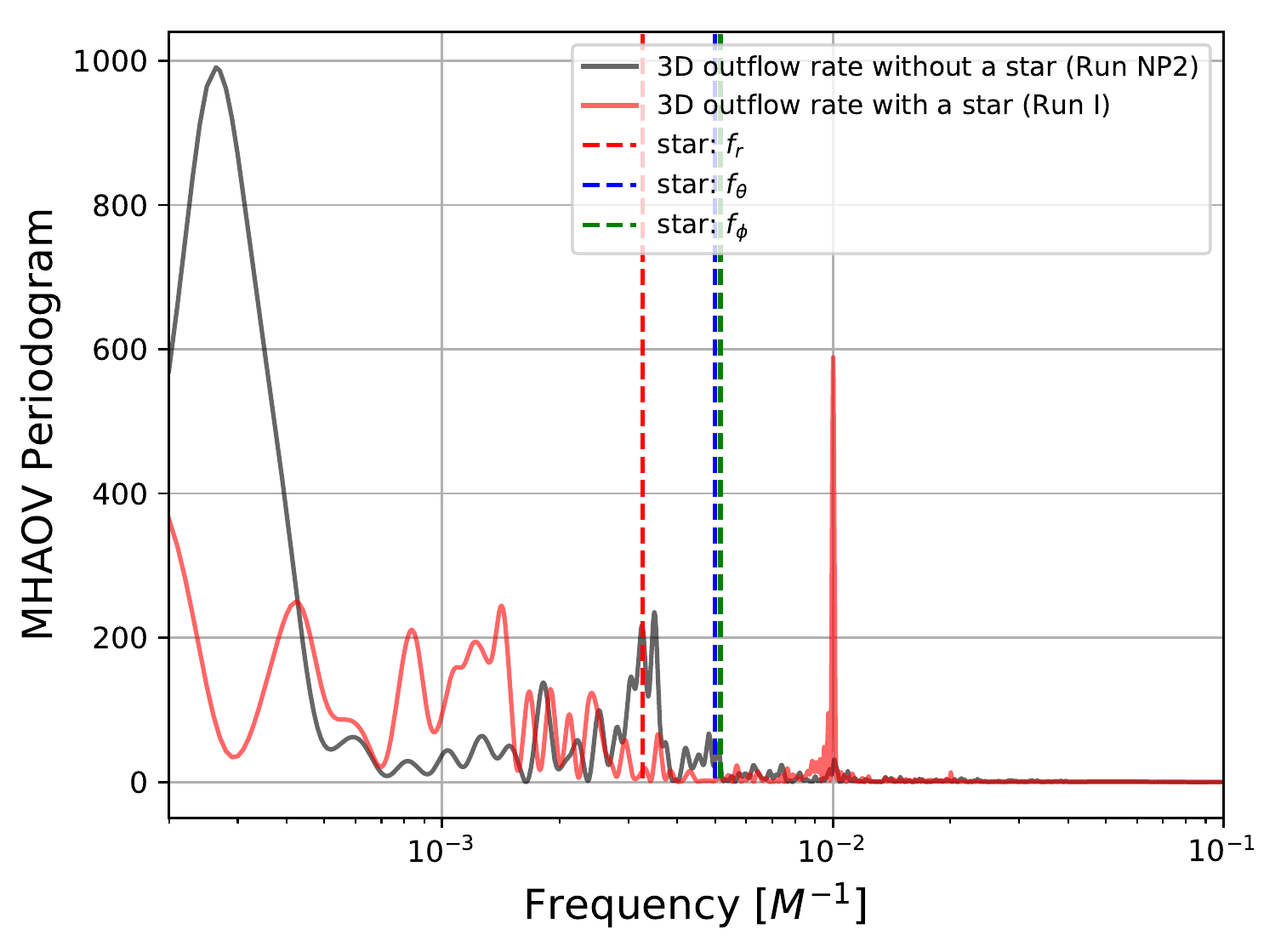}
    \includegraphics[width=\columnwidth]{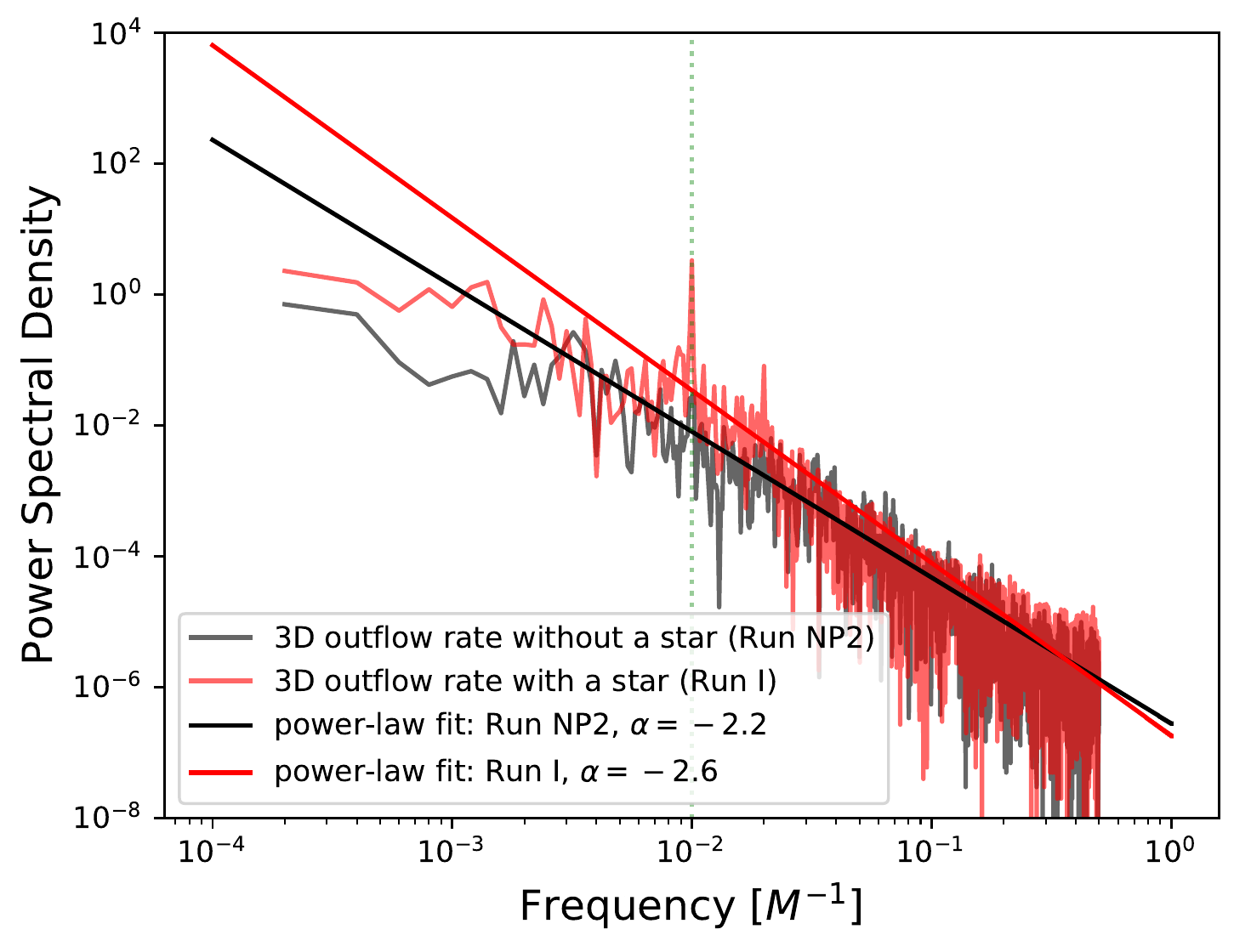}
    \caption{Comparison of the outflow rate for the unperturbed 3D run (Run~NP2) with the 3D run perturbed by the star (Run~I). {\bf Left panel:} Comparison of the MHAOV periodogram between the 3D run without a star (NP2; black solid line) and the 3D run with a star (Run I; red solid line). The perturbed Run~I exhibits a clear peak at $10^{-2}\,M^{-1}$, which corresponds to the twice the vertical orbital frequency $f_\theta$. For comparison, we also depict by vertical dashed lines characteristic stellar frequencies $f_r$, $f_{\theta}$, and $f_{\phi}$ according to Table~\ref{tab_star_periodicity}. {\bf Right panel:} The power spectral density (PSD) for Run~NP2 (black solid line) and Run~I (red solid line). At $\nu>10^{-2}\,M^{-1}$, both PSDs can be described by a simple power-law function with the slope of $\alpha=-2.2$ and $\alpha=-2.6$ for Run~NP2 and Run~I, respectively. The PSD for Run~I has a clear peak at $10^{-2}\,M^{-1}$, which corresponds to twice the orbital frequency of the star.}
    \label{fig_comparison_runI_NP_outflow}
\end{figure*}

The periodic behaviour is also clearly visible in the outflow rate. In Fig.~\ref{fig_comparison_runI_NP_outflow} (left panel), we compare the outflow rate for the 3D unperturbed run (run NP2; black solid line) with the 3D run perturbed by the star (run I; red solid line). In the periodogram, we can see a clear peak at $10^{-2}\,M^{-1}$ for the perturbed run I, which corresponds to twice the orbital frequency of the star.

We also analysed the periodicity properties of $\dot{\mathcal{M}}_{\rm out}$ for all 2D runs, see the most prominent periodicity peaks in Table~\ref{tab_periodicity}. In general, we see that the peaks are consistent with the peaks for $\dot{\mathcal{M}}$. There are, however, also clear differences. For instance, for run~A, the periodicity peak of $\dot{\mathcal{M}}_{\rm out}$ is half of the inflow-rate best frequency, i.e. being the same as the orbital frequency of a perturbing star. In general, the frequency peaks for $\dot{\mathcal{M}}_{\rm out}$ are more clearly defined in the periodogram than for the inflow rate, even for the runs where we do not see a clear peak for the inflow rate, in particular runs D and E. This is due to the fact that the outflow rate is calculated for a limited range of $\theta$ and restricted outflow velocities greater than $0.5c$. In this way, the denser and the slower material of the torus is effectively filtered out.  

\subsubsection{Power spectral density of accretion rate}

The power spectral densities (PSD) of the $\dot{\mathcal{M}}$ time series exhibit a power-law profile, with the change in a slope towards the smallest and the highest frequencies, i.e. the overall profile is a broken power law. In the intermediate range between $10^{-3}\,M^{-1}$ and $10^{-1}\,M^{-1}$, the PSD is described well with a single power-law, $p(\nu)\propto \nu^{\alpha}$, with the slope between $\alpha=-1$ and $\alpha=-4$. 

In Fig.~\ref{fig_comparison_runI_NP} (right panel), we compare PSDs of the unperturbed 3D Run~NP2 (black line) with the perturbed 3D Run~I (red line). In both cases, the fitted power law has the slope close to $\alpha=-2$, $\alpha=-1.8$ for Run~NP2 and $\alpha=-1.7$ for Run~I,  which is between the pure red-noise variability ($\alpha=-1$) and the random-walk process ($\alpha=-2$), see e.g. \citet{1995A&A...300..707T}. The 2D runs have a comparable PSD power-law slope for $\nu>10^{-2}\,M^{-1}$: Runs A, B, C, D, and H between $\alpha=-1$ and $\alpha=-2$, while Runs E, F, and G exhibit a steeper high-frequency slope between $\alpha=-2$ and $\alpha=-4$. When Run~I and Run~NP2 are restricted to higher frequencies ($\nu>10^{-2}\,M^{-1}$), the power laws are also steeper with $\alpha=-2.41\pm 0.09$ and $\alpha=-1.83 \pm 0.01$, respectively. The power-law slopes of $\dot{\mathcal{M}}$ PSDs for all the runs in the frequency range $\nu>10^{-2}\,M^{-1}$ are summarized in Table~\ref{tab_periodicity} (third column). 

The perturbed Run~I exhibits a clear peak in the PSD at $10^{-2}\,M^{-1}$, see Fig.~\ref{fig_comparison_runI_NP} (right panel), which corresponds to twice the orbital frequency of the star, i.e. the star perturbs the flow twice during one orbit around the SMBH. This peak is also present in the MHAOV periodogram in the left panel of Fig.~\ref{fig_comparison_runI_NP}. This comparison shows that the presence of a star in the proximity of the SMBH does not affect the general red-noise and random-walk variability of SMBHs (exhibited generally by AGN). However, it can clearly introduce periodic and quasi-periodic features that can be revealed via the PSD and periodogram analysis of light curves.


\subsubsection{Power spectral density of outflow rate}

Power spectral densities of the outflow rate $\dot{\mathcal{M}}_{\rm out}$ also generally have a power-law character. In Table~\ref{tab_periodicity} (fifth column), we list high-frequency ($\nu>10^{-2}\,M^{-1}$) power-law slopes of PSDs for all the runs. The mean value of the slopes is comparable with the one for the inflow rate, however, the standard deviation is half of the one for the inflow rate, $\overline{\alpha}_{\rm out}=-2.09 \pm 0.46$ vs. $\overline{\alpha}_{\rm in}=-2.18 \pm 0.94$. The lower dispersion of the power-law slopes for the outflow-rate PSDs can be interpreted by restricting the outflow-rate to the material with the velocity larger than $0.5c$. The slopes of the inflow and the outflow PSDs are consistent with the inferred slopes between $-2.6$ and $-3.3$ for the nearby AGN monitored in the optical domain by the Kepler mission \citep{2011ApJ...743L..12M}, as well as the PSD slope for Sgr~A* in the infrared domain, which is between $-2$ and $-3$ \citep{2009ApJ...691.1021D}. This shows that steeper slopes detected for both AGN and quiescent SMBHs are overall consistent with the accretion driven by the magneto-rotational instability.  

In Fig.~\ref{fig_comparison_runI_NP_outflow} (right panel), we calculate PSDs for the 3D runs I and NP2 denoted by a solid red line and a solid black line, respectively. The PSD is fitted well with a single power-law and there is no noticeable break towards the higher frequencies in comparison with the inflow PSDs in Fig.~\ref{fig_comparison_runI_NP}. However, in comparison with the power laws fitted to the 3D inflow rates, the outflow-rate PSDs have steeper power laws with $\alpha=-2.64 \pm 0.09$ and $\alpha=-2.23 \pm 0.08$ for Run~I and Run~NP2, respectively. 

For the Run I perturbed with a star, we can detect a clear narrow peak in the PSD at $\nu=10^{-2}\,M^{-1}$ in comparison with the unperturbed run NP2. This frequency peak, which corresponds to twice the orbital frequency of the star (two passages through the accretion flow per orbital period), is also clearly revealed in the MHAOV periodogram in the left panel of Fig.~\ref{fig_comparison_runI_NP_outflow}. Furthermore, the frequency peak of $10^{-2}\,M^{-1}$ is better defined and more significant according to the periodogram value than for the inflow rate (Fig.~\ref{fig_comparison_runI_NP}, left panel).   


\subsubsection{Matching to orbital period of the star}

A free test particle in the Kerr field has three, generally independent frequencies of motion $f_r,f_\theta,f_\phi$ \citep{schmidt2002celestial} and this is also true for the stellar orbit. All of the three frequencies can appear in the periodogram of the 3D run with the perturbing star. However, only the $f_r,f_\theta$ frequencies can in principle appear in the 2D runs, since the dynamics in the azimuthal direction are suppressed there both for the star and for the flow.  

\begin{figure*}
    \centering
    \includegraphics[width=\columnwidth]{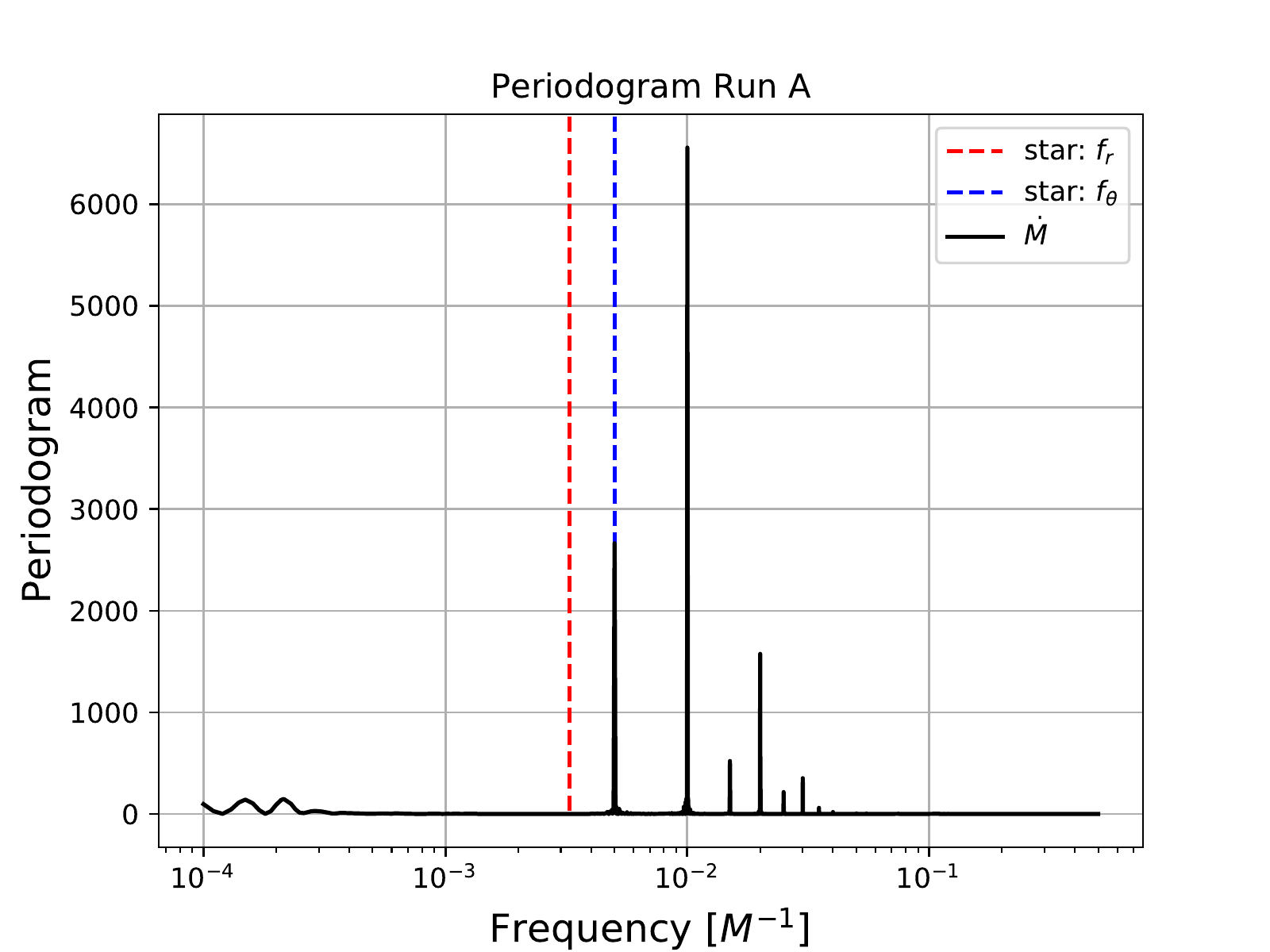}
    \includegraphics[width=\columnwidth]{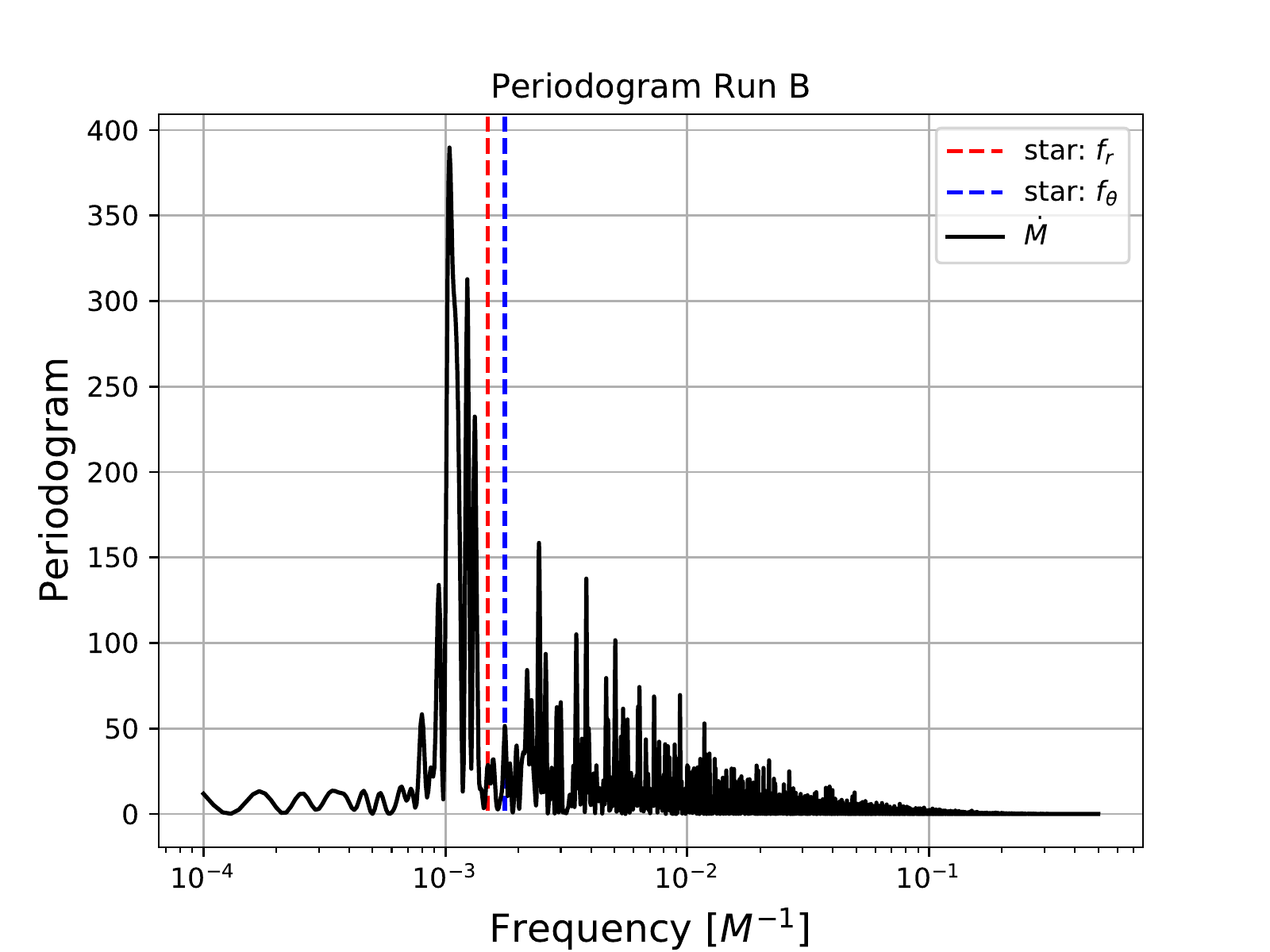}
    \caption{Periodograms for the accretion rate as well as $r$ and $\theta$ directions of the stellar motion. {\bf Left panel:} Run A with the accretion-rate periodogram plotted with a black line, while the characteristic stellar frequencies in the $r$ and the $\theta$ directions, $f_r$ and $f_{\theta}$, are plotted with the red and blue dashed lines, respectively. All the sharp peaks correspond to multiples of $f_\theta$, the highest peak to $2 f_\theta$. {\bf Right panel:} The same as in the left panel, but for Run B. Here the relation of the peaks to the orbital frequencies is unclear.} 
    \label{fig_periodogram_AB}
\end{figure*}

Let us start with the 2D runs. For Run A, we obtained $f_{r}=3.264 \times 10^{-3}\,M^{-1}$ and $f_{\theta}=5\times 10^{-3}\,M^{-1}$, while for Run B, we got $f_{r}=1.5 \times 10^{-3}\,M^{-1}$ and $f_{\theta}=1.756\times 10^{-3}\,M^{-1}$. In Fig.~\ref{fig_periodogram_AB}, we compare the $\dot{M}$-periodogram with the periodicities related to $r$ and $\theta$ directions. For Run A, we find coincident peaks at $5\times 10^{-3}\,M^{-1}$ between $\dot{M}$ and $\theta$ periodograms, which corresponds to the orbital period of the star of $200\,M$. The main peak for $\dot{M}$ periodogram is at $10^{-2}\,M^{-1} = 2 f_\theta$, which is due to the fact that the star is passing twice through the accretion flow during its orbital period. 

For Run B, the periodogram comparison is not trivial, since the star is located generally farther away ($15-25\,M$) on an elliptical orbit within the disk. The characteristic orbital frequencies for the $r$ and $\theta$ coordinates are larger than the highest peak for $\dot{M}$,  $f_{\dot{M}}=1.04\times 10^{-3}\,M^{-1}$. In fact, there is a gap in the periodogram at the orbital frequencies of the perturber, which would suggest some sort of destructive interference with the waves in the disk. However, at this point it is not clear whether this is simply a coincidence or a more generic effect.

In general, the episodic accretion, when it develops within the perturbed flow, shows longer periodicity than the orbital frequencies of the star, but its temporal properties depend on the discretization of the equations, as was discussed in Section \ref{sssec:resolution}. This is in particular the case for Run~H, where the dominant period for both the inflow and the outflow rate is $1300-1400\,M$, while the stellar period is shorter, see Table~\ref{tab_periodicity} and \ref{tab_star_periodicity} for comparison. 

Run~I (3D) shares similarities to its 2D-analog Run~A, in particular we can detect the periodicity $\sim 10^{-2}\,M^{-1}$, see Fig.~\ref{fig_comparison_runI_NP} and \ref{fig_comparison_runI_NP_outflow}, which is clearly more significant for the outflow rate than for the inflow rate. This period is clearly equal to twice the orbital frequency $f_\theta$. For the inflow rate, there is a prominent peak at lower frequencies at $9.01\times 10^{-4}\,M^{-1}$\,$(1110\,M)$, which could be related to less frequent quasiperiodic inflows regulated by the disappearance of the MAD state. On the other hand, frequencies below $\sim 10^{-3} M^{-1}$ correspond to phenomena with periods comparable to the entire length of the run, so they cannot be studied precisely here. For an overview of prominent inflow and outflow periodicities for all runs, see Table~\ref{tab_periodicity}, while for the list of stellar frequencies $f_r$, $f_{\theta}$, and $f_{\phi}$, see Table~\ref{tab_star_periodicity}.

\section{Discussion} \label{sec:discussion}


We have shown that the transit of a star through the accretion flow has a significant effect on the closest neighbourhood of the black hole for a broad range of stellar orbits. We focused on stars that move in the innermost part of the flow at the radii between 10 - 50 gravitational radii. Such orbits are on the edge of the tidal radius of the star, depending on the mass of the black hole and both the stellar radius and the mass. On the other hand, the orbital period in this region is sufficiently small, so that we can follow the long-term evolution of the flow, which is repeatedly perturbed by the successive passages of the star. 

As the star moves through the accretion flow, a bow shock is formed along its path, which is expanding into the torus both towards and outwards from the black hole in the form of density waves. 
At first, these waves enhance the accretion rate, pushing some of the gas into the black hole. The waves are partially reflected on the torus-funnel boundary and bounce on the strongly ordered magnetic field in the funnel back into the torus, which is accompanied by ripples of the boundary expanding from the center. Moreover, the inclined stars push matter into the funnel along their path as they penetrate the empty region. The gas is then accelerated by the magnetic field and departs in the form of blobs along the boundary. 

Later, the matter is depleted from the region where the star moves, and the accretion rate consequently drops and settles at lower levels. The decrease is higher for inclined stars at smaller radii, where we have observed a drop by three orders of magnitude, which was confirmed by computations with a high as well as a ultra-high resolution.
The influence of the grid resolution and the used approximation of the star has been studied for Runs A, B and G in detail in a separate paper \citet{RAGtime2020}. There we have shown that the results do not depend strongly on the particular shape of the perturbed region, which approximates the star, nor on the resolution (fiducial resolution versus high resolution), except of the case of episodic accretion and the MAD state.


\subsection{Emergence of the MAD state} \label{subsec:MADdiscuss}
For several cases of an embedded star, a state of episodic accretion develops, where the accretion proceeds only through spitting smaller amounts of matter into the black hole here and there. 
This state is characterized by very pronounced peaks and dips in the accretion rate, which can be as high as three orders of magnitudes. The duration of both the peaks and the dips can be significantly longer than the orbital frequency of the star. However, their temporal properties, in particular the duration of the dips, depend on the resolution of the grid. 

During the episodic accretion the torus evolves into the so-called MAD state (Magnetically Arrested Disc)
\citep{2003ApJ...592.1042I,Narayan-MAD-2003} and  the matter is released from the inner edge of the torus by the magnetic reconnection of the field lines and the interchange instability of the gas, which is held from falling into the black hole by the strong vertical magnetic field.
Therefore, the proper description of the mechanism of this effect should include resistivity of the material. Our code, however, works with ideal MHD equations and thus relies only on the numerical resistivity (and hence on the properties of the grid) to deliver such effects. 
We have shown the dependence of the exact temporal profile of the accretion and the outflowing rate on the discretization in Section \ref{sssec:resolution}. 

There are two ways how to deal with this issue in the numerical simulations. Either one can include the proper description of resistivity into the equations, thus employ a resistive version of HARM, or the numerical properties of the grid can be chosen such that the value of numerical resistivity corresponds to the physical one.
The evaluation of numerical viscosity and resistivity of the code is, however, a complex issue. \citet{Rembiasz_2017} have suggested an ansatz for the form of the numerical viscosity and resistivity of Eulerian codes, which depend on the spatial and the temporal discretization of the equations, generally with different exponents given by the order of the numerical schemes, and also on the characteristic length and the characteristic velocity of the system. However, finding the corresponding numerical parameters requires extensive testing of the code on some well-known problems and subsequent fitting of the results. 

A more thorough testing of the code should still be carried out in future, nevertheless, based on the order-of-magnitude comparisons between our setup and that of \citet{Rembiasz_2017}, we deduce that in our case the value of numerical resistivity exceeds the physical one. One of possible ways to lower the numerical resistivity without the need for excessively high grid resolution would be to employ ultra-high-order reconstruction schemes \citep{Rembiasz_2017}, which appears to to be capable of capturing the MRI already when the typical length of the system is covered by only 10 zones for 9th order reconstruction scheme.

On the other hand, if we understand our simulation as a large-eddy simulation, we are neglecting the turbulent gradients occuring on sub-grid scales and thus also a good deal of dissipation \citep[see e.g.][]{2015SSRv..194...97M}. Consequently, our code may be too dissipative in the ordered parts of the simulation domain (such as the funnel), but not sufficiently dissipative in the turbulent parts (such as the disk interior). Additionally, there is quite a number of studies of the MAD accretion state with a similar resolution as ours, which also do not take the physical resistivity into account and rely only on the numerical resistivity similarly as we do \citep[e.g.][]{2008ApJ...677..317I,Sasha-smart-grid,10.1111/j.1365-2966.2012.21074.x,10.1111/j.1365-2966.2012.22002.x,10.1093/mnras/stt1881,10.1093/mnras/stt1860}. The conservative scheme using the total energy equation ensures that the energy released by the reconnection is accounted for in the form of heat. \citet{2003ApJ...592.1042I} include a prescription for resistivity, which however also depends on the grid spacing and is employed to account for the heat production for the same reason.

The development of the MAD state of the accretion torus depends on the initial distribution of the magnetic field. In particular, the large-scale ordered magnetic field, consisting of one big loop covering the whole torus (used in our case) leads to the accumulation of magnetic field near the horizon and a quick evolution of the MAD state, while the configuration of small loops with changing polarity rather produces a SANE (Standard And Normal Evolution) accretion state, where episodic accretion is not seen \citep[e.g.][]{10.1093/mnras/stt1881}. On the other hand, it is expected that the MAD state can be obtained in astrophysical systems by dragging in the magnetic field of the material from larger distances from the center ($\sim10^5\,M$). This process takes much longer time than we can afford to cover by the numerical simulations, hence the initial conditions which prefer the quick establishment of the MAD state in this sense serve as a shortcut to achieve an evolved accretion flow with reasonable numerical demands \citep{Sasha-smart-grid}. 

Taking into account all these remarks, it is a matter of further study to address the existence of the episodic accretion state and its temporal properties, where the resistivity of the material as well as the influence of the initial configuration will be taken into account in a more accurate way.


\subsection{Effect of black hole spin and radiative cooling} \label{subsec:radiative-cooling}

In our simulations, we have seen that the amount of expelled gas was usually around $10\%$ of the amount of accreted gas, with the exception of Run A, where in the settled state, about 20 times more gas was expelled than accreted. Here we should however point out that we use the fiducial value of the spin parameter $a=0.5$. Because it is expected that with increasing value of spin, the acceleration of the material in the funnel region is more effective, the ratio of the accreted versus expelled material is probably spin-dependent. The spin dependence of the efficiency of the jet launched in the MAD state was shown e.g. by \citet{Sasha-smart-grid, 10.1111/j.1365-2966.2012.21074.x}, where the efficiency raised from about $30\%$ for $a=0.5$ to $130-150\%$ for $a=0.99$. 

Another disadvantages of our approach is that we do not take into account the radiative transfer in our simulations. Such an approach is appropriate for the low-luminous sources, where the accretion rate is several orders of magnitude below the Eddington luminosity, in the so-called advection-dominated accretion flows (ADAFs) or more generally radiation-inefficient accretion flows (RIAFs), where the low efficiency can be attributed also to other physical phenomena than just advection \citep{2014ARA&A..52..529Y}. Sgr~A* in the center of our Galaxy is considered as a prime example of such low luminous galactic nuclei containing a hot and diluted inflow \citep[see][for reviews]{2010RvMP...82.3121G,2017FoPh...47..553E}. 

However, recent GRMHD simulations with radiative cooling have shown that even in the case of an accretion rate as low as $10^{-7} M_{\rm Edd}$, the disc shape and accretion rate can be altered by the radiation. \citet{2012MNRAS.426.1928D} showed that by neglecting the dynamical effects of the radiative cooling for a Sgr-A$^*$-like black hole, they obtained significantly different observable spectra of the disk for $\dot{\mathcal{M}}>10^{-7} \dot{M}_{\rm Edd}$. Additionally, \citet{2017ApJ...844L..24R}, using a more sophisticated radiation model and simulations near an M87-like black hole, observed that the total radiative efficiency started deviating noticeably due to the dynamical effects of the radiation when $\dot{\mathcal{M}} > 10^{-5} \dot{M}_{\rm Edd}$. Finally, \citet{Sasha-radiative-cooling} compared the geometry of the cooled and the non-cooled state of the accretion torus and for $\dot{M} \gtrsim 10^{-7} \dot{M}_{\rm Edd}$ they found that cooling led to an increase in the mid-plane density and a pile-up of matter close to the black hole horizon along with the strengthening of the magnetic field, while the turbulence and MRI were suppressed. As a result, the scale-height of the torus was smaller and the effective opening angle of the funnel a little bit larger, which would make more orbits in our simulations inclined rather than embedded in the disk. Nevertheless, the overall structure of the magnetic field with a strongly ordered field along the axis in the funnel remained unchanged. Hence, we can presume that the outcome of the stellar passage through the accretion flow would have qualitatively similar features also for accretion flows with accretion rates up to $\dot{\mathcal{M}} \sim 10^{-7} \dot{M}_{\rm Edd}$ or even $\dot{\mathcal{M}} \sim 10^{-5} \dot{M}_{\rm Edd}$. 

A more substantial change of the picture can be assumed for the case of a cold thin disc accretion regime with the accretion rates of the order of a few to a few tens per cents of the Eddington accretion rate $\dot{M}_{\rm Edd}$, which is supposed to be found in many luminous AGNs. In order to study the perturbation of the Keplerian disc by the star, the dynamical effect of the radiative cooling on the gas dynamics has to be taken into account, which will be the topic of our following study.

Going to even higher accretion rates that are equal to or even exceed the Eddington accretion rate, the advection again becomes more important, more of the viscously generated heat is advected into the black hole and the standard thin disc transforms in a slim disc solution. 


\subsection{Connection to observational results -- I: Prospects for star-flow interactions} \label{subsec:discussion-observation-I}

The characteristic temporary features in our 2D and 3D runs are plasmoids that appear during episodic accretion events following the reconnection of poloidal magnetic field lines. During the episodic accretion, some plasmoids fall into the black hole, while others escape with mildly relativistic velocities close to the rotational axis along the funnel-torus boundary.


\subsubsection{Case of Sgr A* black hole in Galactic Centre}

The presence of one or more stars close to the SMBH helps to establish the MAD state by the clearance of the flow as well as by dragging magnetic field lines. The MAD state can also be present for the low-luminous Sgr~A* system where the ordered, poloidal magnetic field is present as manifested by the rotation of the polarization vector of near-infared flares \citep{2018A&A...618L..10G}. The period of the polarization rotation is comparable to the orbital period of hot spots detected by the GRAVITY Very Large Telescope interferometer, which can be explained by the dominant poloidal component of the magnetic field \citep{2018A&A...618L..10G}. The observed orbiting hot spots could potentially be analogous to the infalling plasmoids in our simulations that appear after the magnetic field reconnection when the MAD state is temporarily interrupted. The timing properties in some of our runs are also comparable to the observed Sgr~A* flares. For instance, for Run H (fiducial resolution), the flaring state lasts $\sim 300\,M\approx 100\,{\rm min}$ (for $M\sim 4\times 10^6\,M_{\odot}$), while the separation between the accretion peaks is at least $900\,M\approx 5$ hours, which corresponds to $<4$ flares a day. These values are comparable to the statistics of NIR flares of Sgr~A* \citep{2012ApJS..203...18W,2018ApJ...863...15W}. For Sgr~A*, the hot spots or plasmoids are further modulated by the Doppler boosting and lensing on the orbital timescale of $P_{\rm ISCO}\sim 30\,{\rm min}(r/6r_{\rm g})^{3/2}$ scaled to the innermost stable circular orbit for a non-rotating black hole.

\begin{figure}
    \centering
    \includegraphics[width=\columnwidth]{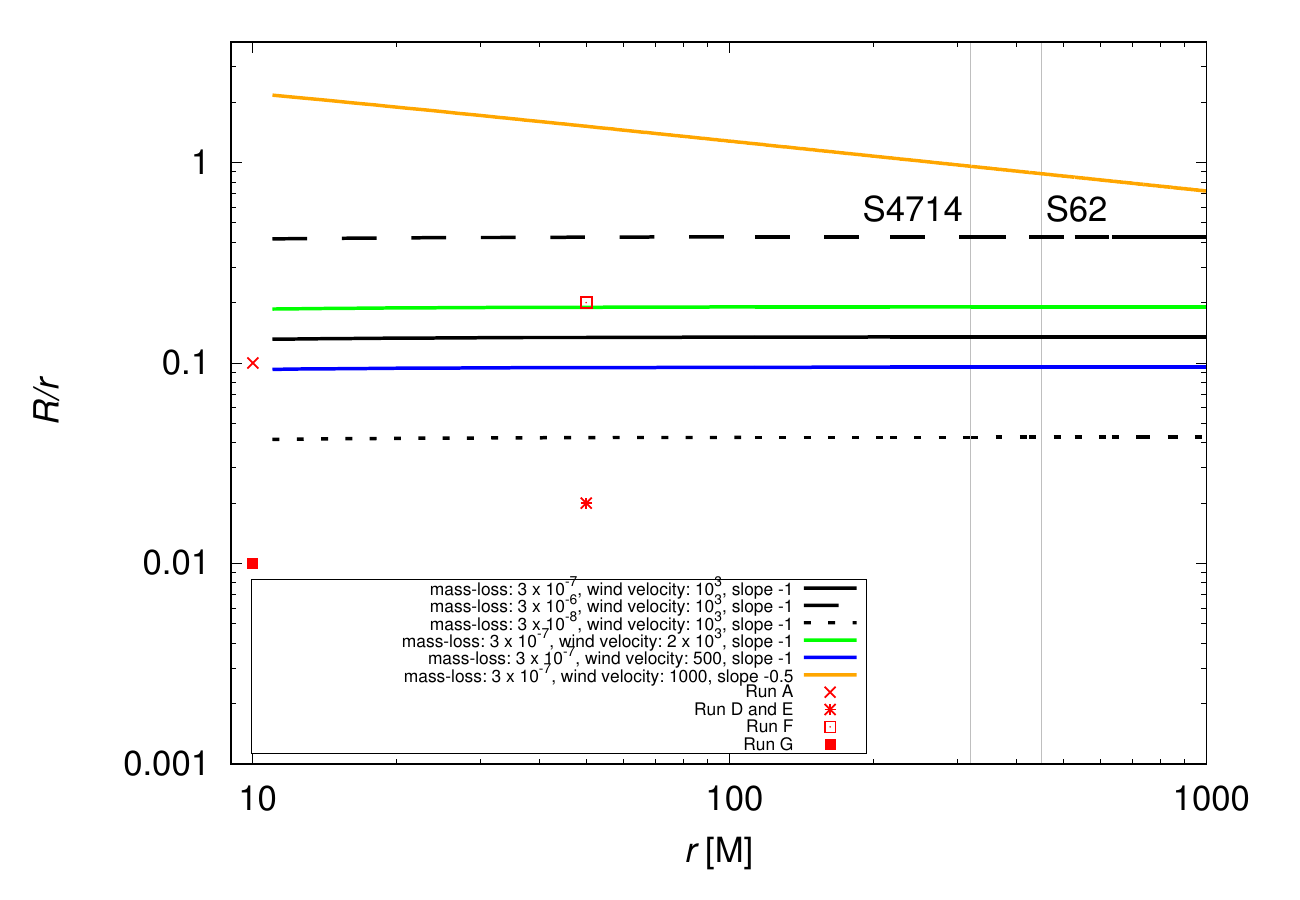}
    \caption{The dependency of the ratio of the effective stellar bow-shock radius and the stellar distance $\mathcal{R}/r$ on the radial distance from Sgr~A*. The lines stand for different stellar parameters with mass-loss rates (in $M_{\odot}\,{\rm yr^{-1}}$) and wind velocities (in ${\rm km/s}$) according to the legend. The black, green, and blue lines stand for the nearly constant $\mathcal{R}/r$ profile for the ambient density slope of $-1$, while the solid orange line depicts the dependency $\mathcal{R}/r\propto r^{-1/4}$ for the ambient density slope of $-0.5$. The red points stand for $\mathcal{R}/r$ values for some of our runs (A, D, E, F, G). Two gray vertical lines mark the pericenter distances for S4714 and S62 stars orbiting Sgr~A* \citep{2020ApJ...899...50P}. }
    \label{fig_ratio_Rr}
\end{figure}

It is currently not clear whether there are stars positioned close enough to affect the MAD state and the associated episodic accretion of Sgr~A*. However, recently the discovery of S62 and a population of faint stars S4711-S4715 \citep{2020ApJ...889...61P,2020ApJ...899...50P} has shown that at least some stars (S62, S4714) can reach pericentre distances of the order of $100\,M$. These stars as well as the bright star S2 interact with the hot flow whose properties are comparable to the accretion flow in our simulations. The cold, thin disk would be inconsistent with the observed constant S2 near-infrared L'-band magnitude (at $3.8\,{\rm \mu m}$) during its orbit around Sgr~A* as was recently shown by \citet{2020A&A...644A.105H}. Although the stars S62 and S4714 with the pericentre distances of the order of $100\,M$ ($\sim 320\,M$ for S4714 and $\sim 451\,M$ for S62) are still relatively far in comparison with the test stars in our simulations, the ratio of their effective bow-shock radii and distances $\mathcal{R}/r$ is likely comparable, as we show in Fig.~\ref{fig_ratio_Rr} where we also depict the ratio values for some of our runs (red points). In addition, $\mathcal{R}/r$ distance profiles are nearly constant from $10\,M$ to $1000\,M$ for the density and the temperature profiles of the hot flow surrounding Sgr~A* as expressed by Eq.~\ref{eq_density_temperature}. The constant profile of $\mathcal{R}/r$ follows from the ambient density profile $n_{\rm a}\propto r^{-1}$, from which $R_{\rm stag}\propto r$ and hence $\mathcal{R}/r=\text{const}$. For the flatter profile of the ambient density, $n_{\rm a}\propto r^{-1/2}$ according to \citet{2013Sci...341..981W}, $R_{\rm stag}\propto r^{3/4}$ and $\mathcal{R}/r$ decreases with the distance as $\mathcal{R}/r \propto r^{-1/4}$, see the solid orange line in Fig.~\ref{fig_ratio_Rr}. This implies that the dynamical impact of nearby S stars on the hot flow around Sgr~A* may share some similarities to our simulations.  This could be revealed by a detailed analysis of the Galactic center flare statistics in comparison with the exact timing and the distance of pericenter passages of S stars. 

Recently, the pericentre passage of the G2 or the Dusty S-cluster object (DSO) in 2014 at $\sim 4000\,M$ from Sgr~A* \citep{2012Natur.481...51G,2013A&A...551A..18E} caught a lot of attention because of its potential effect on the hot flow of Sgr~A*. With an intense monitoring by 8-10 meter class near-infrared adaptive-optics instruments, the source stayed compact in $L'$-band continuum during the pericentre passage \citep{2014ApJ...796L...8W} and revealed a large line width of the order of a few 100 km/s in Br$\gamma$ emission \citep{2015ApJ...800..125V}, which also stayed rather spatially compact with similar flux density and line width before and after the pericentre passage. Its polarized $K_{\rm s}$-band emission implies a general non-spherical geometry, possibly due to the presence of a composite dusty envelope and the development of a bow shock \citep{2016A&A...593A.131S,2017A&A...602A.121Z}, which is more consistent with a stellar source rather than with a core-less, turbulent gas and dust cloud. A likely interpretation is a young stellar object of class I \citep{2013ApJ...768..108S,2017A&A...602A.121Z} or a binary merger product \citep{2016MNRAS.460.3494S,2019ApJ...878...58S} that are both characterized by outflowing dense dusty envelopes. Based on the non-detection of a significant excess emission that could be attributed to a bow shock at lower frequencies, \citet{2015ApJ...802...69B} place an upper limit on the cross-sectional area of the DSO/G2, $\sigma\lesssim 2\times 10^{29}\,{\rm cm^2}$, which implies $\mathcal{R}_{\rm G2}/r\lesssim 0.1$ at the pericentre. This is marginally consistent with Run~A as well as with the $\mathcal{R}/r$ ratio for S2 star for $n_{\rm a} \propto r^{-1}$ density profile, see Fig.~\ref{fig_ratio_Rr}. In case of both S2 and DSO/G2, the association of their pericentre passages with Sgr~A* flares is at the moment difficult because of long orbital timescales involved, $\sim 16$ years for S2 ($2.6\times 10^7\,M$) and $\sim 260$ years for DSO/G2 ($4.2 \times 10^8\,M$). These timescales are at least two orders of magnitude larger than considered in our simulation runs. 

For comparison, for our Runs~D and E with stars at the distance of $50M$, we could not detect any significant periodicity peaks for the inflow rate for $\mathcal{R}/r=0.02$. There are some speculations that X-ray and NIR flares of Sgr~A* are modulated by a pacemaker that could be linked to a star orbiting at $\sim 6.6\,M$ according to \citet{2020ApJ...896...74L}. Hence, Sgr~A* remains a prime testbed for star--accretion flow interactions, especially in case faint stars are discovered at still smaller distances.


\subsubsection{Occurrence of stars in the innermost region}
The general question is what the fraction of galaxies is where we can possibly detect star--accretion flow interactions in the innermost region between $10\,M$ and $50\,M$ as we simulated. For simplicity, we consider galaxies with the fixed SMBH mass of $M=10^7\,M_{\odot}$, which contain $M_{\rm NSC}\sim 10^7\,M_{\odot}$ of stellar mass inside the sphere of influence of the SMBH \citep{2013degn.book.....M}. The sphere of influence has a radius of $r_{\rm h}=(c/\sigma_{\star})^2$ in gravitational radii, where $\sigma_{\star}$ denotes the stellar velocity dispersion in the host bulge. Using the $M-\sigma_{\star}$ relation \citep{2000ApJ...539L...9F,2009ApJ...698..198G}, we get $\sigma_{\star}\sim 110\,{\rm km\,s^{-1}}$ for $M=10^7\,M_{\odot}$, from which $r_{\rm h}\sim 7.44\times 10^6\,M$.

Using the standard initial mass function (IMF) in the form, $\mathrm{d}N_{\star}/\mathrm{d}m=Cm^{-\alpha}$, where $\mathrm{d}N_{\star}$ is the number of stars in the stellar mass bin $(m-\mathrm{d}m/2,m+\mathrm{d}m/2)$ and $\alpha$ is the IMF power-law slope, we can estimate the number of stars heavier than a certain lower-limit $m_{\rm low}$ using the total mass $M_{\rm NSC}$,

\begin{equation}
    N_{\star}(>m_{\rm low})=\frac{2-\alpha}{1-\alpha} M_{\rm NSC} \frac{m_{\rm max}^{1-\alpha}-m_{\rm low}^{1-\alpha}}{m_{\rm max}^{2-\alpha}-m_{\rm min}^{2-\alpha}}\,.
    \label{eq_N_star}
\end{equation}
where $m_{\rm min}$ and $m_{\rm max}$ is the minimum and the maximum stellar mass, which we fix to $0.1\,M_{\odot}$ and $100\,M_{\odot}$, respectively. The power-law slope is fixed to $\alpha=2.35$ in this mass range assuming the standard Salpeter-Kroupa IMF \citep{2001MNRAS.322..231K}.

Next, we focus on stars with the mass larger than or equal to the Solar mass, hence $m_{\rm low}=1\,M_{\odot}$. We consider heavier stars which have a larger potential of developing stronger winds, and hence are expected to have a larger impact on the accretion flow than the lighter and more numerous stars, such as red dwarfs. From Eq.~\eqref{eq_N_star}, we get $N_{\star}=1.3\times 10^6$. We further restrict ourselves to the number of stars $N_{\rm sf}$ that pass through the accretion flow inside a certain radius $r$. Assuming the dynamically relaxed power-law 3D density distribution of stars $n\propto r^{-\gamma}$, which is assumed to extend downwards towards the ISCO, where $\gamma\sim 3/2$ \citep{1976ApJ...209..214B,1977ApJ...216..883B}, we obtain \citep[see e.g.][and references therein]{2018AN....339..324Z},

\begin{equation}
    N_{\rm sf}(<r)=N_{\star}\left(\frac{r}{r_{\rm h}} \right)^{3-\gamma}\,.
    \label{eq_N_sf}
\end{equation}
Using Eq.~\eqref{eq_N_sf}, the fraction of galaxies $f_{\rm g}$ where the stars heavier than $m_{\rm low}$ pass through the flow inside the radius $r$, can be approximated as $f_{\rm g}\sim N_{\rm sf}/(1+N_{\rm sf})\sim N_{\rm sf}$, which is valid under the assumption that all galaxies would be identical and that $N_{\rm sf} \lesssim 0.1$. Here we consider a purely stationary case resulting from the relaxed Bahcall-Wolf-like cusp, i.e. a star inside the radius $r$ is found there at all times. Stars on long-period, high-eccentricity orbits, which result from dynamical scattering into a loss cone, can in principle be located inside $r$ for a short period of time. The probability of detecting such stars in a sparse region close to the SMBH was considered and estimated by \citet{2018AN....339..324Z}.

For $m_{\rm low}=1\,M_{\odot}$ and $r\leq 50M$, we obtain $f_{\rm g}\sim 1/44$ (one in every 44 galaxies). Increasing $m_{\rm low}$ to $2\,M_{\odot}$, which is corresponding better to the inferred masses of Galactic-center S stars, we get $f_{\rm g}\sim 1/116$ (i.e. one in every 116 galaxies). Restricting ourselves to even heavier stars, $m_{\rm low}=10\,M_{\odot}$, i.e. stars that have produced neutron stars, we get $f_{\rm g}=1/1058$. We see that even in this simplistic statistical model of the star--flow occurrence, the fraction of potential sources depends noticeably on the stellar mass spectrum of interest. The occurrence rates estimated above should be considered as crude upper limits as we have not considered any observability criteria at this stage.

We can also estimate the stellar occurrence fraction from the fact that the relaxed cusp is expected to end at a certain distance $r_1$ where the number of enclosed stars drops below one in a statistical sense. At $r\gtrsim r_1$, we expect that every galaxy that has a cuspy NSC hosts at least one star on a tight bound orbit around the SMBH due to the dynamical relaxation process. Below $r_1$ the occurrence of any star is rather determined by the orbital decay due to the emission of gravitational waves.

The radius $r_1$ can be inferred from Eq.~\eqref{eq_N_sf} by setting $N_{\rm sf}(<r)=1$, which gives
\begin{equation}
    r_1=r_{\rm h} N_{\rm h}^{-\frac{1}{3-\gamma}}\,.
    \label{eq_r1}
\end{equation}
Next, we compare the characteristic merger timescale $\tau_{\rm merge}$ between the distance $r_1$ and the interaction distance $r_{\rm int}=50\,M$, which will give us the fraction of galaxies with stars interacting with the accretion flow sufficiently close to the SMBH, $f_{\rm g}\sim \tau_{\rm merge}(r_{\rm int})/\tau_{\rm merge}(r_1)$. For the case when the perturber and the SMBH mass ratio is small enough, i.e. $q=m_{\star}/M\ll 1$, the timescale for the perturber--SMBH merger from the initial distance $r_0$ can be expressed in the following form, assuming a circular orbit \citep{1964PhRv..136.1224P},
\begin{equation}
    \tau_{\rm merge}=\frac{5G}{256c^3}\frac{(1+q)^2}{q}\left(\frac{r_0}{r_{\rm g}}\right)^4 M\,.
    \label{eq_merger_timescale}
\end{equation}
For $M=10^7\,M_{\odot}$, $q=10^{-7}$, $r_0=50\,M$, we obtain $\tau_{\rm merge}\approx 2\,{\rm Myr}$, which is three orders of magnitude less than the two-body relaxation time \citep{2013degn.book.....M} 
\begin{align}
    \tau_{\rm relax}&=1.6\left(\frac{\sigma_{\star}}{110\,{\rm km\,s^{-1}}} \right)^3\left(\frac{\rho_{\star}}{10^6\,M_{\odot}{\rm pc^{-3}}} \right)^{-1}\times\\
    &\times \left(\frac{m_{\star}}{M_{\odot}} \right)^{-1} \left(\frac{\log{\Lambda}}{15} \right)^{-1}\,{\rm Gyr}\,,
\end{align}
where $\sigma_{\star}$ is a one-dimensional stellar velocity dispersion, $\rho_{\star}$ is the stellar density, and $\log{\Lambda}$ is the Coulomb logarithm.
Therefore, in the region of our interest ($10-50\,M$), the stellar dynamics is not determined by dynamical relaxation, but the relaxed stellar cusp still serves as a basic reservoir of perturbers.

Taking into account the lower stellar mass limit $m_{\rm low}=1\,M_{\odot}$, we obtain $N_{\star}=1.3\times 10^6$, which according to Eq.~\eqref{eq_r1} gives $r_1\sim 635\,M$. The fraction of galaxies with at least one star at $50\,M$ or less can be expressed as,
\begin{equation}
    f_{\rm g}\sim \frac{\tau_{\rm merge}(50\,M)}{\tau_{\rm merge}(635\,M)}=\left(\frac{r_{\rm int}}{r_1} \right)^4\sim 3.8\times 10^{-5}\,.
\end{equation}

Taking into account all stars with $m_{\rm low}=0.1\,M_{\odot}$, we get $r_1\sim 80\,M$ and $f_{\rm g}\sim 0.15$. Hence, the estimates are quite uncertain depending on the mass range and characteristic dynamical timescales.

 Finally, a lower limit on the occurrence of stellar and compact perturbers in the inner regions of the NSCs is given by tidal disruption rate $\dot{N}$, which is expected to be inversely proportional to the SMBH mass $M$ for nucleated galaxies \citep{2004ApJ...600..149W}. In particular, for the stellar densities $\rho_{\star}\propto r^{-2}$, we obtain \citep{2004ApJ...600..149W}
\begin{equation}
    \dot{N}\simeq 3.7\times 10^{-4} \left(\frac{M}{10^7\,M_{\odot}} \right)^{-0.25}\,{\rm yr^{-1}}\,.
\end{equation}
For galaxies with the smallest SMBHs of $M=10^5\,M_{\odot}$, the tidal disruption rate is one order of magnitude larger, $\dot{N}\sim 1.2\times 10^{-3}\,{\rm yr^{-1}}$. Therefore the smallest galaxies with the lightest black holes could be the most frequent sites of the perturber--accretion disc interactions, which is observationally hinted by the recent detection of X-ray quasi-periodic eruptions in low-mass black hole systems \citep{2019Natur.573..381M,2020A&A...636L...2G,2021arXiv210413388A}, see also Subsections~\ref{subsub_GSN069} and \ref{subsub_otherQPE} for more details.

In summary, given the current and upcoming large-scale surveys of sources (with at least $10^6$ galaxies involved), there are generally good prospects of finding suitable candidates for star--flow interactions based on the timing properties and the detection of quasiperiodic oscillations/eruptions. In particular, some currently monitored sources can already serve as proper candidates. In the following sections, we will specifically discuss  10 promising sources: OJ 287, J0849$+$5108, RE J1034$+$396, 1ES 1927+65, ESO 253--G003, GSN 069, RX J1301.9$+$2747, eRO-QPE1, and eRO-QPE2. 


\subsection{Connection to observational results -- II: Relation to relativistic radio components} \label{subsec:discussion-observation-II}

Another important observational feature in our studies is the occasional ejection of blobs into the funnel region. The blobs are moving with the Lorentz factor of $\Gamma\gtrsim 1.155$ close to the funnel-torus boundary, which corresponds to at least half of the speed of light. The blobs are injected into the funnel when the MAD state is temporarily interrupted and the accretion-rate increases. For stars on inclined orbits, the blobs can be periodically ejected due to the stellar perturbation also during the MAD state when the accretion rate drops. Here we speculate that depending on the resolution and the numerical resistivity in our simulations, the ejected blobs could be associated with the observed moving radio components in some jetted AGN, in particular those that exhibit periodic behaviour in their light curves. Concerning the time interval between the blob ejections in our simulations, a typical interval is $\sim 900\, M$ for Run~H (FR), which corresponds to $\tau\sim 51 (M/10^9\,M_{\odot})$ days. For the increased numerical resolution, some of the gaps in between blob ejections have a few $1000\,M$ up to even $10^4\,M$, see Fig.~\ref{fig:Mdot-runH}, which already gives timescales of the order of a year.

Therefore our model provides a viable explanation for the ejection of at least some of the radio components that move at relativistic speeds in radio galaxies. Because of the timescales of at least a fraction of the year between the component ejection, this model is especially suited for jetted giant elliptical galaxies at the cores of galaxy clusters that host the massive central black holes, e.g. 3C84 (NGC1275, Perseus A) at the center of the Perseus cluster \citep[see e.g.][and references therein]{2018NatAs...2..472G,2019Galax...7...72B}. The activity in these galaxies is expected to be powered by galaxy mergers \citep{2015ApJ...806..147C}. In giant elliptical galaxies, the jets can pierce through the hot diluted atmosphere and deposit the heat and the mechanical energy at much large distances of $>10\,{\rm kpc}$. Therefore the hot atmosphere can cool down, form colder clouds, and feed the central black hole via the cooling flows or so-called precipitation \citep{2012ApJ...746...94G,2012MNRAS.419.3319M,2014ApJ...789..153L,2015Natur.519..203V,2019MNRAS.488.1917G,2020ApJ...899..159F}. 

On the other hand, the dynamical processes in the nuclear star cluster as well as the dynamical friction of secondary black holes surrounded by their own stellar clusters make giant elliptical galaxies ideal candidates for the interaction of stars and compact remnants with the heated accretion flow. In general, for elliptical galaxies, the tidal disruption event rates are higher than in spiral galaxies, especially at early cosmic times, due to the single initial starburst and large stellar densities \citep{2016ApJ...823..137A}. Giant elliptical galaxies at the cores of galaxy clusters typically exceed the Hills mass for black holes, $m_{\rm Hills}\gtrsim 1.15\times 10^8\,M_{\odot}$, when Solar-type stars are not tidally disrupted and plunge through the event horizon. Therefore these sources provide a proper setting for repetitive star--flow interactions all the way down to the innermost stable circular orbit.   

The ejected blobs of matter that move close to the funnel-torus boundary provide a novel scenario for the interpretation of the observed limb-brightening of jets \citep{1999ApJ...518L..87A,2004ApJ...600..127G,2005MNRAS.356..859P,2016A&A...588L...9B,2016A&A...595A..54M,2018A&A...616A.188K,2018NatAs...2..472G,2021arXiv210107324B}, which is related to the two-fluid models that lead to the jet-stratification and the spine-sheath structure \citep{1989A&A...224...24P,1989MNRAS.237..411S,2001MNRAS.321L...1C,2005A&A...432..401G,2007ApJ...664...26H,2008MNRAS.386..945T,2009ApJ...697..985D,2012RAA....12..817X,2015MNRAS.450.2824M}. The observations generally indicate the faster and the low-density spine section closer to the jet axis, while closer to the brightened edges, there is a slower and a denser material. In theory, this is often modelled by the combination of the Blandford-Znajek mechanism \citep{1977MNRAS.179..433B}, in which the rotational energy of the SMBH is responsible for the narrower and the faster spine part, while the Blandford-Payne mechanism \citep{1982MNRAS.199..883B} invokes the accretion disk and the associated anchored magnetic field to account for the centrifugal launching of the wider, slower, and denser sheath part of the jet \citep{2012RAA....12..817X}. 

In our model, the vertical perturbations of the accretion flow induced by an orbiting star are responsible for the intermittent ejection of denser material along the funnel-torus boundary, i.e. the blobs move in the sheath layer of the nuclear outflow. The ejected plasmoids are always significantly denser than the surrounding funnel material. This can be understood as an alternative and a complementary mechanism to the purely magnetically generated winds anchored in the rotating accretion disc (Blandford-Payne mechanism). The limb-brightening effect within the stellar perturber scenario is caused by the density stratification across the jet width. The blobs outflowing along the funnel-torus boundary are $\sim 10^4$ times denser than the surrounding funnel material located closer to the axis. For the inclined stellar orbits, in particular for Run~A, where the perturber is located at 10\,$M$, we can also notice the spine-sheath structure quite well, i.e. the higher Lorentz-factor material is closer to the jet axis, while the denser blobs with a lower Lorentz factor are at the funnel edges, see Fig.~\ref{fig:RA-slices}. As we have showed in Subsection~\ref{subsec_periodicity_psd}, the power spectral density of this outflow has a power-law shape with the quasiperiodicity peaks due to the periodic perturbations. The PSD of radio light curves could be used to distinguish the perturber model for the sheath structure from the more general Blandford-Payne mechanism. At the moment, it is rather difficult to compare our simulations to observations since the obtained density and the Lorentz-factor profiles are limited to an order of a few hundred of gravitational radii due to our grid size, whereas the jet studies reliably resolve the structures beyond this length-scale. Hence, it will be necessary to extend the higher resolution to larger scales in the future. However, there are good prospects for model--data comparisons as our model can naturally account for the non-uniform limb-brightened structure of jets \citep[see e.g.][for the well-resolved spine-sheath structure of 3C273]{2021arXiv210107324B}.


\subsubsection{ OJ287 quasar vs. QPO phenomenon}
It is appropriate to compare our GRMHD simulations with jetted AGN that possess ADAF-type flows. BL Lac sources do not exhibit broad emission lines, therefore they belong to this category of jetted sources that are powered by hot flows. OJ287 is a typical representative of BL Lac sources, nicknamed as a Rosetta stone of blazars \citep{1994VA.....38...77T,1988ApJ...325..628S,2018MNRAS.478.3199B}, which hosts one or two SMBHs with the total mass of the order of $10^8-10^{10}\,M_{\odot}$ \citep{2002A&A...388L..48L,2016ApJ...819L..37V}. In \citet{2018MNRAS.478.3199B}, they list ejection times of individual moving radio components of OJ287 and these are separated by the variable time interval of $\sim 0.2$ years up to a year and more in the observer's frame, which is consistent with the ejection times in our simulations if the most massive primary black hole from the above-mentioned mass range is assumed. Even more interestingly, OJ287 exhibits a periodicity of $\sim 22$ years in the radio light curve \citep[$14.5$ GHz; ][]{2018MNRAS.478.3199B} and the period of $\sim 11.65$ years in the optical V-band light curve (monitoring since $\sim 1890$; \citeauthor{1988ApJ...325..628S}, \citeyear{1988ApJ...325..628S}). It was suggested that behind the periodic characteristic of the optical light curve is a secondary black hole of mass $\sim 10^8\,M_{\odot}$ that orbits the more massive primary black hole of $10^{10}\,M_{\odot}$ \citep{1988ApJ...325..628S, 2016ApJ...819L..37V}. While orbiting, the secondary black hole plunges through the accretion flow around the primary component and induces shock waves that propagate inwards. The periodic behaviour of the radio light curve was modelled using the bulk jet precession, either caused by the torque of a secondary black hole onto the accretion flow around the primary component or by the Lense-Thirring precession of a misaligned accretion flow \citep{2018MNRAS.478.3199B}.  

In our model, the OJ287-like periodicities, where the radio period is approximately twice as large as the optical period, could be related to the passages of a perturber through the accretion flow on an inclined trajectory. The presence of a secondary massive black hole is not entirely necessary to induce the changes, but very likely, as we estimate below. The longer radio period could be related to the fact that for blazars we detect radio components from the jet side that is oriented towards us close to the line of sight, we do not see the counter-jet components. The flux density from the jet cone oriented towards us is Doppler-boosted as $S_{\nu}=S_0\delta^{p-\alpha}$ \citep{1972MNRAS.157..359M}, where $S_0$ is the intrinsic jet emission, $\delta$ is the Doppler-boosting factor\footnote{The Doppler-boosting factor depends on the Lorentz factor $\Gamma$, the intrinsic velocity of moving components $\beta$, and the angle to the line of sight $\phi$: $\delta=[\Gamma(1-\beta\cos{\phi})]^{-1}$.}, $\alpha$ is the spectral index\footnote{We apply the notation $S_{\nu}\propto \nu^{\alpha}$, with $\alpha<-0.7$ standing for steep sources, $-0.7 \leq \alpha \leq -0.4$ denoting flat sources, and $\alpha>-0.4$ standing for inverted sources, according to the classification by \citet{2019A&A...630A..83Z}.}, and $p$ is the geometrical factor ($p=3$ for spherical jet components, $p=2$ for a cylindrical jet geometry; see \citeauthor{1979ApJ...232...34B}, \citeyear{1979ApJ...232...34B}). The optical outbursts with the period of $\sim 11$ years could simply be twice as frequent since the perturber is passing twice across the flow. The optical outbursts would be related to the accretion, while the radio outburts to the ejected relativistic radio components. For our simulated 3D run~I with an inclined star, the period of $100\,M$ (half of the orbital period) is clearly detectable in both the accretion rate and the outflow rate, although more clearly for the outflow rate, see Figs.~\ref{fig_comparison_runI_NP} and \ref{fig_comparison_runI_NP_outflow}. However, here we have to point out that the outflow rate is computed as the surface integral over the spherical sectors around the axis in \textit{both} directions from the black hole. Hence we follow the matter that is escaping along both sides of the jet. If only one side of the jet is seen and contributes to the radio emission, the peak in the periodogram would be found at half of the value of the frequency for the both-sided outflow.

Our model is applicable only for the scenario with the SMBH of $10^{10}\,M_{\odot}$, when the stellar perturber or a smaller secondary black hole would be located at $r\sim 144 M$ with an orbital period of 22 years ($\sim 17$ years in the rest frame considering $z \sim 0.3$), which is comparable to our studied distance range in case the effective cross-section of the perturber is large enough: $\sim 20\,M$ to achieve the comparable ratio $\mathcal{R}/r$ as for run~F. The stagnation radii of the order of 1$M$--20$M$ correspond to $\sim 20\,000\,R_{\odot}$--$400\,000\,R_{\odot}$, respectively, for a $10^{10}\,M_{\odot}$ black hole. Such large stagnation radii of stellar bow shocks are only conceivable in an extremely low-density environment with number densities of $7.45\,{\rm cm^{-3}}$ and $0.02\,{\rm cm^{-3}}$ (for the stellar parameters of the S2 star and the orbital distance of $r\sim 144 M$), respectively, which is too low for the inner region of the accretion flow. If we consider the ADAF-type flow for OJ287 with the accretion rate of $\dot{\mathcal{M}}=0.08$ in the Eddington units and the viscosity parameter $\alpha=0.26$ \citep{2019ApJ...882...88V}, the number densitities at $r=144M$ are $2.8 \times 10^{11}\,{\rm cm^{-3}}$. The stagnation radius around S2-like stars thus would be much smaller, only a tiny fraction of gravitational radius $M$, $R_{\rm stag}\sim 5\times 10^{-6}\,M$, which is below our current resolution. The present results, however, indicate that the influence of such a small perturber on the accretion rate would be negligible.

Therefore the presence of another massive black hole as a perturber is necessary to induce any observable changes in both the inflow and the outflow rate.
Using eq.~(\ref{eq:Rsync}) we can estimate the synchronization radius of the flow in the setup with $M=10^{10}\,M_{\odot}, \,m_*=10^8\,M_{\odot},\,r=144\,M$ to obtain $R_{\rm sync} \sim 1.4\,M$. The ratio $\mathcal{R}/r$ is thus about one order of magnitude lower then in case of Run~F, but it is roughly comparable with Runs D and E. These two runs, which differ by the one having an inclined star, while the other has an embedded star, have not shown any significant peak in the periodogram of the accretion rate, but a narrow peak is seen in the outflow rate. We can thus speculate that $R_{\rm sync} \sim 1.4\,M$ at the distance of the order of $100\,M$ is too small for an observable effect in the accretion rate to arise, but it is sufficient to expel enough matter into the funnel to influence the radio emission, regardless of the orientation of the perturber's orbit. In this case, the optical burst would have to be caused by some additional effect apart from the direct influence of the perturber on the accretion rate, e.g. by an enhanced emission due to the accretion of the gas on the perturber itself. In particular, if the perturber of $m_*=10^8M_{\odot}$ transits through the accretion disc of the larger SMBH, a small accretion disc would form around the perturber on the length scale $R_{\rm sync}\sim1\,M$, which corresponds to about 100 gravitational radii of the smaller black hole. The accretion disc of the SMBH with such a mass will heat more than the gas around the larger SMBH and the peak of its spectrum falls into the optical band. The repetitive emergence and depletion of the accretion disc thus can cause the detected periodicity in the optical $V$-band.

For the case of less massive primary SMBH of $10^8\,M_{\odot}$ \citep{2002A&A...388L..48L}, the perturber would have to be located at $\sim 3093\,M$ for the same orbital period, which is too far for the perturbations to have a comparable effect as in our simulations.

The case of OJ287 could be a special one in the zoo of AGN quasi-periodic oscillation (QPO) phenomena. There are several sources where the quasi-periodicity has been reported for different energy bands, from the $\gamma$-ray up to the radio domain, see \citet{2021arXiv210404124Z} for a brief overview of QPO sources. The general physical process is unknown and is likely related to different mechanisms for different sources, depending on whether the QPO is transient or stable and what the period of the QPO is. General candidates are accretion-disc instabilities (e.g. radiation-pressure instability), disc-jet precession due to a secondary black hole or the Lense-Thirring effect, the accretion-disc perturbation due to the presence of a bound perturber, or the combinations of these mechanisms.

\citet{2021arXiv210404124Z} report the case of one of the longest transient QPO, lasting for about 10 years, which is associated with the Radio-Loud Narrow-Line Seyfert 1 (NLSy 1) galaxy J0849$+$5108 ($z=0.584$). The QPO is detected with $>5\sigma$ significance in the radio light curve at 15 GHz with the observed period of $P_{\rm QPO}=175.93\pm6.34$ days. The \textit{Fermi}-LAT $\gamma$-ray light curve was found not to possess any significant periodicity. \citet{2021arXiv210404124Z} discuss two scenarios for the origin of the radio QPO: secular disc instabilities, such as the Lightman-Eardley secular instability, or a helical motion of an ejected blob. Here we speculate that the QPO could have been present due to an orbiting perturber in combination with an increase in the accretion rate, which can make the ejected blobs more massive and pronounced, see e.g. Fig.~\ref{fig:Mdot-2D}, in particular Run~F until $t=60\,000\,M$, when the accretion as well as the outflow rates drop. Since the jet is oriented close to the line of sight, we can assume that the QPO period is associated with the orbital period of $P_{\rm orb}=P_{\rm QPO}/(1+z)\sim 111$ days. For the black hole mass of $M\sim 10^{7.4}\,M_{\odot}$, the orbital distance associated with the QPO is $r\sim 536\,M$. Given that J0849$+$5108 is NLSy 1 with a high accretion rate close to the Eddington limit, the perturber is unlikely of a stellar nature since the stagnation radius is expected to be well below $1\,M$. Hence, a secondary massive black hole with the the mass in the range $5\times 10^5-2.5\times 10^6\,M_{\odot}$ is more plausible as the ratio $R_{\rm sync}/r\sim 0.01-0.1$, see Eq.~\eqref{eq:Rsync}, which is comparable to our Runs~A, D, E, and F, for which we can detect significant periodic peaks in the outflow rate. This configuration could also be realistic in terms of the duration and the stability of the system as the merger timescale is in the range $t_{\rm merge}\sim 7.6\times 10^4-3.3\times 10^5$ years \citep{1964PhRv..136.1224P}, see also Eq.~\eqref{eq_merger_timescale}.


\subsection{Repetitive transits vs.\ Tidal Disruption Events} \label{subsec:discussion-TDE}
Transits of a bound star through and across the accretion flow leave repetitive imprints in the observed light curves. This is in contrast with one-time tidal disruption events (TDEs) that are characterized by a single rapid flux density rise, followed by a gradual fall-off proceeding with time typically as $\propto t^{-5/3}$. The mechanism of TDEs near supermassive black holes was introduced in the 1970s \citep{Hills_1975,Rees_1988} and since then it was studied in many works including the mathematical description of the relativistic TDEs near a Kerr black hole \citep{1975ApJ...197..705M,2019GReGr..51...30S}. On the other hand, the vertical transits of a star through the accretion flow could manifest themselves as quasiperiodic oscillations (QPOs) in X-ray light curves \citep{1993MNRAS.265..365V,1999PASJ...51..571S}.


\subsubsection{RE J1034+396}
\label{sec_REJ}
For the narrow-line Seyfert 1 galaxy RE J1034+396 ($z=0.042$), the first significant QPO was found \citep{2008Natur.455..369G} and subsequently confirmed as a recurrent phenomenon over the past 11 years \citep{2010A&A...524A..26C,2020MNRAS.495.3538J,2021MNRAS.500.2475J}. The detected period was initially $P_{\rm QPO}\simeq 3733\,{\rm s}$ based on 2007 data \citep{2008Natur.455..369G}, and shortened to $P_{\rm QPO}=3550\pm 80\,{\rm s}$ based on 2018 monitoring \citep{2020MNRAS.495.3538J}. The mechanism triggering the QPO still remains unknown, although it seems to be related to the soft X-ray component \citep{2020MNRAS.495.3538J}. \citet{2021MNRAS.500.2475J} suggested that the QPO could be triggered by the expansion/contraction of the vertical structure within the flow. 

In this regard, since it is clearly a recurrent phenomenon on the timescale of 10 years with a period that is slightly shortening, it could be related to vertical stellar transits similar to those in our simulations. To estimate the prospects for this hypothesis, we consider an SMBH of $M=2\times 10^6\,M_{\odot}$, which falls in the most likely range of $1-4\times 10^6\,M_{\odot}$ for RE J1034+396  \citep{2016A&A...594A.102C,2020MNRAS.495.3538J}. In that case, the perturbing star would orbit the SMBH with the period of $P_{\rm orb}=2P_{\rm QPO}$, which would be $\sim 7466$ s for the earlier epoch and $\sim 7100$ s for the later epoch. These periods correspond to the orbital distance of the star at $r\sim 24.4\,M$ and $r\sim 23.6\,M$, respectively (for a non-rotating black hole), which corresponds to our studied regime of the star--accretion flow interactions. 

The orbital shrinkage could be attributed to the combination of gravitational radiation and the magnetohydrodynamic drag. Specifically, the period derivative induced by gravitational waves in an extreme mass ratio inspiral is $\dot{P}\sim -10^3 P^{-5/3} G^{5/3} M^{2/3} m_{\ast} c^{-5}$ \citep{peters1963gravitational}. Attributing the period drift in RE J1034+396 only to gravitational waves then yields $m_{\ast}\sim 10^{-3} |\dot{P}| P^{5/3} M^{-2/3} G^{-5/3} c^5 \sim 100 M_\odot$ making the perturber likely a black hole. It should also be noted that the low redshift of RE J1034+396 and the $\sim 0.1 mHz$ orbital frequency would make the event an ideal source for a simultaneous detection by the upcoming space-based gravitational-wave detector LISA \citep{2017arXiv170200786A}. However, if the perturber is a star, the mass is overestimated since hydrodynamical drag also significantly contributes to the period derivative for usual stellar-sized objects passing through the disk \citep{2000ApJ...536..663N}.


\subsubsection{1ES 1927+65}
An exemplary case of how the TDE can transform the innermost region of the disk is the source 1ES 1927+65 \citep{2020ApJ...898L...1R}. It is a type 2 AGN that developed broad Balmer lines after the detected optical outburst (2017 December 23; \citeauthor{2019ApJ...883...94T}, \citeyear{2019ApJ...883...94T}). Therefore, 1ES 1927+65 may be considered as a changing-look AGN, where the transformation was triggered by the TDE, which is manifested by the declining UV/optical flux (following approximately $t^{-5/3}$ law). However, the X-ray light curve behaved differently after the outburst. First, it dropped by three orders of magnitude, which was followed by the rise in the X-ray flux to pre-outburst values by four orders of magnitude in just 100 days. During the dip in X-rays, the high-energy power law disappeared and the 0.5-10 keV X-ray luminosity was dominated by a black-body component. As the X-ray luminosity started to increase, so did the power-law component, which asymptotically dominates 0.5--10 keV luminosity. 

This transformation and rapid changes can be interpreted by the destruction of the innermost part of the accretion disc by a tidal stream  \citep{2020ApJ...898L...1R}. The tidal stream creates shocks that can remove the angular momentum and accelerate the accretion, which is a scenario supported by simulations \citep{2019ApJ...881..113C}. During this process the magnetic field support of the hot corona is destroyed and the corona and its power-law component disappear or are largely diminished. 

Nevertheless, \citet{2020ApJ...898L...1R} also report an additional and persistent X-ray flux variability of 1ES 1927+65 with a characteristic period of 8 hours. Within our model, this would suggest that the TDE was only partial and that there is a remnant core still orbiting the black hole and perturbing the flow. The accretion disk in 1ES 1927+65 is in a cold geometrically thick state with a hot corona, so we cannot compare its observations with our simulations directly. However, the aforementioned scenario of the destruction of the inner part of the disk and the depletion of the corona has some similarities with the destruction of the inner part of the disk and the emergence of the MAD state as observed in our simulations. On the other hand, at our simulation resolutions we observe dips in accretion as well as outflow rates on time-scales $\sim 10^3-10^4 M$, which translates to units of hours to tens of days for a black hole of mass $10^6\,M_{\odot}$--$1.9\times 10^7\,M_{\odot}$ \citep{2020ApJ...898L...1R}. In other words, our dips and peaks related to the MAD state occur on a timescale of at least an order of a magnitude shorter than the dip observed in 1ES 1927+65. Hence, the dip is likely caused by the TDE itself. 

Nevertheless, the persistent 8-hour variability is consistent with a remnant on an orbit of radius $r=21\,M$ for $M=1.9 \times 10^7\,M_{\odot}$ and $r=150\,M$ for $M=10^6\,M_{\odot}$ (assuming the orbital period is twice as large as the variability timescale and a zero SMBH spin), which is comparable to our set-up. We can speculate that the perturber could be, for instance, a white dwarf core after the tidal disruption of an asymptotic giant branch star \citep{2020MNRAS.493L.120K}. However, further monitoring of 1ES 1927+65 is needed to learn more about the triggering mechanism of both its short-term and long-term variability. The distinction between tidal debris and a persistent remnant of the TDE will then be possible to make based on the observation of a stable component of the variability that does not fall off as the usual $\sim t^{-5/3}$ law.


\subsubsection{ESO 253--G003}
\label{sec_ESO}
Another source that exhibits periodic outbursts is the galaxy ESO 253--G003 \citep[$z=0.042489$;][]{1996PASP..108.1117A} where the event ASASSN-14ko was detected with the All-Sky Automated Survey for Supernovae (ASAS-SN) originally considered as a supernova of type IIn close to the nucleus of type II Seyfert galaxy \citep{2014ATel.6732....1H,2017MNRAS.464.2672H}. During the six years of monitoring in the optical $V\!$- and $g$-band by ASAS-SN, in total seventeen outbursts were detected with the nearly equal time separation. Based on the $O$-$C$ (observed minus computed) plot, the mean period of the flares is $P_0=114.2 \pm 0.4$ days with a large period derivative of $\dot{P}=-0.0017 \pm 0.0003$. Two scenarios can explain the main observational features: First, a repeated partial tidal stripping of a star on an eccentric orbit, and second, an intermediate-mass black hole on a near-circular orbit. 

 \textit{Eccentric tidal stripping:} As already proposed (and preferred) by \citet{2020arXiv200903321P}, a partial tidal disruption event is consistent with both the periodicity and the large period derivative. To make the TDE more likely, we assume a highly eccentric orbit with the eccentricity of $e\sim 0.9$. Then we also assume that at the pericenter, the effective length-scale of the TDE is comparable to the pericenter distance, i.e. $\mathcal{R}/r_{\rm p}\approx 1$, which will effectively lead to one disturbance of the flow per orbital period and not two, closely spaced in time, which would be inconsistent with the observations. For the orbital period of $P_{\rm orb}\sim P_0/(1+z)\sim 109.55$ days and the SMBH mass of $M=7.24\times 10^7\,M_{\odot}$ \citep{2020arXiv200903321P}, we obtain the semi-major axis $a_{\rm TDE}\sim 262\,M$ and the pericenter distance of $r_{\rm p}\sim 26\,M$. 
 
 For the star to be at least partially tidally stripped at the pericenter, its radius should be $R_{\star}\sim 10\,R_{\odot}$ so that the tidal radius is approximately equal to the pericenter distance for $m_{\star}=1\,M_{\odot}$. Since the radius of the stellar atmosphere has to be so large, the best prospects for interpreting ASASSN-14ko with a periodic behaviour is the partial tidal disruption of a red giant or an asymptotic giant branch star, for which the large, loose envelope can be tidally stripped while the core is left untouched. Depending on the effective radius of the envelope, the mass loss can be large enough to explain the large $\dot{P}$. Based on the monitoring of the recent flare in May 2020, the X-ray flux dropped by a factor of four at the beginning of the outburst and returned to its pre-outburst level in $\sim 8$ days \citep{2020arXiv200903321P}. This shares some similarities to the abrupt drop in the X-ray flux detected for 1ES 1927$+$65 \citep{2020ApJ...898L...1R}, which was interpreted to be due to the interaction of the tidal stream with the accretion disk and the subsequent depletion of the inner disk portion and the related hard X-ray corona. 
    
 \textit{Near-circular IMBH:} In the second scenario, closer to our simulations, we propose a circular orbit of a star that hits the accretion disk twice during its orbital period, hence $P_{\rm orb}\sim 2P_0/(1+z)\sim 219.09$ days, for which $a_{\star}\sim 415\,M$. In order to reach $\mathcal{R}/a_{\star}\sim 0.01-0.1$ as in our simulations, we require the stagnation radius of $\mathcal{R} \sim 600-6000\,R_{\odot}$, which is only possible for low accretion-flow number densities of $20000\,{\rm cm^{-3}}$ and $200\,{\rm cm^{-3}}$, respectively. This is unlikely at these scales for Type II AGN. We obtain a better consistency for the intermediate-mass black hole (IMBH) as a perturber orbiting on a circular trajectory. Using Eq.~\eqref{eq:Rsync} for our semi-major axis and the SMBH mass, we obtain the synchronization radius of $R_{\rm sync}=0.1-1\,M$ for $m_{\star}=2\times 10^4\,M_{\odot}-2\times 10^5\,M_{\odot}$. However, in this case, it is more difficult to explain the large period derivative as the distance is too large for gravitational radiation to cause these large losses and the IMBH-disc interactions would need to be modelled in detail.

The observed optical outbursts of ESO 253--G003 have similar asymmetric profiles as the outbursts in the accretion rate $\dot{\mathcal{M}}$ for our simulated runs, see Figs.~\ref{fig:Mdot-2D} and \ref{fig:Mdot-runH}. In particular, there is a steep increase and a gradual fall-off, which can be modelled as an exponential decay with time. The duration of the $\dot{\mathcal{M}}$ flares in our runs is $\sim 200$--$300\,M$, which gives $\sim 0.8$--$1.2$ days for the SMBH mass of ESO 253--G003, depending on the resolution to a certain extent as we saw for run H, see Fig.~\ref{fig:Mdot-runH}. This is a similar order of magnitude as the rise time of $5.60\pm 0.05$ days as seen by TESS for the optical outbursts. However, detailed radiative-transfer modelling is necessary for a direct comparison.

\subsubsection{ GSN 069}
\label{subsub_GSN069}
In contrast to RE J1034+396 (Section~\ref{sec_REJ}), which exhibits nearly sinusoidal X-ray flares, and ESO 253--G003 (Section~\ref{sec_ESO}) with broader optical asymmetric flares, Seyfert 2 galaxy GSN 069 (2MASX J0119086--3411305; $z=0.018$) experiences very narrow symmetric flares in the X-ray domain, or rather X-ray quasi-periodic eruptions (QPE), with the recurrence time of approximately nine hours and the flare duration of $75$ minutes \citep{2019Natur.573..381M}. During the eruptions, the count rate increases by a factor of $\sim 10$ up to $\sim 30$, with the QPE amplitude having the flux density by a factor of $\sim 100$ larger than during the quiescent level in the higher energy band of $0.8-1.0\,{\rm keV}$ and decreasing towards softer energy bands. Let us note that currently the origin of QPE and the mechanism of their recurrence are unknown.

\begin{figure*}
    \centering
    \includegraphics[width=0.3\textwidth]{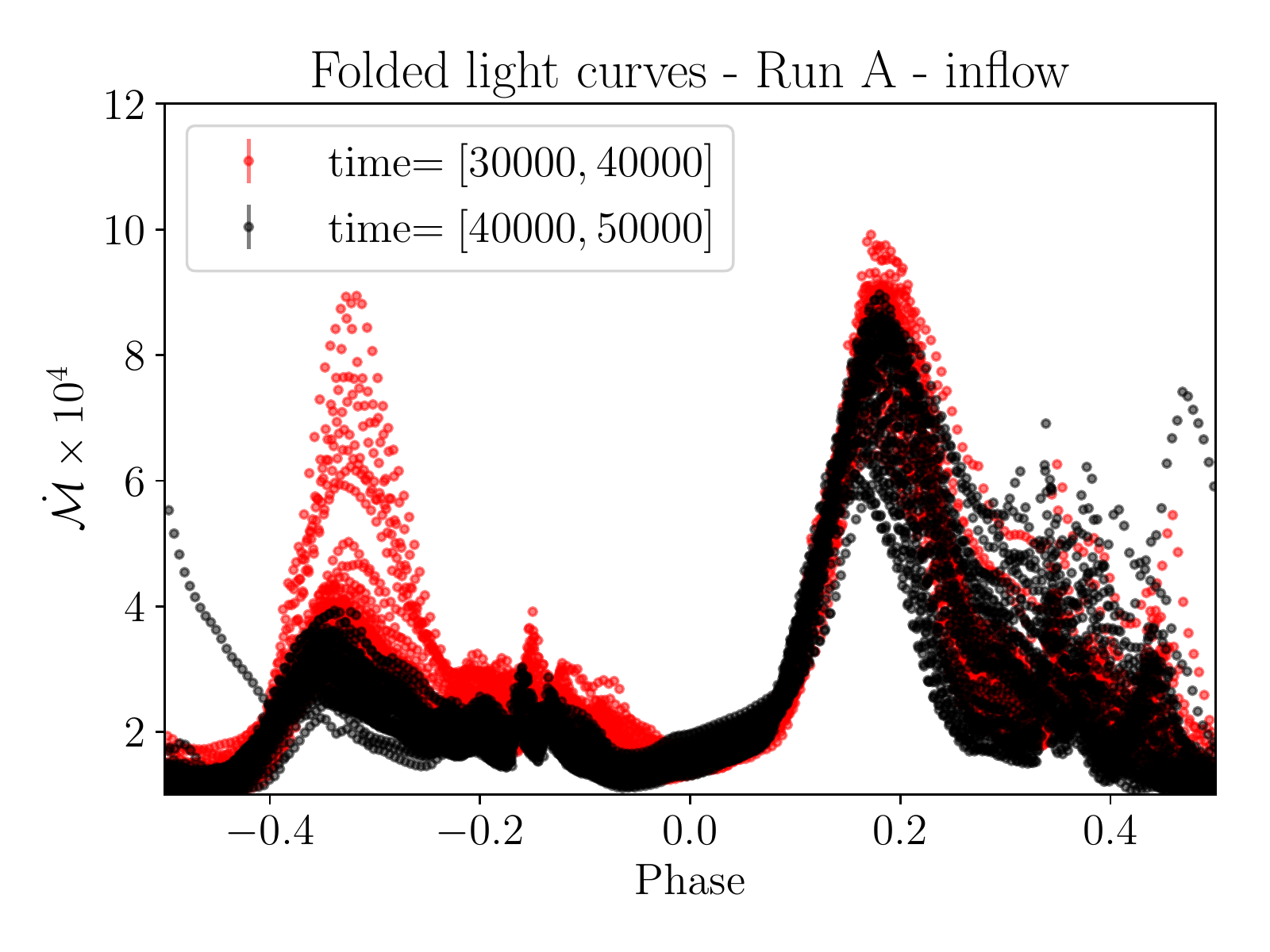}
    \includegraphics[width=0.3\textwidth]{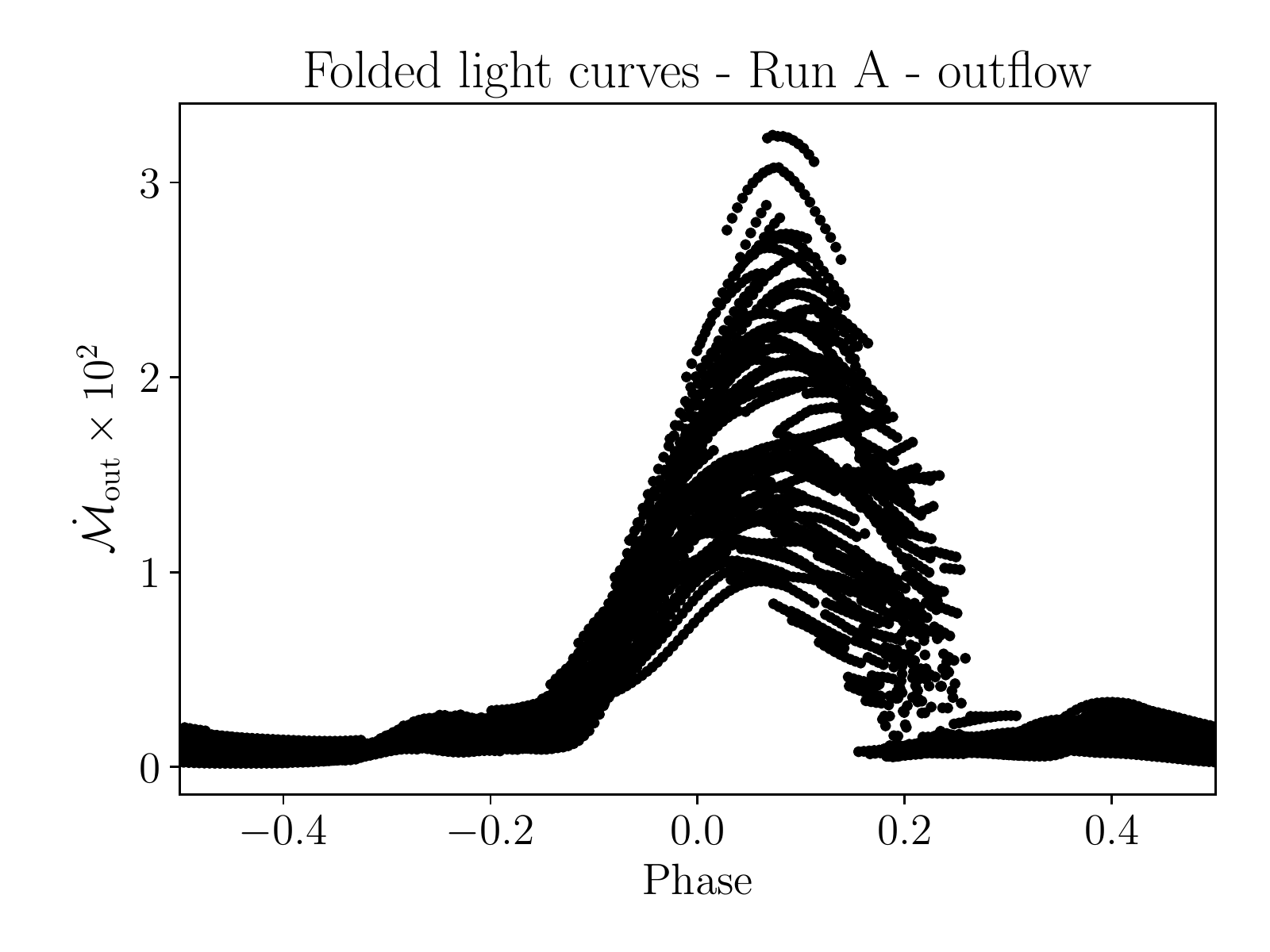}
    \includegraphics[width=0.3\textwidth]{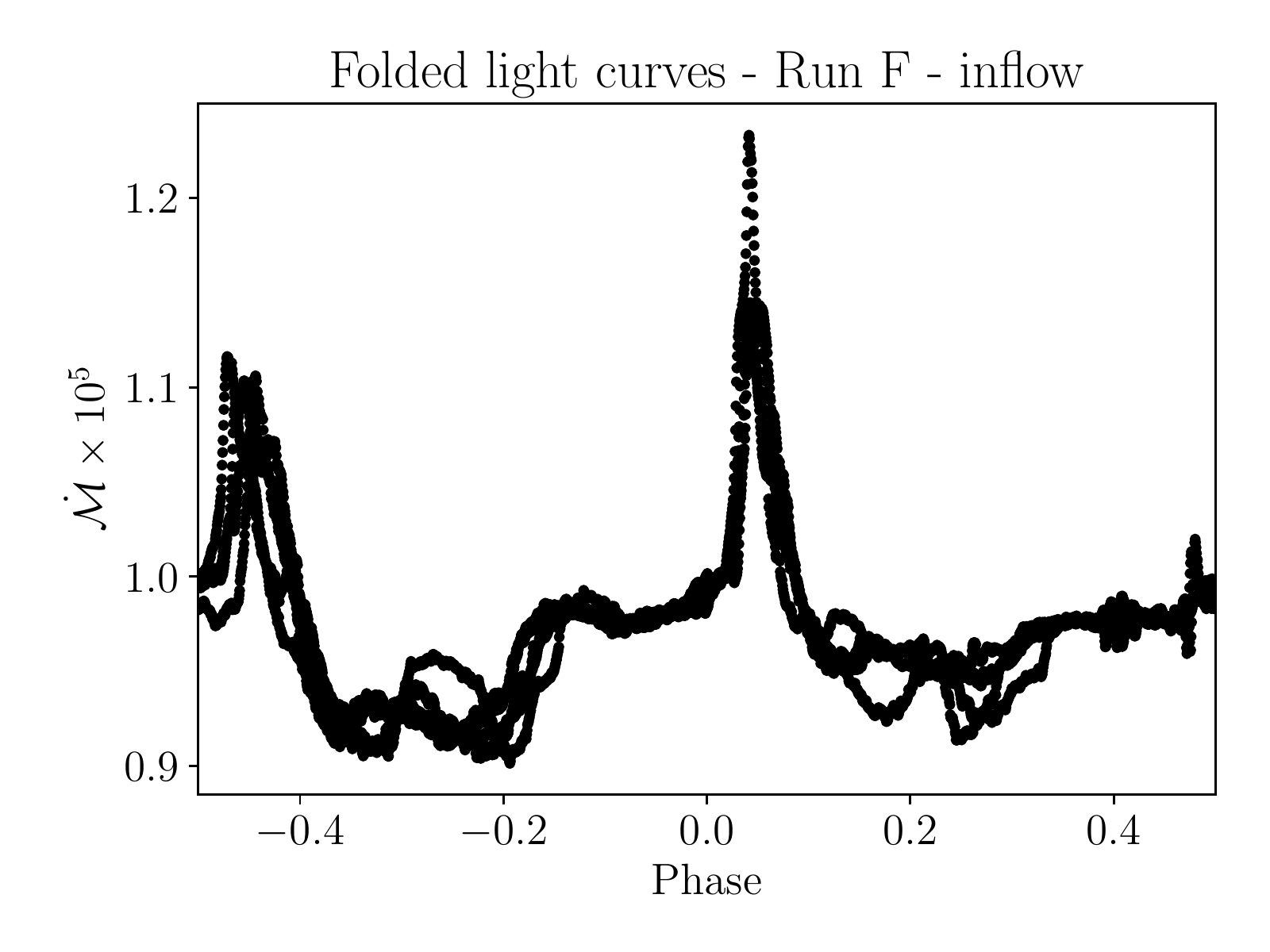}
    \caption{Folded inflow and outflow rates with respect to the orbital period for Runs A ($202\,M$) and F ($2225\,M$). \textbf{Left panel:} A folded inflow rate for Run A, considering the temporal range from $30\,000\,M$ up to $50\,000\,M$. The red points denote the data in the range $[30\,000\,M,40\,000\,M]$, while the black points stand for the time range $[40\,000\,M,50\,000\,M]$. \textbf{Middle panel:} A folded outflow rate for Run A, considering the temporal range from $40\,000\,M$ up to $50\,000\,M$. \textbf{Right panel:} A folded inflow rate for Run F, considering the temporal range from $90\,000\,M$ to $100\,000\,M$.}
    \label{fig_folded_light_curves}
\end{figure*}

The periodicity of eruptions is getting longer, from $P_{\rm erupt}=29\,800\,{\rm s}$ (December 24,  2019), through $32\,150\,{\rm s}$ (January 16/17, 2019), up to $32\,700\,{\rm s}$ (February 14/15, 2019). If we consider a regular vertical perturbation of the accretion flow around the SMBH of $M=4\times 10^5\,M_{\odot}$, then the orbital period of a perturbing body moving on a nearly circular orbit is $P_{\rm orb}\sim 2P_{\rm erupt}\simeq 16.26-17.85$ hours in the co-moving frame, from which we obtain the distance range $r=282.1\,M$, $296.8\,M$, and $300.1\,M$, respectively. We note that the actual true period of the system really appears to be $P_{\rm orb}=2P_{\rm erupt}\sim 18$ hours, since a larger flare is followed by a smaller flare in just something over nine hours, which is again followed by a more luminous flare in just slightly less than nine hours \citep{2021MNRAS.503.1703I}.

First, we consider a wind-blowing star as a perturber. For the estimated bolometric luminosity of $L_{\rm bol}=(0.09-4.8)\times 10^{43}\,{\rm erg\,s^{-1}}$, we obtain the accretion rate of $\dot{M}=10^{22}-5.3\times 10^{23}\,{\rm g\,s^{-1}}$ ($1.6\times 10^{-4}\,{\rm M_{\odot}\,yr^{-1}}$-$8.45\times 10^{-3}\,{\rm M_{\odot}\,yr^{-1}}$), assuming $\sim 10\%$ efficiency. For the S2-like star with $\dot{m}_{\rm w}\sim 3\times 10^{-7}\,M_{\odot}\,{\rm yr^{-1}}$ and $v_{\rm w}\sim 10^3\,{\rm km\,s^{-1}}$, we obtain the stagnation radius of $\mathcal{R}\sim 10^{-4}\,M$ for the distance of $\sim 300\,M$, which results in $\mathcal{R}/r\sim 10^{-7}$ that is six orders of magnitude below the ratios for Run A and F (see Table~\ref{Table:runs}). Hence, a wind-blowing star is not expected to significantly perturb the flow to produce QPEs seen for GSN 069 unless a different mechanism than the mechanical perturbation of the flow is involved. \citet{2020MNRAS.493L.120K} proposed that the accretion from a low-mass white dwarf ($\sim 0.21\,M_{\odot}$) could power the soft X-ray flares via the Roche-lobe overflow during the pericenter passage of a highly eccentric orbit. \citet{2020MNRAS.493L.120K} discusses that the low-mass white dwarf is likely a remnant core of a red giant. Having the orbital period of $\sim 9$ hours, the white dwarf is supposed to have a semi-major axis of $\sim 190\,M$, with the pericenter distance of $\sim 11\,M$, considering the current eccentricity of $e=0.94$.       

A case of the intermediate-mass black hole, which would be comparable in mass or slightly less massive than the primary one, $m_{\rm IMBH}\sim 10^5\,M_{\odot}$, would produce the synchronization radius of $\mathcal{R}\sim R_{\rm sync}\sim 75\,M$ for the distance of $r\sim 300\,M$, see Eq.~\eqref{eq:Rsync}. Hence, the ratio $\mathcal{R}/r\sim 0.25$ is then comparable to our Run~F. The scenario of two orbiting massive black holes has similarities to the self-lensing model of \citet{2021MNRAS.503.1703I}, where the authors provide a possible interpretation of the QPEs via the lensing of the background accretion minidisc by the foreground black hole. However, the authors conclude that this model cannot cover all the features of the flares simultaneously, specifically their width, magnitude as well as the strong dependence of the flare magnitude on the X-ray energy. 

Since our methodology does not include the radiative transfer, we do not calculate the light curves for the comparison with the XMM and the Chandra light curves of GSN 069. However, for the limit of the optically thin approximation, we at least compare qualitatively the inflow- and the outflow-rate flares for Runs A and F, see Fig.~\ref{fig_folded_light_curves}. When $\dot{\mathcal{M}}$ and $\dot{\mathcal{M}}_{\rm out}$ are folded according to the orbital frequency, we see that they are actually two peaks for the inflow rate, see the left panel (Run~A) as well as the right panel (Run~F), while for the outflow rate, only one peak is apparent (middle panel, Run~A). During the flares, the increase in the inflow rate is almost an order of magnitude for the larger peak of Run~A, while for Run~F, we get a factor of only $\sim 1.3$ for the more prominent peak. For the outflow rate of Run~A, the overall increase can exceed an order of magnitude, but the peak is variable, see the middle panel of Fig.~\ref{fig_folded_light_curves}. Overall, we reach a qualitative consistency with the observations in terms of alternating brighter and fainter flares during the true period and the inflow as well as the outflow peaks are relatively symmetric and narrow with respect to the total phase (especially for Run~F). Furthermore, the magnitude of the peaks as well as their ratio depends on the density of the accretion disc. While for Run~A we get a prominent increase by an order of magnitude, for Run~F we detect only mild flares. This is due to the fact that towards the end of Run~F, the flow is already largely depleted at the distance of the perturber. Furthermore, for Run~A, we can also observe the gradual change of the magnitude and the ratio of the peaks between the earlier epochs ($30\,000-40\,000\,M$, red points in Fig.~\ref{fig_folded_light_curves}) and the later epochs ($40\,000-50\,000\,M$, black points in Fig.~\ref{fig_folded_light_curves}), which can be interpreted by the evacuation of the part of the disc by the perturber. For a better quantitative comparison, we would need to extend the orbit of the perturber up to $\sim 300\,M$ and include radiative-transfer calculations.

\subsubsection{ Other QPE sources}
\label{subsub_otherQPE}

\begin{table*}
    \centering
     \caption{Overview of the sources that show the quasiperiodic behaviour in their light curves. This periodic behaviour can be interpreted by the perturbation of the accretion flow by the star or a compact object. In the second column, we specify the perturber type and the third column contains an estimate of the perturber distance from the central SMBH.}
    \begin{tabular}{ccc}
    \hline
    \hline
    Source & Perturber & Distance [$M$] \\
    \hline
    Sgr~A* & stars S4714 and S62  & $\sim 320$--$451$ \\ 
    OJ287  & SMBH $\sim 10^8\,M_{\odot}$  & $\sim 144$ \\ 
    J0849$+$5108 & SMBH $\sim 5\times 10^5-2.5\times 10^6\,M_{\odot}$ & $\sim 536$\\
    RE J1034$+$396 & star/IMBH $\gtrsim 100\,M_{\odot}$  & $\sim 24$ \\ 
    1ES 1927$+$65  & (partial) TDE  & $\sim 21$--$152$ \\ 
    ESO 253-G003  & IMBH $\gtrsim 2\times 10^4\,M_{\odot}$, & $\sim 262$--$415$\\
    & partial TDE&$\sim 26$\\
    GSN 069 & IMBH $\sim 10^5\,M_{\odot}$ & $\sim 300$\\
                 & white dwarf $\sim 0.21\,M_{\odot}$ & $\sim 11-190$\\
    RX J1301.9$+$2747 & IMBH $\sim (0.8-3)\times 10^5\,M_{\odot}$ & $\sim 52-121$\\    
    eRO-QPE1 & SMBH $\sim 10^6-10^4\,M_{\odot}$ & $\sim 55-1190$\\
             & ``S2-type'' star            & $\sim 55-1190$\\
    eRO-QPE2 & SMBH $\sim 10^6-10^4\,M_{\odot}$ & $\sim 15-315$\\
             & ``S2-type'' star            & $\sim 15-315$\\
    \hline
    \end{tabular}
    \label{tab_source_perturber}
\end{table*}

The QPE phenomenon, i.e. repetitive, bright, and relatively narrow X-ray flares, has also been confirmed in a few other sources. One of them is the edge-on AGN RX J1301.9$+$2747 ($z=0.02358$), which has similar features to GSN 069, in particular the soft X-ray spectrum dominated by the thermal emission of the accretion disc, while the flares are described well by the thermal bremsstrahlung or the Comptonization of the thermal disc seed photons \citep{2020A&A...636L...2G}. As for GSN 069, RX J1301 exhibits a similar large/small flare behaviour with a more pronounced long/short recurrence time alternation. The longer recurrence time is $\sim 20\,{\rm ks}$ ($\sim 5.6$ hours) and the shorter one is $\sim 13.5\,{\rm ks}$ ($3.8$ hours), which implies an eccentric orbit for a potential perturbing body. \citet{2021MNRAS.503.1703I} constrain the orbital eccentricity to $e\gtrsim 0.16$. Given the SMBH mass estimate of $(0.8-2.8)\times 10^6\,M_{\odot}$ \citep{2017ApJ...837....3S} and the total rest-frame orbital period of a perturber, $P_{\rm orb}=(20+13.5)\,{\rm ks}/(1+z)\simeq 32728\,{\rm s}=9.09\,{\rm h}$, the semi-major axis of a second body is $a\simeq 52-121\,M$ considering the SMBH mass uncertainties. Within the framework of our simulations, the ratio of the cross-sectional radius to the distance of the order of $\mathcal{R}/r\sim 0.1$ is achieved for a secondary black hole of mass $\sim 8\times 10^4-3\times 10^5\,M_{\odot}$ according to Eq.~\eqref{eq:Rsync}. Again, this scenario has similarities to the self-lensing configuration of two black holes as presented in \citet{2021MNRAS.503.1703I}, but according to our simulations, the perturbations of the accretion flow do not require the close to edge-on geometry.

During the blind search of half of the X-ray sky using \textit{eROSITA}, \citet{2021arXiv210413388A} identified two more QPE sources -- eRO-QPE1 (2MASS02314715-1020112; $z=0.0505$) and eRO-QPE2 (2MASX J02344872-4419325; $z=0.0175$) -- whose hosts are quiescent, non-active galaxies, in contrast with GSN 069 and RX J1301. For eRO-QPE1, the characteristic recurrence timescale is $P_{\rm QPE}=18.5$ hours, which corresponds to the rest-frame orbital timescale of $P_{\rm orb}=2P_{\rm QPE}/(1+z)\sim 35.2$ hours. If we consider the lower limit on the primary black hole mass of $m_1\sim 10^5\,M_{\odot}$, we obtain the distance of a secondary black hole at $r\sim 1190\,M$ and its mass should be $m_2\gtrsim 10^4\,M_{\odot}$ to create the synchronization radius of $R_{\rm sync}\gtrsim 119\,M$ (and hence $\mathcal{R}/r\gtrsim 0.1$). For the upper primary black hole mass of $m_2\sim 10^7\,M_{\odot}$, we analogously obtain the estimates $r\sim 55\,M$, $m_2\gtrsim 10^6\,M_{\odot}$, and $R_{\rm sync}\gtrsim 5.5\,M$ to reach the same ratio $\mathcal{R}/r\gtrsim 0.1$. Concerning eRO-QPE2, the observed recurrence time $P_{\rm QPE}=2.44$ hours implies the rest-frame orbital time-scale of $P_{\rm orb}=2P_{\rm QPE}/(1+z)=4.8$ hours. Considering the same range for the primary black hole mass, $m_1=10^5-10^7\,M_{\odot}$, we obtain the distance range of a perturber, $r=315-15\,M$, and the mass range $m_2=10^4-10^6\,M_{\odot}$ to yield the cross-sectional radius/distance ratio of $\mathcal{R}/r\sim 0.1$. Since both host galaxies are quiescent and in principle the primary black hole could be of Sgr~A* type fueled by an ADAF with a low accretion rate and surrounded by a dense Nuclear Star Cluster, the perturbers in both cases could be massive OB stars with powerful winds, such as the S2 star. For some ADAF solutions, in particular with the density slope of $-1$, the ratio $\mathcal{R}/r$ is effectively constant across a broad range of distances, see Fig.~\ref{fig_ratio_Rr}. For Sgr~A*, the ratio is $\mathcal{R}/r\gtrsim 0.1$ for the mass-loss rate and the wind velocity comparable to the S2 star.     

\smallskip

For clarity, we provide an overview of promising sources, the likely perturbers of their accretion flows, and their distance in Table~\ref{tab_source_perturber}.


\section{Conclusions}
\label{sec:conclusions}
In this paper, we simulated the repetitive transits of a star through a radiatively inefficient accretion flow in the close vicinity of a rotating supermassive black hole with $a=0.5$, at the distance between 10\,$M$ and 50\,$M$. 

We observed the influence of the perturbation with several resulting features:

\begin{itemize}[leftmargin=*]
\item The accretion rate is affected by the presence of the star. Depending on the orbital parameters of the star, the accretion rate can decrease by as much as three orders of magnitude. 

\item The accretion rate time series is modulated by the passages of the star. The accretion-rate periodogram yields significant peaks at frequencies which are related to the motion of the star. In particular, we can detect a clear peak for both 2D and 3D runs at the frequency corresponding to the double of its orbital frequency, which is caused by the fact that the star passes through the accretion flow twice during one orbital period. 

\item The transits of the star can expel blobs of matter into the funnel region, which are then accelerated by the ordered magnetic field to mildly relativistic speeds along the boundary between the funnel and the torus. This process is particularly effective if the star penetrates into the funnel region.

\item The amount of outflowing matter is usually about one tenth of the accreted mass for our chosen spin $a=0.5$. In case of a star that is nearby the horizon and moves close to the rotational axis (Run A) the outflowing rate can be more than one order of magnitude higher than the accretion rate.

\item The formation of the MAD state and the related episodic accretion can be accelerated with respect to the unperturbed torus evolution due to the evacuation of the gas from the inner part of the torus. Moreover, the star drags the magnetic field along its motion, which adds the poloidal component that is essential to the emergence of the MAD state. The MAD state was formed in the flow in case of stars embedded in the torus.

\item In the MAD state, the individual events of the episodic accretion are accompanied by blasts of outflowing matter into the funnel, while during the dips in the accretion rate the outflow is ceased even though there is strong magnetic field along the axis. 

\item The duration of the dips and partially also of the peaks during the MAD state depends on the spatial and temporal discretization. This is due to the fact that the magnetic resistivity, which we expect to play an important role during the MAD cycles, is captured only by means of the numerical resistivity. Hence, the quantification of the temporal properties of the flow in the MAD state require a further, more detailed study.

\item For a very large star ($\mathcal{R}=10\,M$) the accretion structure in the inner part can be completely destroyed and the accretion and outflowing rate drops down abruptly. The long term evolution then depends on the total size of the accretion disc and the supply of gas from the outer regions.

\item Most of our simulations were done in 2D, but the main properties of the resulting flow were affirmed by a 3D computation. The comparison between the outflowing rates for Run~A in 2D and Run~I in 3D with the same star parameters has shown that the effect of the star transit is enhanced by about 2-3 order of magnitudes in the 2D simulations. The question whether the star can induce emergence of the MAD state in 3D will be studied by using longer and more robust 3D simulations in future work.

\item Our estimate of the occurrence of stars in galactic nuclei based on the standard Salpeter--Kroupa IMF and the assumption of a relaxed stellar number-density cusp shows that in about 1 in 50 galaxies a star larger than our Sun can be found in the innermost region. Our scenario is thus viable for comparison with observed variability of SMBHs. We have discussed several promising sources such as  Sgr~A*, OJ 287, J0849$+$5108, RE J1034$+$396, 1ES 1927+65, ESO 253--G003, GSN 069, RX J1301.9$+$2747, eRO-QPE1, and eRO-QPE2 in Section \ref{sec:discussion}.  
\end{itemize}

\acknowledgments
The authors acknowledge the Czech Science Foundation research grant (GA\v{C}R 21-11268S) and the Czech-Polish mobility program (M\v{S}MT 8J20PL037 and PPN/BCZ/2019/1/00069). PS has been partially supported by the fellowship Lumina Quaeruntur No.\ LQ100032102 of the Czech Academy of Sciences. MZ acknowledges the financial support by the National Science Center, Poland, grant No. 2017/26/A/ST9/00756 (Maestro 9) and the NAWA financial support under the agreement PPN/WYM/2019/1/00064 to perform a three-month exchange stay at the Charles University in Prague and the Astronomical Institute of the Czech Academy of Sciences in Prague. VW was supported by European Union’s Horizon 2020 research and innovation programme under grant agreement No 894881. We would like to thank to the anonymous referee for her/his valuable comments, which helped to improve the manuscript.

\software{HARM \citep{0004-637X-589-1-444,2006ApJ...641..626N}; {\tt HARMPI} \citep{2015MNRAS.454.1848R,2007MNRAS.379..469T}; \textmyfont{numpy} \citep{2020NumPy-Array}; \textmyfont{matplotlib} \citep{hunter07}; \textmyfont{scipy} \citep{2020SciPy-NMeth}; \textmyfont{P4J} \citep{2018ApJS..236...12H}; \textmyfont{gnuplot}}; \textmyfont{BHPToolkit} \href{http://bhptoolkit.org/}{bhptoolkit.org} 

\vspace*{15mm}

\bibliography{Stellar-transits}

\begin{thebibliography}{}
\expandafter\ifx\csname natexlab\endcsname\relax\def\natexlab#1{#1}\fi
\providecommand{\url}[1]{\href{#1}{#1}}
\providecommand{\dodoi}[1]{doi:~\href{http://doi.org/#1}{\nolinkurl{#1}}}
\providecommand{\doeprint}[1]{\href{http://ascl.net/#1}{\nolinkurl{http://ascl.net/#1}}}
\providecommand{\doarXiv}[1]{\href{https://arxiv.org/abs/#1}{\nolinkurl{https://arxiv.org/abs/#1}}}

\bibitem[{{Abramowicz} {et~al.}(1978){Abramowicz}, {Jaroszynski}, \&
  {Sikora}}]{1978A&A....63..221A}
{Abramowicz}, M., {Jaroszynski}, M., \& {Sikora}, M. 1978, \aap, 63, 221

\bibitem[{{Aguero} {et~al.}(1996){Aguero}, {Paolantonio}, \&
  {Suarez}}]{1996PASP..108.1117A}
{Aguero}, E.~L., {Paolantonio}, S., \& {Suarez}, F. 1996, \pasp, 108, 1117,
  \dodoi{10.1086/133844}

\bibitem[{{Aharon} {et~al.}(2016){Aharon}, {Mastrobuono Battisti}, \&
  {Perets}}]{2016ApJ...823..137A}
{Aharon}, D., {Mastrobuono Battisti}, A., \& {Perets}, H.~B. 2016, \apj, 823,
  137, \dodoi{10.3847/0004-637X/823/2/137}

\bibitem[{{Amaro-Seoane} {et~al.}(2017){Amaro-Seoane}, {Audley}, {Babak},
  {Baker}, {Barausse}, {Bender}, {Berti}, {Binetruy}, {Born}, {Bortoluzzi},
  {Camp}, {Caprini}, {Cardoso}, {Colpi}, {Conklin}, {Cornish}, {Cutler},
  {Danzmann}, {Dolesi}, {Ferraioli}, {Ferroni}, {Fitzsimons}, {Gair}, {Gesa
  Bote}, {Giardini}, {Gibert}, {Grimani}, {Halloin}, {Heinzel}, {Hertog},
  {Hewitson}, {Holley-Bockelmann}, {Hollington}, {Hueller}, {Inchauspe},
  {Jetzer}, {Karnesis}, {Killow}, {Klein}, {Klipstein}, {Korsakova}, {Larson},
  {Livas}, {Lloro}, {Man}, {Mance}, {Martino}, {Mateos}, {McKenzie},
  {McWilliams}, {Miller}, {Mueller}, {Nardini}, {Nelemans}, {Nofrarias},
  {Petiteau}, {Pivato}, {Plagnol}, {Porter}, {Reiche}, {Robertson},
  {Robertson}, {Rossi}, {Russano}, {Schutz}, {Sesana}, {Shoemaker}, {Slutsky},
  {Sopuerta}, {Sumner}, {Tamanini}, {Thorpe}, {Troebs}, {Vallisneri},
  {Vecchio}, {Vetrugno}, {Vitale}, {Volonteri}, {Wanner}, {Ward}, {Wass},
  {Weber}, {Ziemer}, \& {Zweifel}}]{2017arXiv170200786A}
{Amaro-Seoane}, P., {Audley}, H., {Babak}, S., {et~al.} 2017, arXiv e-prints,
  arXiv:1702.00786.
\newblock \doarXiv{1702.00786}

\bibitem[{{Antonucci}(1993)}]{1993ARA&A..31..473A}
{Antonucci}, R. 1993, \araa, 31, 473,
  \dodoi{10.1146/annurev.aa.31.090193.002353}

\bibitem[{{Arcodia} {et~al.}(2021){Arcodia}, {Merloni}, {Nandra}, {Buchner},
  {Salvato}, {Pasham}, {Remillard}, {Comparat}, {Lamer}, {Ponti}, {Malyali},
  {Wolf}, {Arzoumanian}, {Bogensberger}, {Buckley}, {Gendreau}, {Gromadzki},
  {Kara}, {Krumpe}, {Markwardt}, {Ramos-Ceja}, {Rau}, {Schramm}, \&
  {Schwope}}]{2021arXiv210413388A}
{Arcodia}, R., {Merloni}, A., {Nandra}, K., {et~al.} 2021, arXiv e-prints,
  arXiv:2104.13388.
\newblock \doarXiv{2104.13388}

\bibitem[{{Attridge} {et~al.}(1999){Attridge}, {Roberts}, \&
  {Wardle}}]{1999ApJ...518L..87A}
{Attridge}, J.~M., {Roberts}, D.~H., \& {Wardle}, J. F.~C. 1999, \apjl, 518,
  L87, \dodoi{10.1086/312078}

\bibitem[{{Babak} {et~al.}(2017){Babak}, {Gair}, {Sesana}, {Barausse},
  {Sopuerta}, {Berry}, {Berti}, {Amaro-Seoane}, {Petiteau}, \&
  {Klein}}]{2017PhRvD..95j3012B}
{Babak}, S., {Gair}, J., {Sesana}, A., {et~al.} 2017, \prd, 95, 103012,
  \dodoi{10.1103/PhysRevD.95.103012}

\bibitem[{{Baganoff} {et~al.}(2003){Baganoff}, {Maeda}, {Morris}, {Bautz},
  {Brandt}, {Cui}, {Doty}, {Feigelson}, {Garmire}, {Pravdo}, {Ricker}, \&
  {Townsley}}]{2003ApJ...591..891B}
{Baganoff}, F.~K., {Maeda}, Y., {Morris}, M., {et~al.} 2003, \apj, 591, 891,
  \dodoi{10.1086/375145}

\bibitem[{{Bahcall} \& {Wolf}(1976)}]{1976ApJ...209..214B}
{Bahcall}, J.~N., \& {Wolf}, R.~A. 1976, \apj, 209, 214, \dodoi{10.1086/154711}

\bibitem[{{Bahcall} \& {Wolf}(1977)}]{1977ApJ...216..883B}
---. 1977, \apj, 216, 883, \dodoi{10.1086/155534}

\bibitem[{{Balbus} \& {Hawley}(1991)}]{1991ApJ...376..214B}
{Balbus}, S.~A., \& {Hawley}, J.~F. 1991, \apj, 376, 214,
  \dodoi{10.1086/170270}

\bibitem[{{Ballone} {et~al.}(2018){Ballone}, {Schartmann}, {Burkert},
  {Gillessen}, {Plewa}, {Genzel}, {Pfuhl}, {Eisenhauer}, {Habibi}, {Ott}, \&
  {George}}]{2018MNRAS.479.5288B}
{Ballone}, A., {Schartmann}, M., {Burkert}, A., {et~al.} 2018, \mnras, 479,
  5288, \dodoi{10.1093/mnras/sty1408}

\bibitem[{{Blandford} {et~al.}(2019){Blandford}, {Meier}, \&
  {Readhead}}]{2019ARA&A..57..467B}
{Blandford}, R., {Meier}, D., \& {Readhead}, A. 2019, \araa, 57, 467,
  \dodoi{10.1146/annurev-astro-081817-051948}

\bibitem[{{Blandford} \& {K{\"o}nigl}(1979)}]{1979ApJ...232...34B}
{Blandford}, R.~D., \& {K{\"o}nigl}, A. 1979, \apj, 232, 34,
  \dodoi{10.1086/157262}

\bibitem[{{Blandford} {et~al.}(1990){Blandford}, {Netzer}, {Woltjer}, \& {et
  al.}}]{1990agn..conf.....B}
{Blandford}, R.~D., {Netzer}, H., {Woltjer}, L., \& {et al.} 1990, {Active
  Galactic Nuclei, Saas-Fe Advaced Course (Springer: Berlin)}

\bibitem[{{Blandford} \& {Payne}(1982)}]{1982MNRAS.199..883B}
{Blandford}, R.~D., \& {Payne}, D.~G. 1982, \mnras, 199, 883,
  \dodoi{10.1093/mnras/199.4.883}

\bibitem[{{Blandford} \& {Znajek}(1977)}]{1977MNRAS.179..433B}
{Blandford}, R.~D., \& {Znajek}, R.~L. 1977, \mnras, 179, 433,
  \dodoi{10.1093/mnras/179.3.433}

\bibitem[{{Boccardi} {et~al.}(2016){Boccardi}, {Krichbaum}, {Bach}, {Bremer},
  \& {Zensus}}]{2016A&A...588L...9B}
{Boccardi}, B., {Krichbaum}, T.~P., {Bach}, U., {Bremer}, M., \& {Zensus},
  J.~A. 2016, \aap, 588, L9, \dodoi{10.1051/0004-6361/201628412}

\bibitem[{{Boettcher} {et~al.}(2012){Boettcher}, {Harris}, \&
  {Krawczynski}}]{2012rjag.book.....B}
{Boettcher}, M., {Harris}, D.~E., \& {Krawczynski}, H. 2012, {Relativistic Jets
  from Active Galactic Nuclei (Wiley: Weinheim)}

\bibitem[{Boffi \& Gastaldi(2016)}]{Boffi_2016}
Boffi, D., \& Gastaldi, L. 2016, Numerische Mathematik, 135, 711,
  \dodoi{10.1007/s00211-016-0814-1}

\bibitem[{Boilevin-Kayl {et~al.}(2019)Boilevin-Kayl, Fern{\'{a}}ndez, \&
  Gerbeau}]{Boilevin_Kayl_2019}
Boilevin-Kayl, L., Fern{\'{a}}ndez, M.~A., \& Gerbeau, J.-F. 2019, Computers
  {\&} Fluids, 179, 744, \dodoi{10.1016/j.compfluid.2018.05.024}

\bibitem[{{B{\"o}ker} {et~al.}(2002){B{\"o}ker}, {Laine}, {van der Marel},
  {Sarzi}, {Rix}, {Ho}, \& {Shields}}]{2002AJ....123.1389B}
{B{\"o}ker}, T., {Laine}, S., {van der Marel}, R.~P., {et~al.} 2002, \aj, 123,
  1389, \dodoi{10.1086/339025}

\bibitem[{{Bortolas} {et~al.}(2017){Bortolas}, {Mapelli}, \&
  {Spera}}]{2017MNRAS.469.1510B}
{Bortolas}, E., {Mapelli}, M., \& {Spera}, M. 2017, \mnras, 469, 1510,
  \dodoi{10.1093/mnras/stx930}

\bibitem[{{Bower} {et~al.}(2015){Bower}, {Markoff}, {Dexter}, {Gurwell},
  {Moran}, {Brunthaler}, {Falcke}, {Fragile}, {Maitra}, {Marrone}, {Peck},
  {Rushton}, \& {Wright}}]{2015ApJ...802...69B}
{Bower}, G.~C., {Markoff}, S., {Dexter}, J., {et~al.} 2015, \apj, 802, 69,
  \dodoi{10.1088/0004-637X/802/1/69}

\bibitem[{{Britzen} {et~al.}(2019){Britzen}, {Fendt}, {Zaja{\v{c}}ek}, {Jaron},
  {Pashchenko}, {Aller}, \& {Aller}}]{2019Galax...7...72B}
{Britzen}, S., {Fendt}, C., {Zaja{\v{c}}ek}, M., {et~al.} 2019, Galaxies, 7,
  72, \dodoi{10.3390/galaxies7030072}

\bibitem[{{Britzen} {et~al.}(2018){Britzen}, {Fendt}, {Witzel}, {Qian},
  {Pashchenko}, {Kurtanidze}, {Zajacek}, {Martinez}, {Karas}, {Aller}, {Aller},
  {Eckart}, {Nilsson}, {Ar{\'e}valo}, {Cuadra}, {Subroweit}, \&
  {Witzel}}]{2018MNRAS.478.3199B}
{Britzen}, S., {Fendt}, C., {Witzel}, G., {et~al.} 2018, \mnras, 478, 3199,
  \dodoi{10.1093/mnras/sty1026}

\bibitem[{{Brownsberger} \& {Romani}(2014)}]{2014ApJ...784..154B}
{Brownsberger}, S., \& {Romani}, R.~W. 2014, \apj, 784, 154,
  \dodoi{10.1088/0004-637X/784/2/154}

\bibitem[{{Bruni} {et~al.}(2021){Bruni}, {G{\'o}mez}, {Vega-Garc{\'\i}a},
  {Lobanov}, {Fuentes}, {Savolainen}, {Kovalev}, {Perucho}, {Mart{\'\i}},
  {Edwards}, {Gurvits}, {Lisakov}, {Pushkarev}, \&
  {Sokolovsky}}]{2021arXiv210107324B}
{Bruni}, G., {G{\'o}mez}, J.~L., {Vega-Garc{\'\i}a}, L., {et~al.} 2021, arXiv
  e-prints, arXiv:2101.07324.
\newblock \doarXiv{2101.07324}

\bibitem[{{Carollo} {et~al.}(1998){Carollo}, {Stiavelli}, \&
  {Mack}}]{1998AJ....116...68C}
{Carollo}, C.~M., {Stiavelli}, M., \& {Mack}, J. 1998, \aj, 116, 68,
  \dodoi{10.1086/300407}

\bibitem[{{Celotti} {et~al.}(2001){Celotti}, {Ghisellini}, \&
  {Chiaberge}}]{2001MNRAS.321L...1C}
{Celotti}, A., {Ghisellini}, G., \& {Chiaberge}, M. 2001, \mnras, 321, L1,
  \dodoi{10.1046/j.1365-8711.2001.04160.x}

\bibitem[{{Chan} {et~al.}(2019){Chan}, {Piran}, {Krolik}, \&
  {Saban}}]{2019ApJ...881..113C}
{Chan}, C.-H., {Piran}, T., {Krolik}, J.~H., \& {Saban}, D. 2019, \apj, 881,
  113, \dodoi{10.3847/1538-4357/ab2b40}

\bibitem[{{Chiaberge} {et~al.}(2015){Chiaberge}, {Gilli}, {Lotz}, \&
  {Norman}}]{2015ApJ...806..147C}
{Chiaberge}, M., {Gilli}, R., {Lotz}, J.~M., \& {Norman}, C. 2015, \apj, 806,
  147, \dodoi{10.1088/0004-637X/806/2/147}

\bibitem[{{C{\^o}t{\'e}} {et~al.}(2006){C{\^o}t{\'e}}, {Piatek}, {Ferrarese},
  {Jord{\'a}n}, {Merritt}, {Peng}, {Ha{\textcommabelow s}egan}, {Blakeslee},
  {Mei}, {West}, {Milosavljevi{\'c}}, \& {Tonry}}]{2006ApJS..165...57C}
{C{\^o}t{\'e}}, P., {Piatek}, S., {Ferrarese}, L., {et~al.} 2006, \apjs, 165,
  57, \dodoi{10.1086/504042}

\bibitem[{{Czerny} {et~al.}(2010){Czerny}, {Lachowicz}, {Dov{\v{c}}iak},
  {Karas}, {Pech{\'a}{\v{c}}ek}, \& {Das}}]{2010A&A...524A..26C}
{Czerny}, B., {Lachowicz}, P., {Dov{\v{c}}iak}, M., {et~al.} 2010, \aap, 524,
  A26, \dodoi{10.1051/0004-6361/200913724}

\bibitem[{{Czerny} {et~al.}(2016){Czerny}, {You}, {Kurcz},
  {{\'S}redzi{\'n}ska}, {Hryniewicz}, {Niko{\l}ajuk}, {Krupa}, {Wang}, {Hu}, \&
  {{\.Z}ycki}}]{2016A&A...594A.102C}
{Czerny}, B., {You}, B., {Kurcz}, A., {et~al.} 2016, \aap, 594, A102,
  \dodoi{10.1051/0004-6361/201628103}

\bibitem[{{Dai} {et~al.}(2010){Dai}, {Fuerst}, \&
  {Blandford}}]{2010MNRAS.402.1614D}
{Dai}, L.~J., {Fuerst}, S.~V., \& {Blandford}, R. 2010, \mnras, 402, 1614,
  \dodoi{10.1111/j.1365-2966.2009.16038.x}

\bibitem[{{D'arcangelo} {et~al.}(2009){D'arcangelo}, {Marscher}, {Jorstad},
  {Smith}, {Larionov}, {Hagen-Thorn}, {Williams}, {Gear}, {Clemens}, {Sarcia},
  {Grabau}, {Tollestrup}, {Buie}, {Taylor}, \& {Dunham}}]{2009ApJ...697..985D}
{D'arcangelo}, F.~D., {Marscher}, A.~P., {Jorstad}, S.~G., {et~al.} 2009, \apj,
  697, 985, \dodoi{10.1088/0004-637X/697/2/985}

\bibitem[{{Decin} {et~al.}(2012){Decin}, {Cox}, {Royer}, {Van Marle},
  {Vandenbussche}, {Ladjal}, {Kerschbaum}, {Ottensamer}, {Barlow}, {Blommaert},
  {Gomez}, {Groenewegen}, {Lim}, {Swinyard}, {Waelkens}, \&
  {Tielens}}]{2012A&A...548A.113D}
{Decin}, L., {Cox}, N.~L.~J., {Royer}, P., {et~al.} 2012, \aap, 548, A113,
  \dodoi{10.1051/0004-6361/201219792}

\bibitem[{{Dgani} {et~al.}(1996){Dgani}, {van Buren}, \&
  {Noriega-Crespo}}]{1996ApJ...461..927D}
{Dgani}, R., {van Buren}, D., \& {Noriega-Crespo}, A. 1996, \apj, 461, 927,
  \dodoi{10.1086/177114}

\bibitem[{{Dgani} {et~al.}(1993){Dgani}, {Walder}, \&
  {Nussbaumer}}]{1993A&A...267..155D}
{Dgani}, R., {Walder}, R., \& {Nussbaumer}, H. 1993, \aap, 267, 155

\bibitem[{{Dibi} {et~al.}(2012){Dibi}, {Drappeau}, {Fragile}, {Markoff}, \&
  {Dexter}}]{2012MNRAS.426.1928D}
{Dibi}, S., {Drappeau}, S., {Fragile}, P.~C., {Markoff}, S., \& {Dexter}, J.
  2012, \mnras, 426, 1928, \dodoi{10.1111/j.1365-2966.2012.21857.x}

\bibitem[{{Do} {et~al.}(2009){Do}, {Ghez}, {Morris}, {Yelda}, {Meyer}, {Lu},
  {Hornstein}, \& {Matthews}}]{2009ApJ...691.1021D}
{Do}, T., {Ghez}, A.~M., {Morris}, M.~R., {et~al.} 2009, \apj, 691, 1021,
  \dodoi{10.1088/0004-637X/691/2/1021}

\bibitem[{{Do} {et~al.}(2019){Do}, {Hees}, {Ghez}, {Martinez}, {Chu}, {Jia},
  {Sakai}, {Lu}, {Gautam}, {O'Neil}, {Becklin}, {Morris}, {Matthews},
  {Nishiyama}, {Campbell}, {Chappell}, {Chen}, {Ciurlo}, {Dehghanfar},
  {Gallego-Cano}, {Kerzendorf}, {Lyke}, {Naoz}, {Saida}, {Sch{\"o}del},
  {Takahashi}, {Takamori}, {Witzel}, \& {Wizinowich}}]{2019Sci...365..664D}
{Do}, T., {Hees}, A., {Ghez}, A., {et~al.} 2019, Science, 365, 664,
  \dodoi{10.1126/science.aav8137}

\bibitem[{Donea(1982)}]{Donea1982}
Donea, J. 1982, 33, 689

\bibitem[{Donea {et~al.}(2004)Donea, Huerta, Ponthot, \&
  Rodr\'{\i}guez-Ferrari}]{donea.j.huerta.a.ea:arbitrary}
Donea, J., Huerta, A., Ponthot, J.-P., \& Rodr\'{\i}guez-Ferrari. 2004, in
  Encyclopedia of Computational Mechanics (John Wiley \& Sons),
  \dodoi{10.1002/0470091355.ecm009}

\bibitem[{{Eatough} {et~al.}(2013){Eatough}, {Falcke}, {Karuppusamy}, {Lee},
  {Champion}, {Keane}, {Desvignes}, {Schnitzeler}, {Spitler}, {Kramer},
  {Klein}, {Bassa}, {Bower}, {Brunthaler}, {Cognard}, {Deller}, {Demorest},
  {Freire}, {Kraus}, {Lyne}, {Noutsos}, {Stappers}, \&
  {Wex}}]{2013Natur.501..391E}
{Eatough}, R.~P., {Falcke}, H., {Karuppusamy}, R., {et~al.} 2013, \nat, 501,
  391, \dodoi{10.1038/nature12499}

\bibitem[{{Eckart} {et~al.}(2002){Eckart}, {Genzel}, {Ott}, \&
  {Sch{\"o}del}}]{2002MNRAS.331..917E}
{Eckart}, A., {Genzel}, R., {Ott}, T., \& {Sch{\"o}del}, R. 2002, \mnras, 331,
  917, \dodoi{10.1046/j.1365-8711.2002.05237.x}

\bibitem[{{Eckart} {et~al.}(2012){Eckart}, {Garc{\'\i}a-Mar{\'\i}n}, {Vogel},
  {Teuben}, {Morris}, {Baganoff}, {Dexter}, {Sch{\"o}del}, {Witzel},
  {Valencia-S.}, {Karas}, {Kunneriath}, {Straubmeier}, {Moser}, {Sabha},
  {Buchholz}, {Zamaninasab}, {Mu{\v{z}}i{\'c}}, {Moultaka}, \&
  {Zensus}}]{2012A&A...537A..52E}
{Eckart}, A., {Garc{\'\i}a-Mar{\'\i}n}, M., {Vogel}, S.~N., {et~al.} 2012,
  \aap, 537, A52, \dodoi{10.1051/0004-6361/201117779}

\bibitem[{{Eckart} {et~al.}(2013){Eckart}, {Mu{\v{z}}i{\'c}}, {Yazici},
  {Sabha}, {Shahzamanian}, {Witzel}, {Moser}, {Garcia-Marin}, {Valencia-S.},
  {Jalali}, {Bremer}, {Straubmeier}, {Rauch}, {Buchholz}, {Kunneriath}, \&
  {Moultaka}}]{2013A&A...551A..18E}
{Eckart}, A., {Mu{\v{z}}i{\'c}}, K., {Yazici}, S., {et~al.} 2013, \aap, 551,
  A18, \dodoi{10.1051/0004-6361/201219994}

\bibitem[{{Eckart} {et~al.}(2017){Eckart}, {H{\"u}ttemann}, {Kiefer},
  {Britzen}, {Zaja{\v{c}}ek}, {L{\"a}mmerzahl}, {St{\"o}ckler}, {Valencia-S},
  {Karas}, \& {Garc{\'\i}a-Mar{\'\i}n}}]{2017FoPh...47..553E}
{Eckart}, A., {H{\"u}ttemann}, A., {Kiefer}, C., {et~al.} 2017, Foundations of
  Physics, 47, 553, \dodoi{10.1007/s10701-017-0079-2}

\bibitem[{{Eckart} {et~al.}(2019){Eckart}, {Zajacek}, {Valencia-S}, {Parsa},
  {Hosseini}, {Straubmeier}, {Horrobin}, {Subroweit}, \&
  {Tursunov}}]{2019JPhCS1258a2019E}
{Eckart}, A., {Zajacek}, M., {Valencia-S}, M., {et~al.} 2019, in Journal of
  Physics Conference Series, Vol. 1258, Journal of Physics Conference Series,
  012019, \dodoi{10.1088/1742-6596/1258/1/012019}

\bibitem[{{Edgar}(2004)}]{2004NewAR..48..843E}
{Edgar}, R. 2004, \nar, 48, 843, \dodoi{10.1016/j.newar.2004.06.001}

\bibitem[{{Ferrarese} \& {Merritt}(2000)}]{2000ApJ...539L...9F}
{Ferrarese}, L., \& {Merritt}, D. 2000, \apjl, 539, L9, \dodoi{10.1086/312838}

\bibitem[{{Fishbone} \& {Moncrief}(1976)}]{1976ApJ...207..962F}
{Fishbone}, L.~G., \& {Moncrief}, V. 1976, \apj, 207, 962,
  \dodoi{10.1086/154565}

\bibitem[{{Frisbie} {et~al.}(2020){Frisbie}, {Donahue}, {Voit}, {Connor}, {Li},
  {Sun}, {Lakhchaura}, {Werner}, \& {Grossova}}]{2020ApJ...899..159F}
{Frisbie}, R. L.~S., {Donahue}, M., {Voit}, G.~M., {et~al.} 2020, \apj, 899,
  159, \dodoi{10.3847/1538-4357/aba8a8}

\bibitem[{Gammie {et~al.}(2003)Gammie, McKinney, \&
  Tóth}]{0004-637X-589-1-444}
Gammie, C.~F., McKinney, J.~C., \& Tóth, G. 2003, ApJ, 589, 444.
\newblock \url{http://stacks.iop.org/0004-637X/589/i=1/a=444}

\bibitem[{{Gaspari} {et~al.}(2012){Gaspari}, {Ruszkowski}, \&
  {Sharma}}]{2012ApJ...746...94G}
{Gaspari}, M., {Ruszkowski}, M., \& {Sharma}, P. 2012, \apj, 746, 94,
  \dodoi{10.1088/0004-637X/746/1/94}

\bibitem[{{Genzel} {et~al.}(2010){Genzel}, {Eisenhauer}, \&
  {Gillessen}}]{2010RvMP...82.3121G}
{Genzel}, R., {Eisenhauer}, F., \& {Gillessen}, S. 2010, Reviews of Modern
  Physics, 82, 3121, \dodoi{10.1103/RevModPhys.82.3121}

\bibitem[{{Genzel} {et~al.}(2003){Genzel}, {Sch{\"o}del}, {Ott}, {Eckart},
  {Alexander}, {Lacombe}, {Rouan}, \& {Aschenbach}}]{2003Natur.425..934G}
{Genzel}, R., {Sch{\"o}del}, R., {Ott}, T., {et~al.} 2003, \nat, 425, 934,
  \dodoi{10.1038/nature02065}

\bibitem[{{Ghez} {et~al.}(2003){Ghez}, {Duch{\^e}ne}, {Matthews}, {Hornstein},
  {Tanner}, {Larkin}, {Morris}, {Becklin}, {Salim}, {Kremenek}, {Thompson},
  {Soifer}, {Neugebauer}, \& {McLean}}]{2003ApJ...586L.127G}
{Ghez}, A.~M., {Duch{\^e}ne}, G., {Matthews}, K., {et~al.} 2003, \apjl, 586,
  L127, \dodoi{10.1086/374804}

\bibitem[{{Ghisellini} {et~al.}(2005){Ghisellini}, {Tavecchio}, \&
  {Chiaberge}}]{2005A&A...432..401G}
{Ghisellini}, G., {Tavecchio}, F., \& {Chiaberge}, M. 2005, \aap, 432, 401,
  \dodoi{10.1051/0004-6361:20041404}

\bibitem[{{Gierli{\'n}ski} {et~al.}(2008){Gierli{\'n}ski}, {Middleton}, {Ward},
  \& {Done}}]{2008Natur.455..369G}
{Gierli{\'n}ski}, M., {Middleton}, M., {Ward}, M., \& {Done}, C. 2008, \nat,
  455, 369, \dodoi{10.1038/nature07277}

\bibitem[{{Gillessen} {et~al.}(2012){Gillessen}, {Genzel}, {Fritz}, {Quataert},
  {Alig}, {Burkert}, {Cuadra}, {Eisenhauer}, {Pfuhl}, {Dodds-Eden}, {Gammie},
  \& {Ott}}]{2012Natur.481...51G}
{Gillessen}, S., {Genzel}, R., {Fritz}, T.~K., {et~al.} 2012, \nat, 481, 51,
  \dodoi{10.1038/nature10652}

\bibitem[{{Giovannini} {et~al.}(2018){Giovannini}, {Savolainen}, {Orienti},
  {Nakamura}, {Nagai}, {Kino}, {Giroletti}, {Hada}, {Bruni}, {Kovalev},
  {Anderson}, {D'Ammando}, {Hodgson}, {Honma}, {Krichbaum}, {Lee}, {Lico},
  {Lisakov}, {Lobanov}, {Petrov}, {Sohn}, {Sokolovsky}, {Voitsik}, {Zensus}, \&
  {Tingay}}]{2018NatAs...2..472G}
{Giovannini}, G., {Savolainen}, T., {Orienti}, M., {et~al.} 2018, Nature
  Astronomy, 2, 472, \dodoi{10.1038/s41550-018-0431-2}

\bibitem[{{Giroletti} {et~al.}(2004){Giroletti}, {Giovannini}, {Feretti},
  {Cotton}, {Edwards}, {Lara}, {Marscher}, {Mattox}, {Piner}, \&
  {Venturi}}]{2004ApJ...600..127G}
{Giroletti}, M., {Giovannini}, G., {Feretti}, L., {et~al.} 2004, \apj, 600,
  127, \dodoi{10.1086/379663}

\bibitem[{{Giustini} {et~al.}(2020){Giustini}, {Miniutti}, \&
  {Saxton}}]{2020A&A...636L...2G}
{Giustini}, M., {Miniutti}, G., \& {Saxton}, R.~D. 2020, \aap, 636, L2,
  \dodoi{10.1051/0004-6361/202037610}

\bibitem[{{Gravity Collaboration} {et~al.}(2018{\natexlab{a}}){Gravity
  Collaboration}, {Abuter}, {Amorim}, {Anugu}, {Baub{\"o}ck}, {Benisty},
  {Berger}, {Blind}, {Bonnet}, {Brandner}, {Buron}, {Collin}, {Chapron},
  {Cl{\'e}net}, {Coud{\'e} Du Foresto}, {de Zeeuw}, {Deen},
  {Delplancke-Str{\"o}bele}, {Dembet}, {Dexter}, {Duvert}, {Eckart},
  {Eisenhauer}, {Finger}, {F{\"o}rster Schreiber}, {F{\'e}dou}, {Garcia},
  {Garcia Lopez}, {Gao}, {Gendron}, {Genzel}, {Gillessen}, {Gordo}, {Habibi},
  {Haubois}, {Haug}, {Hau{\ss}mann}, {Henning}, {Hippler}, {Horrobin},
  {Hubert}, {Hubin}, {Jimenez Rosales}, {Jochum}, {Jocou}, {Kaufer}, {Kellner},
  {Kendrew}, {Kervella}, {Kok}, {Kulas}, {Lacour}, {Lapeyr{\`e}re}, {Lazareff},
  {Le Bouquin}, {L{\'e}na}, {Lippa}, {Lenzen}, {M{\'e}rand}, {M{\"u}ler},
  {Neumann}, {Ott}, {Palanca}, {Paumard}, {Pasquini}, {Perraut}, {Perrin},
  {Pfuhl}, {Plewa}, {Rabien}, {Ram{\'\i}rez}, {Ramos}, {Rau},
  {Rodr{\'\i}guez-Coira}, {Rohloff}, {Rousset}, {Sanchez-Bermudez},
  {Scheithauer}, {Sch{\"o}ller}, {Schuler}, {Spyromilio}, {Straub},
  {Straubmeier}, {Sturm}, {Tacconi}, {Tristram}, {Vincent}, {von Fellenberg},
  {Wank}, {Waisberg}, {Widmann}, {Wieprecht}, {Wiest}, {Wiezorrek}, {Woillez},
  {Yazici}, {Ziegler}, \& {Zins}}]{2018A&A...615L..15G}
{Gravity Collaboration}, {Abuter}, R., {Amorim}, A., {et~al.}
  2018{\natexlab{a}}, \aap, 615, L15, \dodoi{10.1051/0004-6361/201833718}

\bibitem[{{Gravity Collaboration} {et~al.}(2018{\natexlab{b}}){Gravity
  Collaboration}, {Abuter}, {Amorim}, {Baub{\"o}ck}, {Berger}, {Bonnet}, {Brand
  ner}, {Cl{\'e}net}, {Coud{\'e} Du Foresto}, {de Zeeuw}, {Deen}, {Dexter},
  {Duvert}, {Eckart}, {Eisenhauer}, {F{\"o}rster Schreiber}, {Garcia}, {Gao},
  {Gendron}, {Genzel}, {Gillessen}, {Guajardo}, {Habibi}, {Haubois}, {Henning},
  {Hippler}, {Horrobin}, {Huber}, {Jim{\'e}nez-Rosales}, {Jocou}, {Kervella},
  {Lacour}, {Lapeyr{\`e}re}, {Lazareff}, {Le Bouquin}, {L{\'e}na}, {Lippa},
  {Ott}, {Panduro}, {Paumard}, {Perraut}, {Perrin}, {Pfuhl}, {Plewa}, {Rabien},
  {Rodr{\'\i}guez-Coira}, {Rousset}, {Sternberg}, {Straub}, {Straubmeier},
  {Sturm}, {Tacconi}, {Vincent}, {von Fellenberg}, {Waisberg}, {Widmann},
  {Wieprecht}, {Wiezorrek}, {Woillez}, \& {Yazici}}]{2018A&A...618L..10G}
---. 2018{\natexlab{b}}, \aap, 618, L10, \dodoi{10.1051/0004-6361/201834294}

\bibitem[{{Grossov{\'a}} {et~al.}(2019){Grossov{\'a}}, {Werner}, {Rajpurohit},
  {Mernier}, {Lakhchaura}, {Gab{\'a}nyi}, {Canning}, {Nulsen}, {Massaro},
  {Sun}, {Connor}, {King}, {Allen}, {Frisbie}, {Donahue}, \&
  {Fabian}}]{2019MNRAS.488.1917G}
{Grossov{\'a}}, R., {Werner}, N., {Rajpurohit}, K., {et~al.} 2019, \mnras, 488,
  1917, \dodoi{10.1093/mnras/stz1728}

\bibitem[{{G{\"u}ltekin} {et~al.}(2009){G{\"u}ltekin}, {Richstone}, {Gebhardt},
  {Lauer}, {Tremaine}, {Aller}, {Bender}, {Dressler}, {Faber}, {Filippenko},
  {Green}, {Ho}, {Kormendy}, {Magorrian}, {Pinkney}, \&
  {Siopis}}]{2009ApJ...698..198G}
{G{\"u}ltekin}, K., {Richstone}, D.~O., {Gebhardt}, K., {et~al.} 2009, \apj,
  698, 198, \dodoi{10.1088/0004-637X/698/1/198}

\bibitem[{{Gvaramadze} {et~al.}(2014){Gvaramadze}, {Menten}, {Kniazev},
  {Langer}, {Mackey}, {Kraus}, {Meyer}, \&
  {Kami{\'n}ski}}]{2014MNRAS.437..843G}
{Gvaramadze}, V.~V., {Menten}, K.~M., {Kniazev}, A.~Y., {et~al.} 2014, \mnras,
  437, 843, \dodoi{10.1093/mnras/stt1943}

\bibitem[{{Habibi} {et~al.}(2017){Habibi}, {Gillessen}, {Martins},
  {Eisenhauer}, {Plewa}, {Pfuhl}, {George}, {Dexter}, {Waisberg}, {Ott}, {von
  Fellenberg}, {Baub{\"o}ck}, {Jimenez-Rosales}, \&
  {Genzel}}]{2017ApJ...847..120H}
{Habibi}, M., {Gillessen}, S., {Martins}, F., {et~al.} 2017, \apj, 847, 120,
  \dodoi{10.3847/1538-4357/aa876f}

\bibitem[{{Hardee}(2007)}]{2007ApJ...664...26H}
{Hardee}, P.~E. 2007, \apj, 664, 26, \dodoi{10.1086/518409}

\bibitem[{Harris {et~al.}(2020)Harris, Millman, van~der Walt, Gommers,
  Virtanen, Cournapeau, Wieser, Taylor, Berg, Smith, Kern, Picus, Hoyer, van
  Kerkwijk, Brett, Haldane, Fernández~del Río, Wiebe, Peterson,
  Gérard-Marchant, Sheppard, Reddy, Weckesser, Abbasi, Gohlke, \&
  Oliphant}]{2020NumPy-Array}
Harris, C.~R., Millman, K.~J., van~der Walt, S.~J., {et~al.} 2020, Nature, 585,
  357–362, \dodoi{10.1038/s41586-020-2649-2}

\bibitem[{Hawley {et~al.}(2011)Hawley, Guan, \& Krolik}]{Hawley_2011}
Hawley, J.~F., Guan, X., \& Krolik, J.~H. 2011, The Astrophysical Journal, 738,
  84, \dodoi{10.1088/0004-637x/738/1/84}

\bibitem[{{Hills}(1975)}]{Hills_1975}
{Hills}, J.~G. 1975, \nat, 254, 295, \dodoi{10.1038/254295a0}

\bibitem[{{Holoien} {et~al.}(2014){Holoien}, {Kiyota}, {Brimacombe},
  {Schneider}, {Sheppard}, {Prieto}, {Grupe}, {Stanek}, {Kochanek}, {Davis},
  {Simonian}, {Basu}, {Beacom}, {Shappee}, {Bersier}, {Szczygiel}, {Pojmanski},
  {Conseil}, {Nicholls}, \& {Nicolas}}]{2014ATel.6732....1H}
{Holoien}, T.~W.~S., {Kiyota}, S., {Brimacombe}, J., {et~al.} 2014, The
  Astronomer's Telegram, 6732, 1

\bibitem[{{Holoien} {et~al.}(2017){Holoien}, {Stanek}, {Kochanek}, {Shappee},
  {Prieto}, {Brimacombe}, {Bersier}, {Bishop}, {Dong}, {Brown}, {Danilet},
  {Simonian}, {Basu}, {Beacom}, {Falco}, {Pojmanski}, {Skowron}, {Wo{\'z}niak},
  {{\'A}vila}, {Conseil}, {Contreras}, {Cruz}, {Fern{\'a}ndez}, {Koff}, {Guo},
  {Herczeg}, {Hissong}, {Hsiao}, {Jose}, {Kiyota}, {Long}, {Monard},
  {Nicholls}, {Nicolas}, \& {Wiethoff}}]{2017MNRAS.464.2672H}
{Holoien}, T.~W.~S., {Stanek}, K.~Z., {Kochanek}, C.~S., {et~al.} 2017, \mnras,
  464, 2672, \dodoi{10.1093/mnras/stw2273}

\bibitem[{Holz \& Hughes(2005)}]{holz2005using}
Holz, D.~E., \& Hughes, S.~A. 2005, The Astrophysical Journal, 629, 15

\bibitem[{{Hosseini} {et~al.}(2020){Hosseini}, {Zaja{\v{c}}ek}, {Eckart},
  {Sabha}, \& {Labadie}}]{2020A&A...644A.105H}
{Hosseini}, S.~E., {Zaja{\v{c}}ek}, M., {Eckart}, A., {Sabha}, N.~B., \&
  {Labadie}, L. 2020, \aap, 644, A105, \dodoi{10.1051/0004-6361/202037724}

\bibitem[{{Huijse} {et~al.}(2018){Huijse}, {Est{\'e}vez}, {F{\"o}rster},
  {Daniel}, {Connolly}, {Protopapas}, {Carrasco}, \&
  {Pr{\'\i}ncipe}}]{2018ApJS..236...12H}
{Huijse}, P., {Est{\'e}vez}, P.~A., {F{\"o}rster}, F., {et~al.} 2018, \apjs,
  236, 12, \dodoi{10.3847/1538-4365/aab77c}

\bibitem[{{Hunter}(2007)}]{hunter07}
{Hunter}, J.~D. 2007, Computing in Science and Engineering, 9, 90,
  \dodoi{10.1109/MCSE.2007.55}

\bibitem[{{Igumenshchev}(2008)}]{2008ApJ...677..317I}
{Igumenshchev}, I.~V. 2008, \apj, 677, 317, \dodoi{10.1086/529025}

\bibitem[{{Igumenshchev} {et~al.}(2003){Igumenshchev}, {Narayan}, \&
  {Abramowicz}}]{2003ApJ...592.1042I}
{Igumenshchev}, I.~V., {Narayan}, R., \& {Abramowicz}, M.~A. 2003, \apj, 592,
  1042, \dodoi{10.1086/375769}

\bibitem[{{Ingram} {et~al.}(2021){Ingram}, {Motta}, {Aigrain}, \&
  {Karastergiou}}]{2021MNRAS.503.1703I}
{Ingram}, A., {Motta}, S.~E., {Aigrain}, S., \& {Karastergiou}, A. 2021,
  \mnras, 503, 1703, \dodoi{10.1093/mnras/stab609}

\bibitem[{{Jin} {et~al.}(2020){Jin}, {Done}, \& {Ward}}]{2020MNRAS.495.3538J}
{Jin}, C., {Done}, C., \& {Ward}, M. 2020, \mnras, 495, 3538,
  \dodoi{10.1093/mnras/staa1356}

\bibitem[{{Jin} {et~al.}(2021){Jin}, {Done}, \& {Ward}}]{2021MNRAS.500.2475J}
---. 2021, \mnras, 500, 2475, \dodoi{10.1093/mnras/staa3386}

\bibitem[{{Karas} {et~al.}(2017){Karas}, {Kop{\'a}{\v{c}}ek}, {Kunneriath},
  {Zaja{\v{c}}ek}, {Araudo}, {Eckart}, \&
  {Kov{\'a}{\v{r}}}}]{2017CoSka..47..124K}
{Karas}, V., {Kop{\'a}{\v{c}}ek}, O., {Kunneriath}, D., {et~al.} 2017,
  Contributions of the Astronomical Observatory Skalnate Pleso, 47, 124.
\newblock \doarXiv{1705.09820}

\bibitem[{{Karas} \& {{\v{S}}ubr}(2001)}]{2001A&A...376..686K}
{Karas}, V., \& {{\v{S}}ubr}, L. 2001, \aap, 376, 686,
  \dodoi{10.1051/0004-6361:20011009}

\bibitem[{{Karssen} {et~al.}(2017){Karssen}, {Bursa}, {Eckart}, {Valencia-S},
  {Dov{\v{c}}iak}, {Karas}, \& {Hor{\'a}k}}]{2017MNRAS.472.4422K}
{Karssen}, G.~D., {Bursa}, M., {Eckart}, A., {et~al.} 2017, \mnras, 472, 4422,
  \dodoi{10.1093/mnras/stx2312}

\bibitem[{{Kelly} {et~al.}(2009){Kelly}, {Bechtold}, \&
  {Siemiginowska}}]{2009ApJ...698..895K}
{Kelly}, B.~C., {Bechtold}, J., \& {Siemiginowska}, A. 2009, \apj, 698, 895,
  \dodoi{10.1088/0004-637X/698/1/895}

\bibitem[{{Kennedy} {et~al.}(2016){Kennedy}, {Meiron}, {Shukirgaliyev},
  {Panamarev}, {Berczik}, {Just}, \& {Spurzem}}]{2016MNRAS.460..240K}
{Kennedy}, G.~F., {Meiron}, Y., {Shukirgaliyev}, B., {et~al.} 2016, \mnras,
  460, 240, \dodoi{10.1093/mnras/stw908}

\bibitem[{{Kim} {et~al.}(2018){Kim}, {Krichbaum}, {Lu}, {Ros}, {Bach},
  {Bremer}, {de Vicente}, {Lindqvist}, \& {Zensus}}]{2018A&A...616A.188K}
{Kim}, J.~Y., {Krichbaum}, T.~P., {Lu}, R.~S., {et~al.} 2018, \aap, 616, A188,
  \dodoi{10.1051/0004-6361/201832921}

\bibitem[{{King}(2020)}]{2020MNRAS.493L.120K}
{King}, A. 2020, \mnras, 493, L120, \dodoi{10.1093/mnrasl/slaa020}

\bibitem[{{Komissarov}(2006)}]{2006MNRAS.368..993K}
{Komissarov}, S.~S. 2006, \mnras, 368, 993,
  \dodoi{10.1111/j.1365-2966.2006.10183.x}

\bibitem[{{Komissarov} \& {Barkov}(2007)}]{2007MNRAS.382.1029K}
{Komissarov}, S.~S., \& {Barkov}, M.~V. 2007, \mnras, 382, 1029,
  \dodoi{10.1111/j.1365-2966.2007.12485.x}

\bibitem[{{Komossa}(2006)}]{2006MmSAI..77..733K}
{Komossa}, S. 2006, \memsai, 77, 733

\bibitem[{{Kozlowski} {et~al.}(1978){Kozlowski}, {Jaroszynski}, \&
  {Abramowicz}}]{1978A&A....63..209K}
{Kozlowski}, M., {Jaroszynski}, M., \& {Abramowicz}, M.~A. 1978, \aap, 63, 209

\bibitem[{{Koz{\l}owski} {et~al.}(2010){Koz{\l}owski}, {Kochanek}, {Udalski},
  {Wyrzykowski}, {Soszy{\'n}ski}, {Szyma{\'n}ski}, {Kubiak}, {Pietrzy{\'n}ski},
  {Szewczyk}, {Ulaczyk}, {Poleski}, \& {OGLE
  Collaboration}}]{2010ApJ...708..927K}
{Koz{\l}owski}, S., {Kochanek}, C.~S., {Udalski}, A., {et~al.} 2010, \apj, 708,
  927, \dodoi{10.1088/0004-637X/708/2/927}

\bibitem[{{Krolik}(1999)}]{1999agnc.book.....K}
{Krolik}, J.~H. 1999, {Active galactic nuclei: from the central black hole to
  the galactic environment (Princeton University Press: Princeton)}

\bibitem[{{Kroupa}(2001)}]{2001MNRAS.322..231K}
{Kroupa}, P. 2001, \mnras, 322, 231, \dodoi{10.1046/j.1365-8711.2001.04022.x}

\bibitem[{{Leibowitz}(2020)}]{2020ApJ...896...74L}
{Leibowitz}, E. 2020, \apj, 896, 74, \dodoi{10.3847/1538-4357/ab93c5}

\bibitem[{{Li} \& {Bryan}(2014)}]{2014ApJ...789..153L}
{Li}, Y., \& {Bryan}, G.~L. 2014, \apj, 789, 153,
  \dodoi{10.1088/0004-637X/789/2/153}

\bibitem[{{Liu} \& {Wu}(2002)}]{2002A&A...388L..48L}
{Liu}, F.~K., \& {Wu}, X.~B. 2002, \aap, 388, L48,
  \dodoi{10.1051/0004-6361:20020566}

\bibitem[{{MacLeod} {et~al.}(2010){MacLeod}, {Ivezi{\'c}}, {Kochanek},
  {Koz{\l}owski}, {Kelly}, {Bullock}, {Kimball}, {Sesar}, {Westman}, {Brooks},
  {Gibson}, {Becker}, \& {de Vries}}]{2010ApJ...721.1014M}
{MacLeod}, C.~L., {Ivezi{\'c}}, {\v{Z}}., {Kochanek}, C.~S., {et~al.} 2010,
  \apj, 721, 1014, \dodoi{10.1088/0004-637X/721/2/1014}

\bibitem[{{MacLeod} \& {Lin}(2020)}]{2020ApJ...889...94M}
{MacLeod}, M., \& {Lin}, D. N.~C. 2020, \apj, 889, 94,
  \dodoi{10.3847/1538-4357/ab64db}

\bibitem[{{Manchester} {et~al.}(2005){Manchester}, {Hobbs}, {Teoh}, \&
  {Hobbs}}]{2005AJ....129.1993M}
{Manchester}, R.~N., {Hobbs}, G.~B., {Teoh}, A., \& {Hobbs}, M. 2005, \aj, 129,
  1993, \dodoi{10.1086/428488}

\bibitem[{{Martins} {et~al.}(2008){Martins}, {Gillessen}, {Eisenhauer},
  {Genzel}, {Ott}, \& {Trippe}}]{2008ApJ...672L.119M}
{Martins}, F., {Gillessen}, S., {Eisenhauer}, F., {et~al.} 2008, \apjl, 672,
  L119, \dodoi{10.1086/526768}

\bibitem[{{Mashhoon}(1975)}]{1975ApJ...197..705M}
{Mashhoon}, B. 1975, \apj, 197, 705, \dodoi{10.1086/153560}

\bibitem[{{McCourt} {et~al.}(2012){McCourt}, {Sharma}, {Quataert}, \&
  {Parrish}}]{2012MNRAS.419.3319M}
{McCourt}, M., {Sharma}, P., {Quataert}, E., \& {Parrish}, I.~J. 2012, \mnras,
  419, 3319, \dodoi{10.1111/j.1365-2966.2011.19972.x}

\bibitem[{{McCrea}(1972)}]{1972MNRAS.157..359M}
{McCrea}, W.~H. 1972, \mnras, 157, 359, \dodoi{10.1093/mnras/157.4.359}

\bibitem[{McKinney {et~al.}(2012)McKinney, Tchekhovskoy, \&
  Blandford}]{10.1111/j.1365-2966.2012.21074.x}
McKinney, J.~C., Tchekhovskoy, A., \& Blandford, R.~D. 2012, Monthly Notices of
  the Royal Astronomical Society, 423, 3083,
  \dodoi{10.1111/j.1365-2966.2012.21074.x}

\bibitem[{{Merritt}(2013)}]{2013degn.book.....M}
{Merritt}, D. 2013, {Dynamics and Evolution of Galactic Nuclei (Princeton:
  Princeton Univ. Press)}

\bibitem[{{Mertens} {et~al.}(2016){Mertens}, {Lobanov}, {Walker}, \&
  {Hardee}}]{2016A&A...595A..54M}
{Mertens}, F., {Lobanov}, A.~P., {Walker}, R.~C., \& {Hardee}, P.~E. 2016,
  \aap, 595, A54, \dodoi{10.1051/0004-6361/201628829}

\bibitem[{{Meyer} {et~al.}(2008){Meyer}, {Do}, {Ghez}, {Morris}, {Witzel},
  {Eckart}, {B{\'e}langer}, \& {Sch{\"o}del}}]{2008ApJ...688L..17M}
{Meyer}, L., {Do}, T., {Ghez}, A., {et~al.} 2008, \apjl, 688, L17,
  \dodoi{10.1086/593147}

\bibitem[{{Meyer} {et~al.}(2006){Meyer}, {Eckart}, {Sch{\"o}del}, {Duschl},
  {Mu{\v{z}}i{\'c}}, {Dov{\v{c}}iak}, \& {Karas}}]{2006A&A...460...15M}
{Meyer}, L., {Eckart}, A., {Sch{\"o}del}, R., {et~al.} 2006, \aap, 460, 15,
  \dodoi{10.1051/0004-6361:20065925}

\bibitem[{{Miesch} {et~al.}(2015){Miesch}, {Matthaeus}, {Brandenburg},
  {Petrosyan}, {Pouquet}, {Cambon}, {Jenko}, {Uzdensky}, {Stone}, {Tobias},
  {Toomre}, \& {Velli}}]{2015SSRv..194...97M}
{Miesch}, M., {Matthaeus}, W., {Brandenburg}, A., {et~al.} 2015, \ssr, 194, 97,
  \dodoi{10.1007/s11214-015-0190-7}

\bibitem[{{Miles} {et~al.}(2020){Miles}, {Coughlin}, \&
  {Nixon}}]{2020ApJ...899...36M}
{Miles}, P.~R., {Coughlin}, E.~R., \& {Nixon}, C.~J. 2020, \apj, 899, 36,
  \dodoi{10.3847/1538-4357/ab9c9f}

\bibitem[{{Mimica} {et~al.}(2015){Mimica}, {Giannios}, {Metzger}, \&
  {Aloy}}]{2015MNRAS.450.2824M}
{Mimica}, P., {Giannios}, D., {Metzger}, B.~D., \& {Aloy}, M.~A. 2015, \mnras,
  450, 2824, \dodoi{10.1093/mnras/stv825}

\bibitem[{{Miniutti} {et~al.}(2019){Miniutti}, {Saxton}, {Giustini},
  {Alexander}, {Fender}, {Heywood}, {Monageng}, {Coriat}, {Tzioumis}, {Read},
  {Knigge}, {Gandhi}, {Pretorius}, \&
  {Ag{\'\i}s-Gonz{\'a}lez}}]{2019Natur.573..381M}
{Miniutti}, G., {Saxton}, R.~D., {Giustini}, M., {et~al.} 2019, \nat, 573, 381,
  \dodoi{10.1038/s41586-019-1556-x}

\bibitem[{{Morris} {et~al.}(2012){Morris}, {Meyer}, \&
  {Ghez}}]{2012RAA....12..995M}
{Morris}, M.~R., {Meyer}, L., \& {Ghez}, A.~M. 2012, Research in Astronomy and
  Astrophysics, 12, 995, \dodoi{10.1088/1674-4527/12/8/007}

\bibitem[{{Mushotzky} {et~al.}(2011){Mushotzky}, {Edelson}, {Baumgartner}, \&
  {Gandhi}}]{2011ApJ...743L..12M}
{Mushotzky}, R.~F., {Edelson}, R., {Baumgartner}, W., \& {Gandhi}, P. 2011,
  \apjl, 743, L12, \dodoi{10.1088/2041-8205/743/1/L12}

\bibitem[{{Mu{\v{z}}i{\'c}} {et~al.}(2010){Mu{\v{z}}i{\'c}}, {Eckart},
  {Sch{\"o}del}, {Buchholz}, {Zamaninasab}, \& {Witzel}}]{2010A&A...521A..13M}
{Mu{\v{z}}i{\'c}}, K., {Eckart}, A., {Sch{\"o}del}, R., {et~al.} 2010, \aap,
  521, A13, \dodoi{10.1051/0004-6361/200913087}

\bibitem[{{Narayan}(2000)}]{2000ApJ...536..663N}
{Narayan}, R. 2000, \apj, 536, 663, \dodoi{10.1086/308956}

\bibitem[{Narayan {et~al.}(2003)Narayan, Igumenshchev, \&
  Abramowicz}]{Narayan-MAD-2003}
Narayan, R., Igumenshchev, I.~V., \& Abramowicz, M.~A. 2003, Publications of
  the Astronomical Society of Japan, 55, L69, \dodoi{10.1093/pasj/55.6.L69}

\bibitem[{Narayan {et~al.}(2012)Narayan, Sądowski, Penna, \&
  Kulkarni}]{10.1111/j.1365-2966.2012.22002.x}
Narayan, R., Sądowski, A., Penna, R.~F., \& Kulkarni, A.~K. 2012, Monthly
  Notices of the Royal Astronomical Society, 426, 3241,
  \dodoi{10.1111/j.1365-2966.2012.22002.x}

\bibitem[{{Neumayer} {et~al.}(2011){Neumayer}, {Walcher}, {Andersen},
  {S{\'a}nchez}, {B{\"o}ker}, \& {Rix}}]{2011MNRAS.413.1875N}
{Neumayer}, N., {Walcher}, C.~J., {Andersen}, D., {et~al.} 2011, \mnras, 413,
  1875, \dodoi{10.1111/j.1365-2966.2011.18266.x}

\bibitem[{{Noble} {et~al.}(2006){Noble}, {Gammie}, {McKinney}, \& {Del
  Zanna}}]{2006ApJ...641..626N}
{Noble}, S.~C., {Gammie}, C.~F., {McKinney}, J.~C., \& {Del Zanna}, L. 2006,
  \apj, 641, 626, \dodoi{10.1086/500349}

\bibitem[{{Ostriker}(1999)}]{1999ApJ...513..252O}
{Ostriker}, E.~C. 1999, \apj, 513, 252, \dodoi{10.1086/306858}

\bibitem[{{Parsa} {et~al.}(2017){Parsa}, {Eckart}, {Shahzamanian}, {Karas},
  {Zaja{\v{c}}ek}, {Zensus}, \& {Straubmeier}}]{2017ApJ...845...22P}
{Parsa}, M., {Eckart}, A., {Shahzamanian}, B., {et~al.} 2017, \apj, 845, 22,
  \dodoi{10.3847/1538-4357/aa7bf0}

\bibitem[{{Payne} {et~al.}(2020){Payne}, {Shappee}, {Hinkle}, {Vallely},
  {Kochanek}, {Holoien}, {Auchettl}, {Stanek}, {Thompson}, {Neustadt},
  {Tucker}, {Armstrong}, {Brimacombe}, {Cacella}, {Cornect}, {Denneau},
  {Fausnaugh}, {Flewelling}, {Grupe}, {Heinze}, {Lopez}, {Monard}, {Prieto},
  {Schneider}, {Sheppard}, {Tonry}, \& {Weiland}}]{2020arXiv200903321P}
{Payne}, A.~V., {Shappee}, B.~J., {Hinkle}, J.~T., {et~al.} 2020, arXiv
  e-prints, arXiv:2009.03321.
\newblock \doarXiv{2009.03321}

\bibitem[{{Pei{\ss}ker} {et~al.}(2020{\natexlab{a}}){Pei{\ss}ker}, {Eckart}, \&
  {Parsa}}]{2020ApJ...889...61P}
{Pei{\ss}ker}, F., {Eckart}, A., \& {Parsa}, M. 2020{\natexlab{a}}, \apj, 889,
  61, \dodoi{10.3847/1538-4357/ab5afd}

\bibitem[{{Pei{\ss}ker} {et~al.}(2020{\natexlab{b}}){Pei{\ss}ker}, {Eckart},
  {Zaja{\v{c}}ek}, {Ali}, \& {Parsa}}]{2020ApJ...899...50P}
{Pei{\ss}ker}, F., {Eckart}, A., {Zaja{\v{c}}ek}, M., {Ali}, B., \& {Parsa}, M.
  2020{\natexlab{b}}, \apj, 899, 50, \dodoi{10.3847/1538-4357/ab9c1c}

\bibitem[{{Pei{\ss}ker} {et~al.}(2019){Pei{\ss}ker}, {Zaja{\v{c}}ek}, {Eckart},
  {Sabha}, {Shahzamanian}, \& {Parsa}}]{2019A&A...624A..97P}
{Pei{\ss}ker}, F., {Zaja{\v{c}}ek}, M., {Eckart}, A., {et~al.} 2019, \aap, 624,
  A97, \dodoi{10.1051/0004-6361/201834947}

\bibitem[{{Pei{\ss}ker} {et~al.}(2021{\natexlab{a}}){Pei{\ss}ker}, {Ali},
  {Zaja{\v{c}}ek}, {Eckart}, {Elaheh Hosseini}, {Karas}, {Cl{\'e}net}, {Sabha},
  {Labadie}, \& {Subroweit}}]{2021arXiv210102077P}
{Pei{\ss}ker}, F., {Ali}, B., {Zaja{\v{c}}ek}, M., {et~al.} 2021{\natexlab{a}},
  arXiv e-prints, arXiv:2101.02077.
\newblock \doarXiv{2101.02077}

\bibitem[{{Pei{\ss}ker} {et~al.}(2021{\natexlab{b}}){Pei{\ss}ker}, {Ali},
  {Zajacek}, {Eckart}, {Hosseini}, {Karas}, {Clenet}, {Sabha}, {Labadie}, \&
  {Subroweit}}]{2021ATel14306....1P}
{Pei{\ss}ker}, F., {Ali}, B., {Zajacek}, M., {et~al.} 2021{\natexlab{b}}, The
  Astronomer's Telegram, 14306, 1

\bibitem[{{Pelletier} \& {Roland}(1989)}]{1989A&A...224...24P}
{Pelletier}, G., \& {Roland}, J. 1989, \aap, 224, 24

\bibitem[{Penna {et~al.}(2013)Penna, Narayan, \&
  Sądowski}]{10.1093/mnras/stt1860}
Penna, R.~F., Narayan, R., \& Sądowski, A. 2013, Monthly Notices of the Royal
  Astronomical Society, 436, 3741, \dodoi{10.1093/mnras/stt1860}

\bibitem[{{Perets} {et~al.}(2007){Perets}, {Hopman}, \&
  {Alexander}}]{2007ApJ...656..709P}
{Perets}, H.~B., {Hopman}, C., \& {Alexander}, T. 2007, \apj, 656, 709,
  \dodoi{10.1086/510377}

\bibitem[{{Peters}(1964)}]{1964PhRv..136.1224P}
{Peters}, P.~C. 1964, Physical Review, 136, 1224,
  \dodoi{10.1103/PhysRev.136.B1224}

\bibitem[{Peters \& Mathews(1963)}]{peters1963gravitational}
Peters, P.~C., \& Mathews, J. 1963, Physical Review, 131, 435

\bibitem[{{Peterson}(1997)}]{1997iagn.book.....P}
{Peterson}, B.~M. 1997, {An Introduction to Active Galactic Nuclei}

\bibitem[{{Punsly}(2008)}]{2001bhgh.book.....P}
{Punsly}, B. 2008, {Black hole gravitohydromagnetics (Springer: Berlin)}

\bibitem[{{Pushkarev} {et~al.}(2005){Pushkarev}, {Gabuzda}, {Vetukhnovskaya},
  \& {Yakimov}}]{2005MNRAS.356..859P}
{Pushkarev}, A.~B., {Gabuzda}, D.~C., {Vetukhnovskaya}, Y.~N., \& {Yakimov},
  V.~E. 2005, \mnras, 356, 859, \dodoi{10.1111/j.1365-2966.2004.08535.x}

\bibitem[{{Rees}(1988)}]{Rees_1988}
{Rees}, M.~J. 1988, \nat, 333, 523, \dodoi{10.1038/333523a0}

\bibitem[{Rembiasz {et~al.}(2017)Rembiasz, Obergaulinger,
  Cerd{\'{a}}-Dur{\'{a}}n, Aloy, \& Müller}]{Rembiasz_2017}
Rembiasz, T., Obergaulinger, M., Cerd{\'{a}}-Dur{\'{a}}n, P., Aloy,
  M.-{\'{A}}., \& Müller, E. 2017, The Astrophysical Journal Supplement
  Series, 230, 18, \dodoi{10.3847/1538-4365/aa6254}

\bibitem[{{Ressler} {et~al.}(2015){Ressler}, {Tchekhovskoy}, {Quataert},
  {Chandra}, \& {Gammie}}]{2015MNRAS.454.1848R}
{Ressler}, S.~M., {Tchekhovskoy}, A., {Quataert}, E., {Chandra}, M., \&
  {Gammie}, C.~F. 2015, \mnras, 454, 1848, \dodoi{10.1093/mnras/stv2084}

\bibitem[{{Ricci} {et~al.}(2020){Ricci}, {Kara}, {Loewenstein}, {Trakhtenbrot},
  {Arcavi}, {Remillard}, {Fabian}, {Gendreau}, {Arzoumanian}, {Li}, {Ho},
  {MacLeod}, {Cackett}, {Altamirano}, {Gandhi}, {Kosec}, {Pasham}, {Steiner},
  \& {Chan}}]{2020ApJ...898L...1R}
{Ricci}, C., {Kara}, E., {Loewenstein}, M., {et~al.} 2020, \apjl, 898, L1,
  \dodoi{10.3847/2041-8213/ab91a1}

\bibitem[{Richter(2017)}]{Richter2017}
Richter, T. 2017, Lecture notes in computational science and engineering, Vol.
  118, Fluid-structure Interactions. Models, Analysis and Finite Elements
  (Springer)

\bibitem[{{Ryan} {et~al.}(2017){Ryan}, {Ressler}, {Dolence}, {Tchekhovskoy},
  {Gammie}, \& {Quataert}}]{2017ApJ...844L..24R}
{Ryan}, B.~R., {Ressler}, S.~M., {Dolence}, J.~C., {et~al.} 2017, \apjl, 844,
  L24, \dodoi{10.3847/2041-8213/aa8034}

\bibitem[{{Schartmann} {et~al.}(2018){Schartmann}, {Burkert}, \&
  {Ballone}}]{2018A&A...616L...8S}
{Schartmann}, M., {Burkert}, A., \& {Ballone}, A. 2018, \aap, 616, L8,
  \dodoi{10.1051/0004-6361/201833156}

\bibitem[{Schmidt(2002)}]{schmidt2002celestial}
Schmidt, W. 2002, Classical and Quantum Gravity, 19, 2743

\bibitem[{{Schneider}(2006)}]{2006eac..book.....S}
{Schneider}, P. 2006, {Extragalactic Astronomy and Cosmology (Springer:
  Berlin)}

\bibitem[{{Sch{\"o}del} {et~al.}(2014){Sch{\"o}del}, {Feldmeier}, {Neumayer},
  {Meyer}, \& {Yelda}}]{2014CQGra..31x4007S}
{Sch{\"o}del}, R., {Feldmeier}, A., {Neumayer}, N., {Meyer}, L., \& {Yelda}, S.
  2014, Classical and Quantum Gravity, 31, 244007,
  \dodoi{10.1088/0264-9381/31/24/244007}

\bibitem[{{Sch{\"o}del} {et~al.}(2002){Sch{\"o}del}, {Ott}, {Genzel},
  {Hofmann}, {Lehnert}, {Eckart}, {Mouawad}, {Alexander}, {Reid}, {Lenzen},
  {Hartung}, {Lacombe}, {Rouan}, {Gendron}, {Rousset}, {Lagrange}, {Brandner},
  {Ageorges}, {Lidman}, {Moorwood}, {Spyromilio}, {Hubin}, \&
  {Menten}}]{2002Natur.419..694S}
{Sch{\"o}del}, R., {Ott}, T., {Genzel}, R., {et~al.} 2002, \nat, 419, 694,
  \dodoi{10.1038/nature01121}

\bibitem[{{Schwarzenberg-Czerny}(1996)}]{1996ApJ...460L.107S}
{Schwarzenberg-Czerny}, A. 1996, \apjl, 460, L107, \dodoi{10.1086/309985}

\bibitem[{{Scoville} \& {Burkert}(2013)}]{2013ApJ...768..108S}
{Scoville}, N., \& {Burkert}, A. 2013, \apj, 768, 108,
  \dodoi{10.1088/0004-637X/768/2/108}

\bibitem[{{Semer{\'a}k} {et~al.}(1999){Semer{\'a}k}, {Karas}, \& {de
  Felice}}]{1999PASJ...51..571S}
{Semer{\'a}k}, O., {Karas}, V., \& {de Felice}, F. 1999, \pasj, 51, 571,
  \dodoi{10.1093/pasj/51.5.571}

\bibitem[{{Shahzamanian} {et~al.}(2016){Shahzamanian}, {Eckart},
  {Zaja{\v{c}}ek}, {Valencia-S.}, {Sabha}, {Moser}, {Parsa}, {Peissker}, \&
  {Straubmeier}}]{2016A&A...593A.131S}
{Shahzamanian}, B., {Eckart}, A., {Zaja{\v{c}}ek}, M., {et~al.} 2016, \aap,
  593, A131, \dodoi{10.1051/0004-6361/201628994}

\bibitem[{{Shu}(1992)}]{1992pavi.book.....S}
{Shu}, F.~H. 1992, {The physics of astrophysics. Volume II: Gas dynamics.}

\bibitem[{{Shu} {et~al.}(2017){Shu}, {Wang}, {Jiang}, {Wang}, {Sun}, \&
  {Zhou}}]{2017ApJ...837....3S}
{Shu}, X.~W., {Wang}, T.~G., {Jiang}, N., {et~al.} 2017, \apj, 837, 3,
  \dodoi{10.3847/1538-4357/aa5eb3}

\bibitem[{{Sillanpaa} {et~al.}(1988){Sillanpaa}, {Haarala}, {Valtonen},
  {Sundelius}, \& {Byrd}}]{1988ApJ...325..628S}
{Sillanpaa}, A., {Haarala}, S., {Valtonen}, M.~J., {Sundelius}, B., \& {Byrd},
  G.~G. 1988, \apj, 325, 628, \dodoi{10.1086/166033}

\bibitem[{{Sol} {et~al.}(1989){Sol}, {Pelletier}, \&
  {Asseo}}]{1989MNRAS.237..411S}
{Sol}, H., {Pelletier}, G., \& {Asseo}, E. 1989, \mnras, 237, 411,
  \dodoi{10.1093/mnras/237.2.411}

\bibitem[{{Stephan} {et~al.}(2016){Stephan}, {Naoz}, {Ghez}, {Witzel},
  {Sitarski}, {Do}, \& {Kocsis}}]{2016MNRAS.460.3494S}
{Stephan}, A.~P., {Naoz}, S., {Ghez}, A.~M., {et~al.} 2016, \mnras, 460, 3494,
  \dodoi{10.1093/mnras/stw1220}

\bibitem[{{Stephan} {et~al.}(2019){Stephan}, {Naoz}, {Ghez}, {Morris},
  {Ciurlo}, {Do}, {Breivik}, {Coughlin}, \& {Rodriguez}}]{2019ApJ...878...58S}
---. 2019, \apj, 878, 58, \dodoi{10.3847/1538-4357/ab1e4d}

\bibitem[{{Stone} {et~al.}(2019){Stone}, {Kesden}, {Cheng}, \& {van
  Velzen}}]{2019GReGr..51...30S}
{Stone}, N.~C., {Kesden}, M., {Cheng}, R.~M., \& {van Velzen}, S. 2019, General
  Relativity and Gravitation, 51, 30, \dodoi{10.1007/s10714-019-2510-9}

\bibitem[{{Sukov\'{a}} {et~al.}(2020){Sukov\'{a}}, {Zaja\v{c}ek}, {Witzany}, \&
  {Karas}}]{RAGtime2020}
{Sukov\'{a}}, P., {Zaja\v{c}ek}, M., {Witzany}, V., \& {Karas}, V. 2020, in
  Proceedings of RAGtime 20–22: Workshops on black holes and neutron stars,
  15–19 Oct., 16–20 Sept., 19–23 Oct. 2018/2019/2020, Opava, Czech
  Republic, ed. Z.~{Stuchlík}, G.~{Török}, \& V.~{Karas}, 299--315

\bibitem[{{Syer} {et~al.}(1991){Syer}, {Clarke}, \&
  {Rees}}]{1991MNRAS.250..505S}
{Syer}, D., {Clarke}, C.~J., \& {Rees}, M.~J. 1991, \mnras, 250, 505,
  \dodoi{10.1093/mnras/250.3.505}

\bibitem[{Sądowski {et~al.}(2013)Sądowski, Narayan, Penna, \&
  Zhu}]{10.1093/mnras/stt1881}
Sądowski, A., Narayan, R., Penna, R., \& Zhu, Y. 2013, Monthly Notices of the
  Royal Astronomical Society, 436, 3856, \dodoi{10.1093/mnras/stt1881}

\bibitem[{{Takalo}(1994)}]{1994VA.....38...77T}
{Takalo}, L.~O. 1994, Vistas in Astronomy, 38, 77,
  \dodoi{10.1016/0083-6656(94)90004-3}

\bibitem[{{Tavecchio} \& {Ghisellini}(2008)}]{2008MNRAS.386..945T}
{Tavecchio}, F., \& {Ghisellini}, G. 2008, \mnras, 386, 945,
  \dodoi{10.1111/j.1365-2966.2008.13072.x}

\bibitem[{{Tchekhovskoy} {et~al.}(2007){Tchekhovskoy}, {McKinney}, \&
  {Narayan}}]{2007MNRAS.379..469T}
{Tchekhovskoy}, A., {McKinney}, J.~C., \& {Narayan}, R. 2007, \mnras, 379, 469,
  \dodoi{10.1111/j.1365-2966.2007.11876.x}

\bibitem[{Tchekhovskoy {et~al.}(2011)Tchekhovskoy, Narayan, \&
  McKinney}]{Sasha-smart-grid}
Tchekhovskoy, A., Narayan, R., \& McKinney, J.~C. 2011, Monthly Notices of the
  Royal Astronomical Society: Letters, 418, L79,
  \dodoi{10.1111/j.1745-3933.2011.01147.x}

\bibitem[{{Timmer} \& {Koenig}(1995)}]{1995A&A...300..707T}
{Timmer}, J., \& {Koenig}, M. 1995, \aap, 300, 707

\bibitem[{{Trakhtenbrot} {et~al.}(2019){Trakhtenbrot}, {Arcavi}, {MacLeod},
  {Ricci}, {Kara}, {Graham}, {Stern}, {Harrison}, {Burke}, {Hiramatsu},
  {Hosseinzadeh}, {Howell}, {Smartt}, {Rest}, {Prieto}, {Shappee}, {Holoien},
  {Bersier}, {Filippenko}, {Brink}, {Zheng}, {Li}, {Remillard}, \&
  {Loewenstein}}]{2019ApJ...883...94T}
{Trakhtenbrot}, B., {Arcavi}, I., {MacLeod}, C.~L., {et~al.} 2019, \apj, 883,
  94, \dodoi{10.3847/1538-4357/ab39e4}

\bibitem[{T\r{u}ma {et~al.}(2018)T\r{u}ma, Stein, Pr\r{u}\v{s}a, \&
  Friedmann}]{TuStPrFr:2018}
T\r{u}ma, K., Stein, J., Pr\r{u}\v{s}a, V., \& Friedmann, E. 2018, Appl. Math.
  Comp., 335, 50 , \dodoi{https://doi.org/10.1016/j.amc.2018.04.030}

\bibitem[{{Urry} \& {Padovani}(1995)}]{1995PASP..107..803U}
{Urry}, C.~M., \& {Padovani}, P. 1995, \pasp, 107, 803, \dodoi{10.1086/133630}

\bibitem[{{Valencia-S.} {et~al.}(2015){Valencia-S.}, {Eckart}, {Zaja{\v{c}}ek},
  {Peissker}, {Parsa}, {Grosso}, {Mossoux}, {Porquet}, {Jalali}, {Karas},
  {Yazici}, {Shahzamanian}, {Sabha}, {Saalfeld}, {Smajic}, {Grellmann},
  {Moser}, {Horrobin}, {Borkar}, {Garc{\'\i}a-Mar{\'\i}n}, {Dov{\v{c}}iak},
  {Kunneriath}, {Karssen}, {Bursa}, {Straubmeier}, \&
  {Bushouse}}]{2015ApJ...800..125V}
{Valencia-S.}, M., {Eckart}, A., {Zaja{\v{c}}ek}, M., {et~al.} 2015, \apj, 800,
  125, \dodoi{10.1088/0004-637X/800/2/125}

\bibitem[{Valkov {et~al.}(2015)Valkov, Rycroft, \& Kamrin}]{Valkov2015}
Valkov, B., Rycroft, C., \& Kamrin, K. 2015, Journal of Applied Mechanics, 82

\bibitem[{{Valtonen} {et~al.}(2016){Valtonen}, {Zola}, {Ciprini}, {Gopakumar},
  {Matsumoto}, {Sadakane}, {Kidger}, {Gazeas}, {Nilsson}, {Berdyugin},
  {Piirola}, {Jermak}, {Baliyan}, {Alicavus}, {Boyd}, {Campas Torrent},
  {Campos}, {Carrillo G{\'o}mez}, {Caton}, {Chavushyan}, {Dalessio}, {Debski},
  {Dimitrov}, {Drozdz}, {Er}, {Erdem}, {Escartin P{\'e}rez}, {Fallah Ramazani},
  {Filippenko}, {Ganesh}, {Garcia}, {G{\'o}mez Pinilla}, {Gopinathan},
  {Haislip}, {Hudec}, {Hurst}, {Ivarsen}, {Jelinek}, {Joshi}, {Kagitani},
  {Kaur}, {Keel}, {LaCluyze}, {Lee}, {Lindfors}, {Lozano de Haro}, {Moore},
  {Mugrauer}, {Naves Nogues}, {Neely}, {Nelson}, {Ogloza}, {Okano}, {Pandey},
  {Perri}, {Pihajoki}, {Poyner}, {Provencal}, {Pursimo}, {Raj}, {Reichart},
  {Reinthal}, {Sadegi}, {Sakanoi}, {Salto Gonz{\'a}lez}, {Sameer}, {Schweyer},
  {Siwak}, {Sold{\'a}n Alfaro}, {Sonbas}, {Steele}, {Stocke}, {Strobl},
  {Takalo}, {Tomov}, {Tremosa Espasa}, {Valdes}, {Valero P{\'e}rez},
  {Verrecchia}, {Webb}, {Yoneda}, {Zejmo}, {Zheng}, {Telting}, {Saario},
  {Reynolds}, {Kvammen}, {Gafton}, {Karjalainen}, {Harmanen}, \&
  {Blay}}]{2016ApJ...819L..37V}
{Valtonen}, M.~J., {Zola}, S., {Ciprini}, S., {et~al.} 2016, \apjl, 819, L37,
  \dodoi{10.3847/2041-8205/819/2/L37}

\bibitem[{{Valtonen} {et~al.}(2019){Valtonen}, {Zola}, {Pihajoki}, {Enestam},
  {Lehto}, {Dey}, {Gopakumar}, {Drozdz}, {Ogloza}, {Zejmo}, {Gupta}, {Pursimo},
  {Ciprini}, {Kidger}, {Nilsson}, {Berdyugin}, {Piirola}, {Jermak}, {Hudec}, \&
  {Laine}}]{2019ApJ...882...88V}
{Valtonen}, M.~J., {Zola}, S., {Pihajoki}, P., {et~al.} 2019, \apj, 882, 88,
  \dodoi{10.3847/1538-4357/ab3573}

\bibitem[{{van Marle} {et~al.}(2014){van Marle}, {Decin}, \&
  {Meliani}}]{2014A&A...561A.152V}
{van Marle}, A.~J., {Decin}, L., \& {Meliani}, Z. 2014, \aap, 561, A152,
  \dodoi{10.1051/0004-6361/201321968}

\bibitem[{{Vink} \& {Yamazaki}(2014)}]{2014ApJ...780..125V}
{Vink}, J., \& {Yamazaki}, R. 2014, \apj, 780, 125,
  \dodoi{10.1088/0004-637X/780/2/125}

\bibitem[{Virtanen {et~al.}(2020)Virtanen, Gommers, Oliphant, Haberland, Reddy,
  Cournapeau, Burovski, Peterson, Weckesser, Bright, {van der Walt}, Brett,
  Wilson, Millman, Mayorov, Nelson, Jones, Kern, Larson, Carey, Polat, Feng,
  Moore, {VanderPlas}, Laxalde, Perktold, Cimrman, Henriksen, Quintero, Harris,
  Archibald, Ribeiro, Pedregosa, {van Mulbregt}, \& {SciPy 1.0
  Contributors}}]{2020SciPy-NMeth}
Virtanen, P., Gommers, R., Oliphant, T.~E., {et~al.} 2020, Nature Methods, 17,
  261, \dodoi{10.1038/s41592-019-0686-2}

\bibitem[{{Vishniac}(1994)}]{1994ApJ...428..186V}
{Vishniac}, E.~T. 1994, \apj, 428, 186, \dodoi{10.1086/174231}

\bibitem[{{Voit} {et~al.}(2015){Voit}, {Donahue}, {Bryan}, \&
  {McDonald}}]{2015Natur.519..203V}
{Voit}, G.~M., {Donahue}, M., {Bryan}, G.~L., \& {McDonald}, M. 2015, \nat,
  519, 203, \dodoi{10.1038/nature14167}

\bibitem[{{Vokrouhlick\'y} \& {Karas}(1993)}]{1993MNRAS.265..365V}
{Vokrouhlick\'y}, D., \& {Karas}, V. 1993, \mnras, 265, 365,
  \dodoi{10.1093/mnras/265.2.365}

\bibitem[{{Wang} \& {Merritt}(2004)}]{2004ApJ...600..149W}
{Wang}, J., \& {Merritt}, D. 2004, \apj, 600, 149, \dodoi{10.1086/379767}

\bibitem[{{Wang} {et~al.}(2013){Wang}, {Nowak}, {Markoff}, {Baganoff},
  {Nayakshin}, {Yuan}, {Cuadra}, {Davis}, {Dexter}, {Fabian}, {Grosso},
  {Haggard}, {Houck}, {Ji}, {Li}, {Neilsen}, {Porquet}, {Ripple}, \&
  {Shcherbakov}}]{2013Sci...341..981W}
{Wang}, Q.~D., {Nowak}, M.~A., {Markoff}, S.~B., {et~al.} 2013, Science, 341,
  981, \dodoi{10.1126/science.1240755}

\bibitem[{{Wang} \& {Loeb}(2014)}]{2014MNRAS.441..809W}
{Wang}, X., \& {Loeb}, A. 2014, \mnras, 441, 809, \dodoi{10.1093/mnras/stu600}

\bibitem[{{Wilkin}(1996)}]{1996ApJ...459L..31W}
{Wilkin}, F.~P. 1996, \apjl, 459, L31, \dodoi{10.1086/309939}

\bibitem[{{Witzany, V.} \& {Jefremov, P.}(2018)}]{Witzany_Jefremov-tori}
{Witzany, V.}, \& {Jefremov, P.} 2018, A\&A, 614, A75,
  \dodoi{10.1051/0004-6361/201732361}

\bibitem[{{Witzel} {et~al.}(2012){Witzel}, {Eckart}, {Bremer}, {Zamaninasab},
  {Shahzamanian}, {Valencia-S.}, {Sch{\"o}del}, {Karas}, {Lenzen}, {Marchili},
  {Sabha}, {Garcia-Marin}, {Buchholz}, {Kunneriath}, \&
  {Straubmeier}}]{2012ApJS..203...18W}
{Witzel}, G., {Eckart}, A., {Bremer}, M., {et~al.} 2012, \apjs, 203, 18,
  \dodoi{10.1088/0067-0049/203/2/18}

\bibitem[{{Witzel} {et~al.}(2014){Witzel}, {Ghez}, {Morris}, {Sitarski},
  {Boehle}, {Naoz}, {Campbell}, {Becklin}, {Canalizo}, {Chappell}, {Do}, {Lu},
  {Matthews}, {Meyer}, {Stockton}, {Wizinowich}, \&
  {Yelda}}]{2014ApJ...796L...8W}
{Witzel}, G., {Ghez}, A.~M., {Morris}, M.~R., {et~al.} 2014, \apjl, 796, L8,
  \dodoi{10.1088/2041-8205/796/1/L8}

\bibitem[{{Witzel} {et~al.}(2018){Witzel}, {Martinez}, {Hora}, {Willner},
  {Morris}, {Gammie}, {Becklin}, {Ashby}, {Baganoff}, {Carey}, {Do}, {Fazio},
  {Ghez}, {Glaccum}, {Haggard}, {Herrero-Illana}, {Ingalls}, {Narayan}, \&
  {Smith}}]{2018ApJ...863...15W}
{Witzel}, G., {Martinez}, G., {Hora}, J., {et~al.} 2018, \apj, 863, 15,
  \dodoi{10.3847/1538-4357/aace62}

\bibitem[{{Xie} {et~al.}(2012){Xie}, {Lei}, {Zou}, {Wang}, {Wu}, \&
  {Wang}}]{2012RAA....12..817X}
{Xie}, W., {Lei}, W.-H., {Zou}, Y.-C., {et~al.} 2012, Research in Astronomy and
  Astrophysics, 12, 817, \dodoi{10.1088/1674-4527/12/7/010}

\bibitem[{Yoon {et~al.}(2020)Yoon, Chatterjee, Markoff, van Eijnatten, Younsi,
  Liska, \& Tchekhovskoy}]{Sasha-radiative-cooling}
Yoon, D., Chatterjee, K., Markoff, S.~B., {et~al.} 2020, Monthly Notices of the
  Royal Astronomical Society, 499, 3178, \dodoi{10.1093/mnras/staa3031}

\bibitem[{{Yuan} {et~al.}(2002){Yuan}, {Markoff}, \&
  {Falcke}}]{2002A&A...383..854Y}
{Yuan}, F., {Markoff}, S., \& {Falcke}, H. 2002, \aap, 383, 854,
  \dodoi{10.1051/0004-6361:20011709}

\bibitem[{{Yuan} \& {Narayan}(2014)}]{2014ARA&A..52..529Y}
{Yuan}, F., \& {Narayan}, R. 2014, ARA\&A, 52, 529,
  \dodoi{10.1146/annurev-astro-082812-141003}

\bibitem[{{Yusef-Zadeh} {et~al.}(2020){Yusef-Zadeh}, {Royster}, {Wardle},
  {Cotton}, {Kunneriath}, {Heywood}, \& {Michail}}]{2020MNRAS.499.3909Y}
{Yusef-Zadeh}, F., {Royster}, M., {Wardle}, M., {et~al.} 2020, \mnras, 499,
  3909, \dodoi{10.1093/mnras/staa2399}

\bibitem[{{Zaja{\v{c}}ek} {et~al.}(2016){Zaja{\v{c}}ek}, {Eckart}, {Karas},
  {Kunneriath}, {Shahzamanian}, {Sabha}, {Mu{\v{z}}i{\'c}}, \&
  {Valencia-S.}}]{2016MNRAS.455.1257Z}
{Zaja{\v{c}}ek}, M., {Eckart}, A., {Karas}, V., {et~al.} 2016, \mnras, 455,
  1257, \dodoi{10.1093/mnras/stv2357}

\bibitem[{{Zaja{\v{c}}ek} {et~al.}(2014){Zaja{\v{c}}ek}, {Karas}, \&
  {Eckart}}]{2014A&A...565A..17Z}
{Zaja{\v{c}}ek}, M., {Karas}, V., \& {Eckart}, A. 2014, \aap, 565, A17,
  \dodoi{10.1051/0004-6361/201322713}

\bibitem[{{Zaja\v{c}ek} {et~al.}(2015){Zaja\v{c}ek}, {Karas}, \&
  {Kunneriath}}]{2015AcPol..55..203Z}
{Zaja\v{c}ek}, M., {Karas}, V., \& {Kunneriath}, D. 2015, Acta Polytechnica,
  55, 203, \dodoi{10.14311/AP.2015.55.0203}

\bibitem[{{Zaja{\v{c}}ek} \& {Tursunov}(2018)}]{2018AN....339..324Z}
{Zaja{\v{c}}ek}, M., \& {Tursunov}, A.~A. 2018, Astronomische Nachrichten, 339,
  324, \dodoi{10.1002/asna.201813499}

\bibitem[{{Zaja{\v{c}}ek} {et~al.}(2017){Zaja{\v{c}}ek}, {Britzen}, {Eckart},
  {Shahzamanian}, {Busch}, {Karas}, {Parsa}, {Peissker}, {Dov{\v{c}}iak},
  {Subroweit}, {Dinnbier}, \& {Zensus}}]{2017A&A...602A.121Z}
{Zaja{\v{c}}ek}, M., {Britzen}, S., {Eckart}, A., {et~al.} 2017, \aap, 602,
  A121, \dodoi{10.1051/0004-6361/201730532}

\bibitem[{{Zaja{\v{c}}ek} {et~al.}(2019){Zaja{\v{c}}ek}, {Busch},
  {Valencia-S.}, {Eckart}, {Britzen}, {Fuhrmann}, {Schneeloch}, {Fazeli},
  {Harrington}, \& {Zensus}}]{2019A&A...630A..83Z}
{Zaja{\v{c}}ek}, M., {Busch}, G., {Valencia-S.}, M., {et~al.} 2019, \aap, 630,
  A83, \dodoi{10.1051/0004-6361/201833388}

\bibitem[{{Zhang} \& {Wang}(2021)}]{2021arXiv210404124Z}
{Zhang}, P., \& {Wang}, Z. 2021, arXiv e-prints, arXiv:2104.04124.
\newblock \doarXiv{2104.04124}

\bibitem[{{Zu} {et~al.}(2013){Zu}, {Kochanek}, {Koz{\l}owski}, \&
  {Udalski}}]{2013ApJ...765..106Z}
{Zu}, Y., {Kochanek}, C.~S., {Koz{\l}owski}, S., \& {Udalski}, A. 2013, \apj,
  765, 106, \dodoi{10.1088/0004-637X/765/2/106}

\end{thebibliography}
\bibliographystyle{aasjournal}

\appendix
\section{Numerical background}
\label{App_numerics}
The studied problem belongs into a wide class of problems generally denoted as a fluid--structure interaction.
However, when talking about the fluid--structure interaction, usually one has in mind an elastic/deformable body, whose deformation is given by the flow and which is exerting a force on the fluid as an elastic response. 
For such purposes a variety of numerical codes and methods have been developed with various description of the fluid and the solid body. 
The level-set method \citep{Valkov2015} or an initial point set method \citep{Richter2017} tracks the interface between the fluid and the solid body, both described in Eulerian setting. However, the fluid is naturally described in the Eulerian (current) configuration, while the solid parts are described in the Lagrangian (reference) setting. Leaving both in their natural configuration leads to a wide range of immersed boundary methods \citep{Boilevin_Kayl_2019} or fictitious domain methods \citep{Boffi_2016} where the fluid is described in the Eulerian setting with its own mesh and the solid is defined in the Lagrangian setting on another mesh that overlaps with the fluid mesh.
A mixture between these approaches represent the arbitrary Lagrangian-Eulerian (ALE) methods, in particular updated ALE method \citep{Donea1982} or total ALE method \citep{donea.j.huerta.a.ea:arbitrary,TuStPrFr:2018}.

However, focusing on the problem of a star moving in the accretion flow, the density contrast is huge, many orders of magnitude. Therefore, the star itself will not be deformed by the interaction with the neighbouring gas and we do not need to take into account the (in)elastic deformation of the star. 
The problem is thus not a classical fluid-structure interaction problem, but rather a motion of a solid body in the fluid.
To describe this situation, we implement an ``internal boundary condition'', imposing the star's velocity to the gas occupying a certain volume, while leaving the other quantities to evolve freely, which can be seen as a special case of the fictitious boundary method.

We modify the publicly available GRMHD code \texttt{HARMPI}, which is equipped with a spherical grid specially tailored to the black hole accretion problem  with a logarithmic and super-logarithmic grid in the $r$-direction and a non-equidistant grid in the $\theta$-direction.
Thanks to these functionalities, the grid can be focused on the most important regions with large space-time curvature close to the horizon, large gas densities along the equatorial plane and large magnetization along the symmetry axis, where jets are produced.
The problem then can be computed to a sufficiently high precision without the need of mesh refinement, which largely speeds up the computations and lower the computational demand, therefore it allows to perform long-term simulations essential for our study.
The disadvantage of this setting is, that the grid is not adapted to the geometric shape of the perturbing body moving in the gas (see Fig.~\ref{fig_star_shape} and the discussion in Appendix \ref{App_test}).

Moreover, in reality the gas does not interact with the star on its solid surface. 
In fact, there are several complicated ways as to how the interaction works. 
Stars with strong stellar winds will not interact with the gas on their surface, but the energy will be transported via the wind at the so-called stagnation radius, where the pressure of the stellar wind will be balanced by the pressure of the accreting gas (see Section \ref{sec:bowshock}). 
The perturbed region in our simulations then corresponds more to this stagnation radius, rather than to the physical size of the star.
In this case, the region occupied by the star and its wind can change its shape when moving in the medium and it is possible that the ambient matter can penetrate ``the object'' (i.e. the star + its wind) to some degree.
Stars and compact objects with a strong magnetic field will influence the plasma through the electromagnetic pressure (Section \ref{sec:pulsar}).
The black holes, on the other hand, affect the gas via the gravitational interaction (Section \ref{sec:CO}).
Our chosen approach does not distinguish between these scenarios, as it is only a rough approximation of the situation. 
In our future work, we aim to focus on the details of the interaction to account for these effects and to quantify, whether it is possible to distinguish between different perturbers based on the observed temporal and spectral properties.
One of the possible modification of the approach would be to consider a softer interaction potential, which would allow the gas to partially penetrate into the moving object, hence making the fluid-structure interface (i.e. the stellar wind -- accretion gas boundary) smeary.
In such a case, different methods apart from internal boundary condition can be implemented, e.g. adding the force term in the momentum equation.

In addition, there are also analytical and numerical studies of the star-fluid interaction under different circumstances. 
In particular, with respect to the situation in our Galactic center, the interaction of dust-enshrouded objects with the circumnuclear matter on the scales $\sim 10^3 M$ has been studied by \citet{2014A&A...565A..17Z}, while the role of the nuclear outflows on the possible bow-shock asymmetrical evolution along the stellar orbit was shown by \citet{2016MNRAS.455.1257Z}. In these cases, the stagnation radius depends on the ambient density as well as the relative stellar velocity as $R_{\rm stag}\propto (\rho_{\rm a}v_{\rm rel})^{-1/2}$, see also Eq.~\eqref{eq_stagnation_radius}, which can be implemented into the code by means of both hard-surface and soft-interaction potential method. 
Regarding the objects in our Galactic center, massive Newtonian 3D simulations with adaptive mesh refinement were used to study the pericentre passage of the near-infrared excess object G2 \citep{2018MNRAS.479.5288B} and the S2 star \citep{2018A&A...616L...8S}, the former was focusing on the fate of the dusty object during the tight fly-by, while the latter was exploring the possibility to probe the Sgr A* accretion profile by an enhanced X-ray emission due to the S2 approach. 
However, such detailed simulations are computationally very demanding even in the Newtonian regime and do not allow for a long-term global accretion flow simulation in the near-horizon region as we have done here.

\section{Verification of the code}
\label{App_test}
\begin{figure}
    \centering
    \includegraphics[width=0.32\columnwidth]{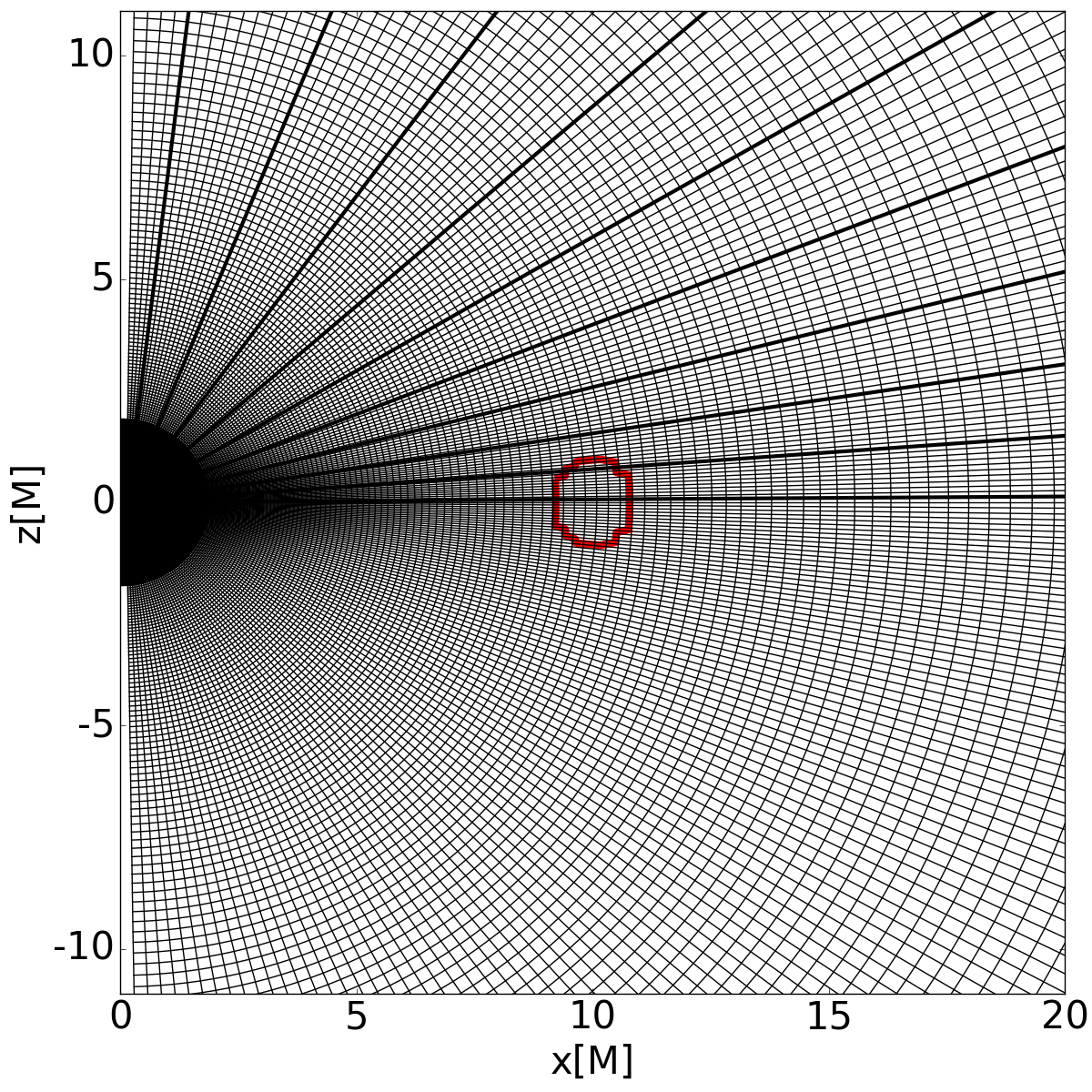}
    \includegraphics[width=0.32\columnwidth]{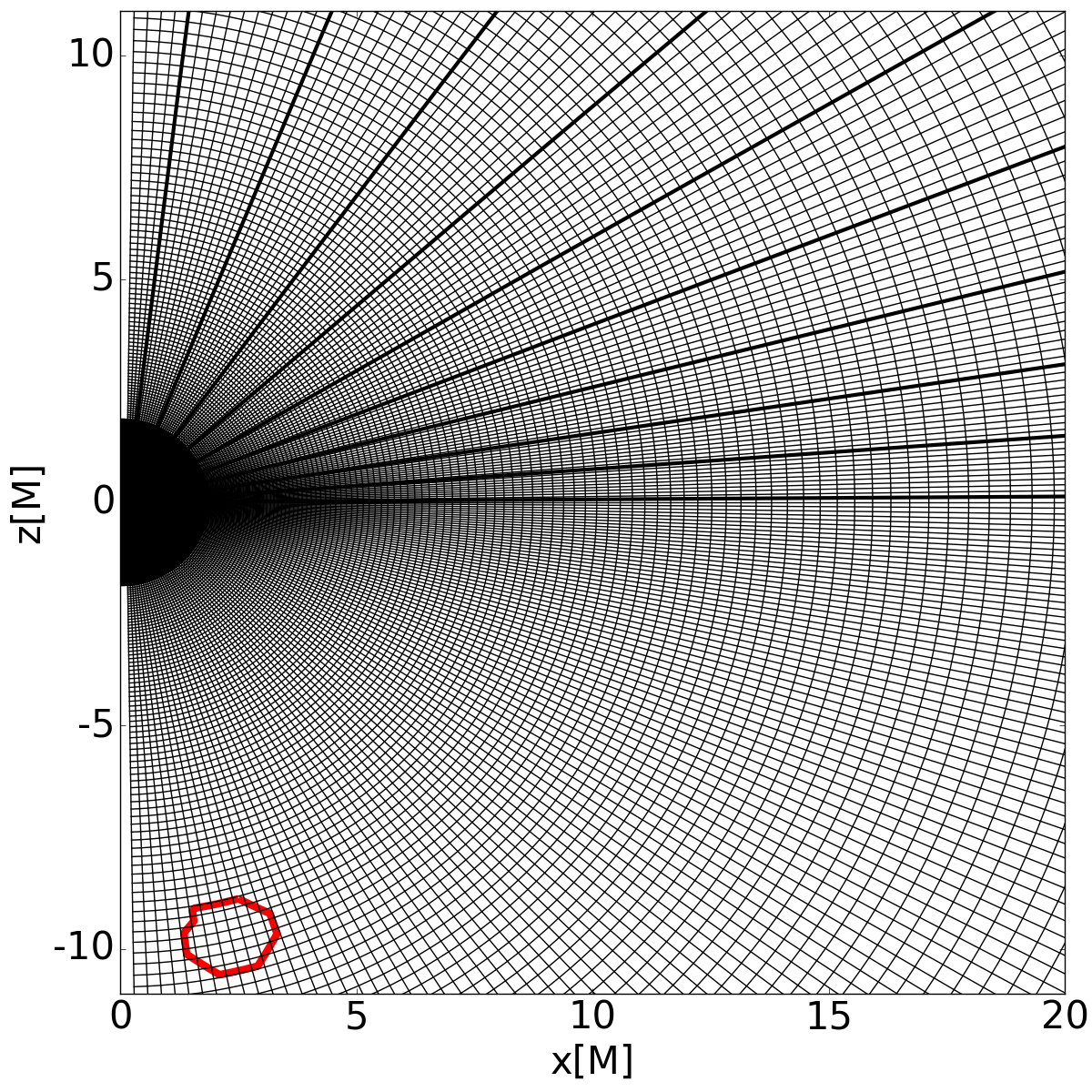}
    \includegraphics[width=0.32\columnwidth]{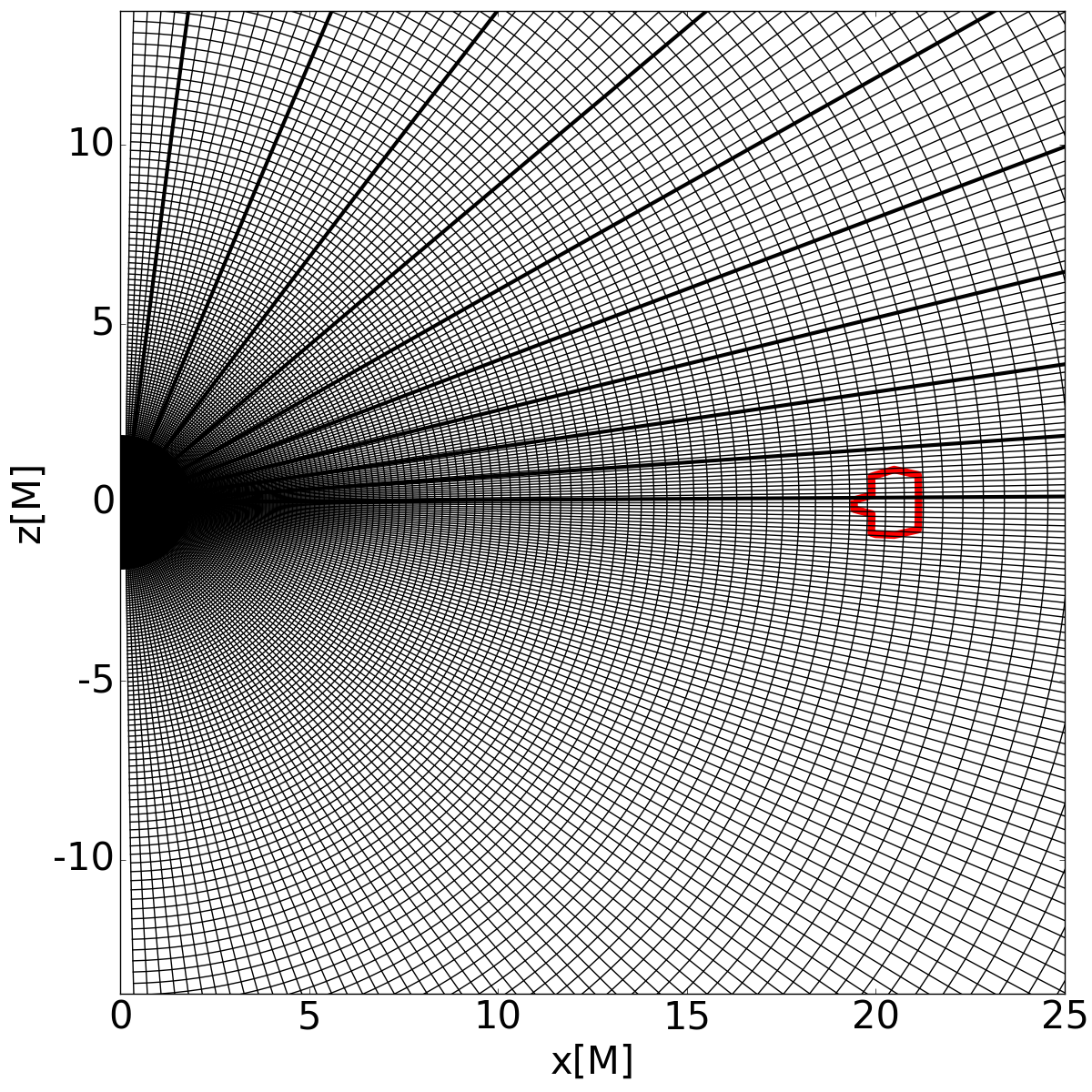}
    \caption{Shape of the internal boundary region for run~A at two different times (left and middle panel) and run~B (right). The thick solid black lines mark  every tenth grid line in $\theta$-direction.}
    \label{fig_star_shape}
\end{figure}

 In this work, we have added the functionality of a moving object in the gaseous environment into the publicly available GRMHD code \texttt{HARMPI}. The effect of the motion of the body is accounted for by imposing the velocity of the star to the gas within the volume of the star. Specifically, we find the distance between the center of any given cell and the center of the star, and if it is smaller than the star radius, we set the velocity to that of the star. Therefore, the shape of the star, or in other words the region, in which the boundary condition is imposed, depends on the actual grid spacing, hence it is changing slightly during the motion of the star. 
The boundary of the perturber is partially smeared out by the fact, that the gas can freely go inside and outside the region, when we only impose the velocity condition to the gas, which is in the region at the given moment. 
The examples of the internal boundary shape for run~A and run~B are given in Fig.~\ref{fig_star_shape}. However, we have seen in \citet{RAGtime2020} that the results of the simulations do not depend strongly on the exact shape of the perturbing object. In particular, it seems that the flow is not sensitive to the full 3D shape of the perturber and that it is sufficient to have the same 2D cross-section of the object in the direction of motion in order to get similar results.

We have tested the new functionality using the case of a supersonic motion of a body (star) through a constant density ($\rho_1=1$) fluid with an adiabatic index $\gamma$ at rest without gravity. In such a case, the Rankine-Hugoniot jump conditions imply that a shock will be formed in the medium, whose downstream density $\rho_2$ is given by
\begin{equation}
    \rho_2 = \frac{\gamma + 1}{\gamma - 1} \rho_1 \label{eg:rho2}
\end{equation}
in the strong-shock limit, i.e for high values of Mach number $\mathcal{M}=v_{\rm s}/c_{\rm s}$, where $v_{\rm s}$ is the velocity of the body and $c_{\rm s}$ is the local sound speed of the gas \citep{2014MNRAS.441..809W}.

For our chosen adiabatic index $\gamma=13/9$ we have run a series of simulations with increasing $\mathcal{M}$ and we have followed the maximum density achieved in the shock front. The grid spans the region $x\in(0,1), y \in(0,1)$ with resolution 512 x 512 cells. The radius of the star $\mathcal{R}=0.01$ and its initial position is $x(0) = 0.2, y(0) = 0.5$. The star is moving with the velocity $\vec{v}_s=(0.01,0)$ along the $y$-symmetry line. We set the internal gas energy density $u$ in such a way so that the sound speed satisfies $c_{\rm s} = \sqrt{\gamma(\gamma-1)u/\rho_1} = v_{\rm s}/\mathcal{M}$ with a chosen value of $\mathcal{M}$.

\begin{figure}
    \centering
    \includegraphics[width=0.32\columnwidth]{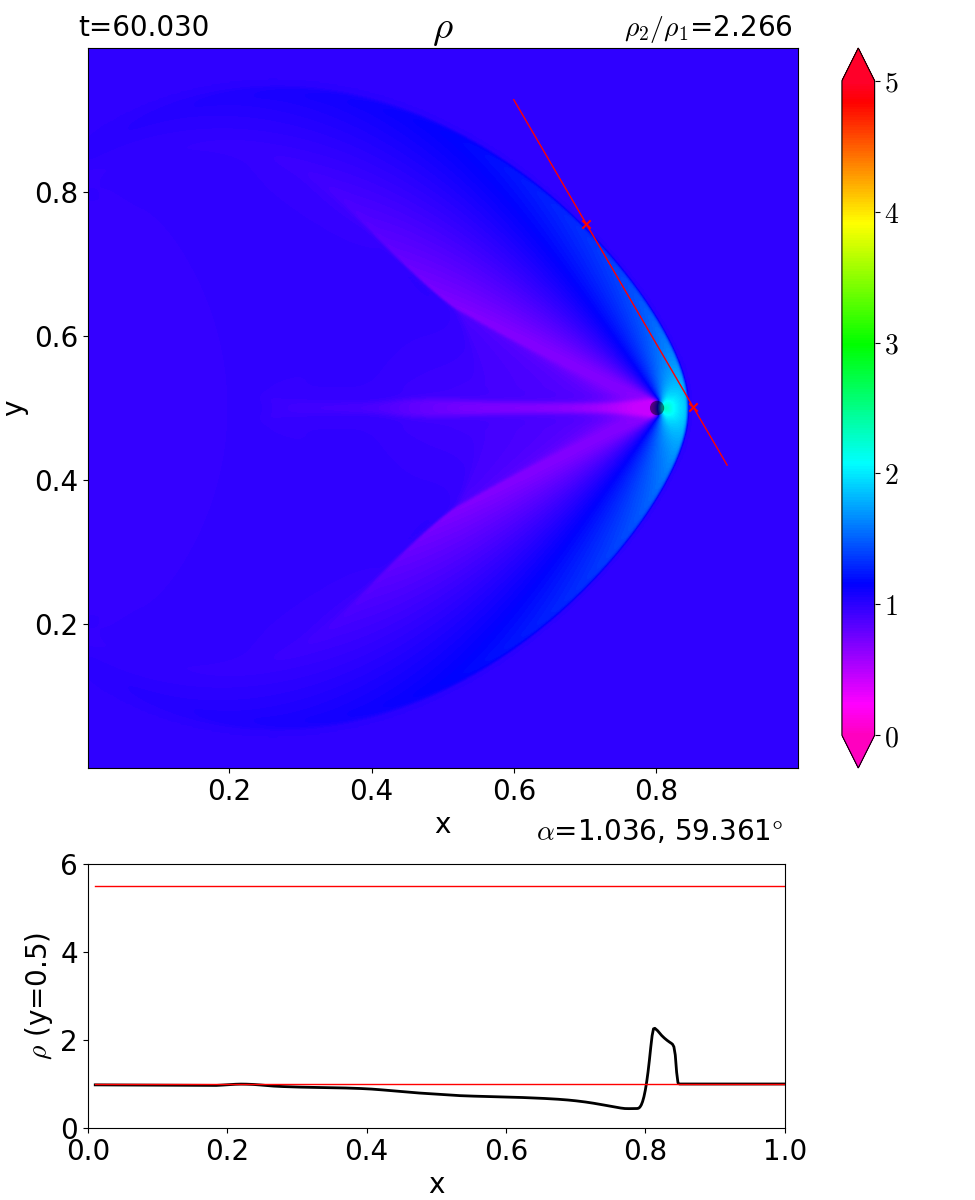}
    \includegraphics[width=0.32\columnwidth]{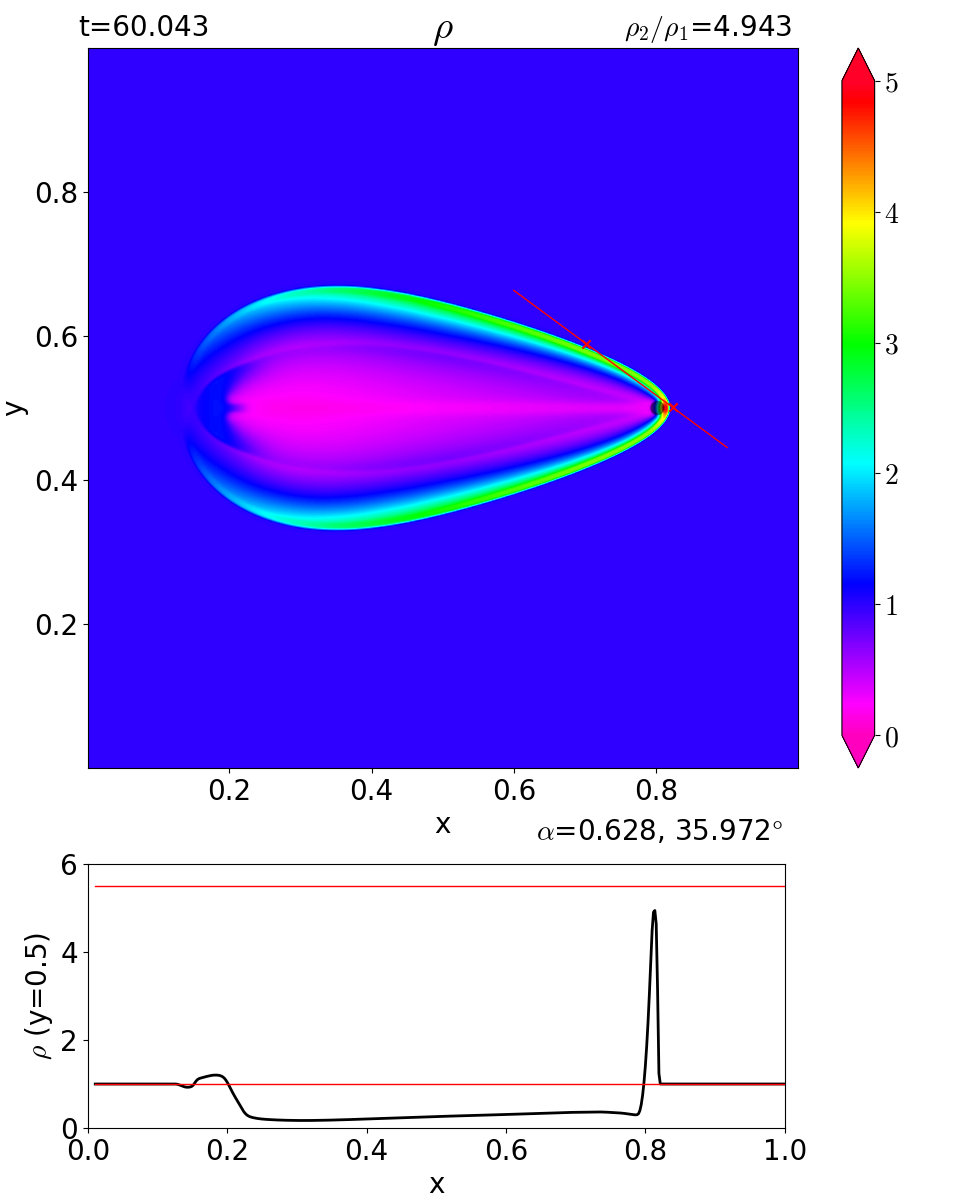}
    \includegraphics[width=0.32\columnwidth]{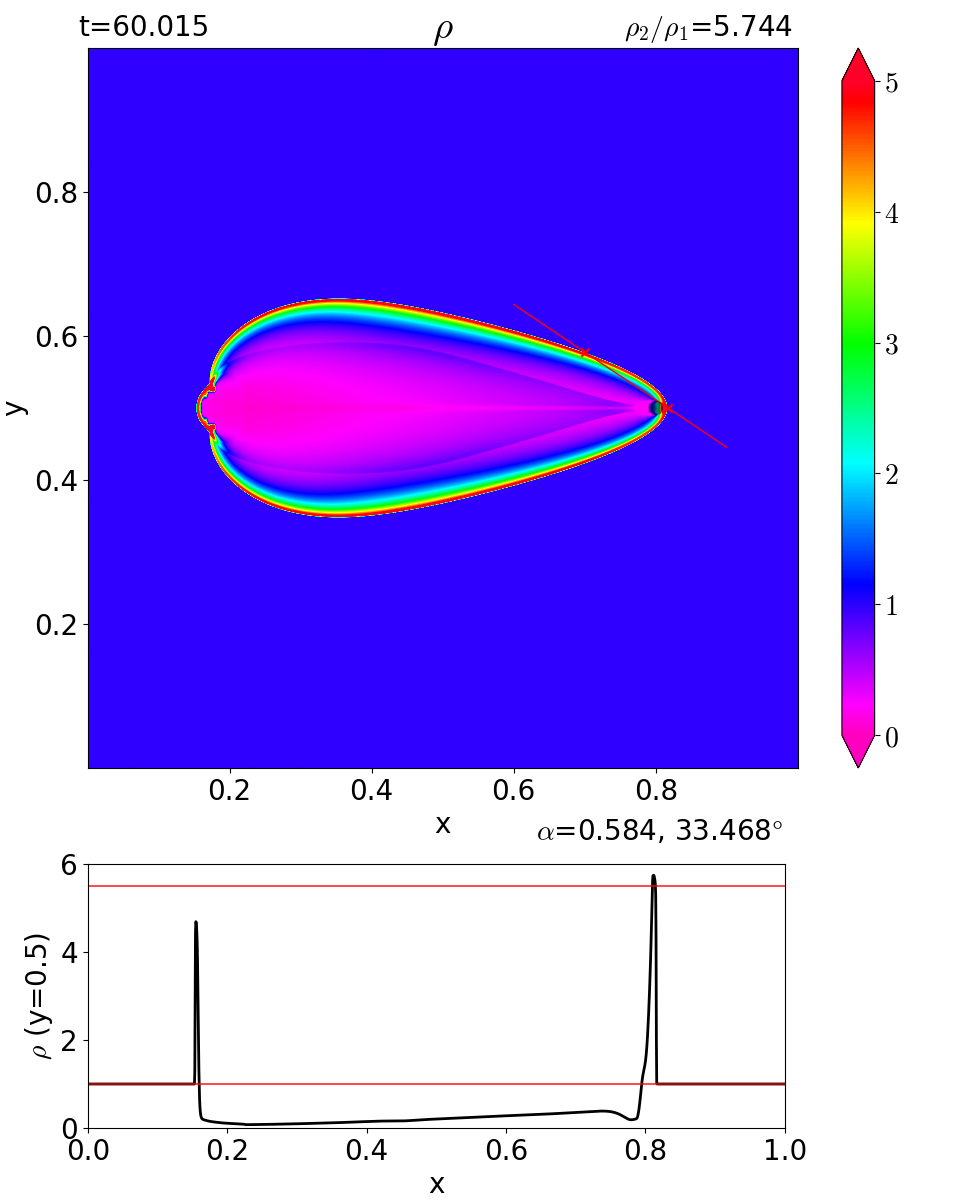}
    
    \includegraphics[width=0.32\columnwidth]{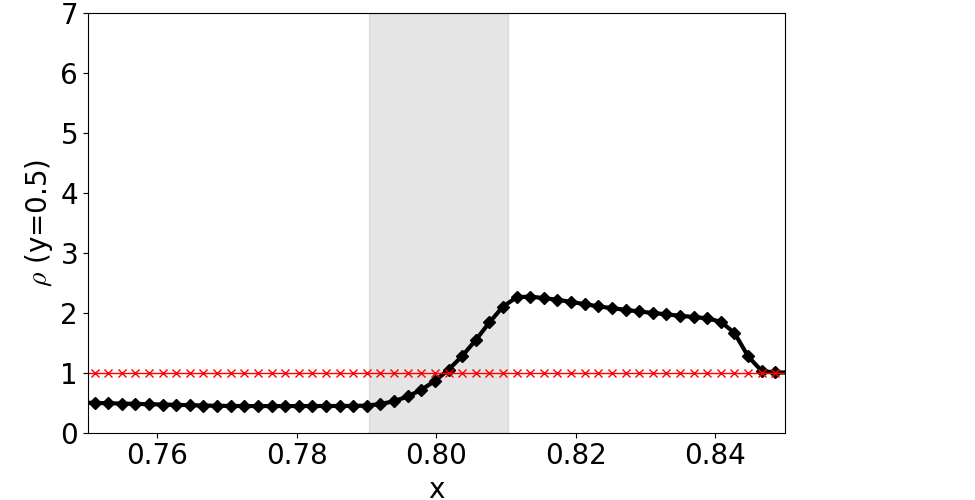}
    \includegraphics[width=0.32\columnwidth]{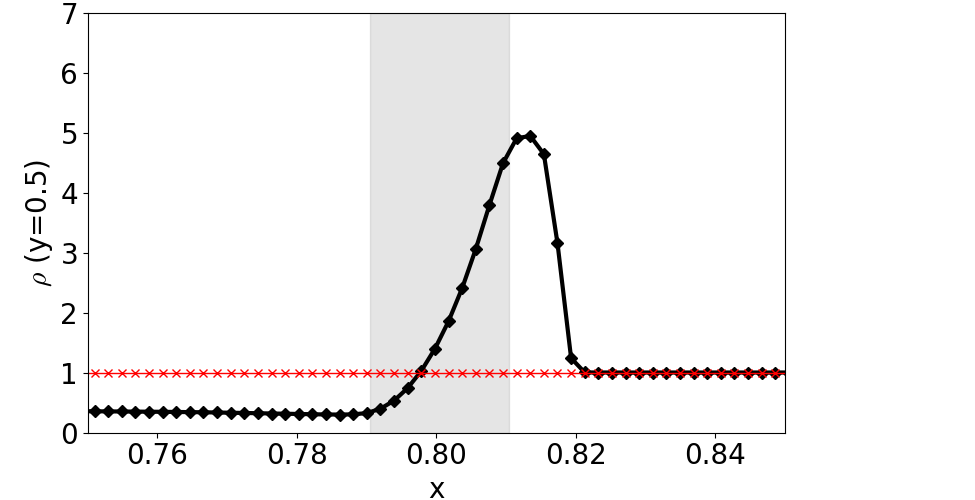}
    \includegraphics[width=0.32\columnwidth]{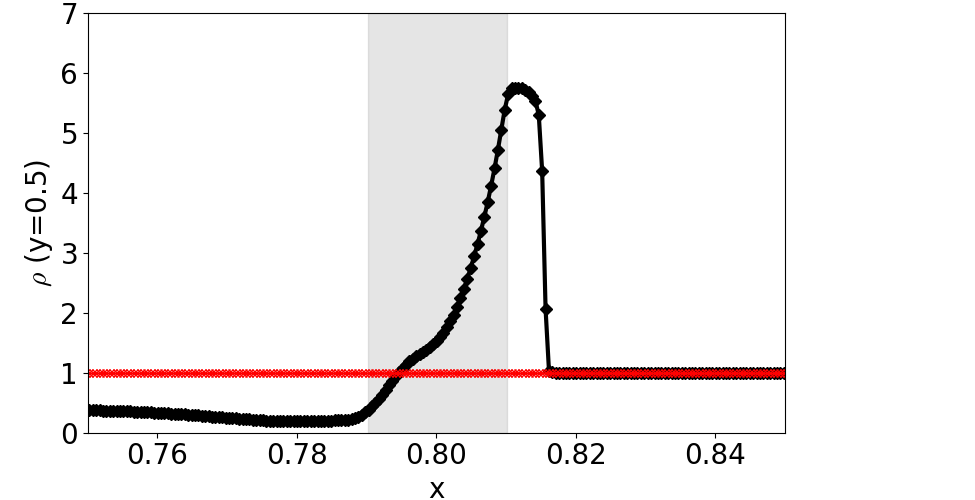}
    \caption{Simulations of a star moving in the constant-density gas with $\mathcal{M}=1.5$ (left), $\mathcal{M}=10$ (middle) and $\mathcal{M}=100$ (right) at $t=60$. The first two runs have the resolution 512 x 512, while the last on the right has the resolution 2048~x~2048. The radius of the star $\mathcal{R}=0.01$ and its velocity is $\vec{v}_s=(0.01,0)$. First row: map of density with the opening half-angle of the shock, second row: the $y=0.5$ density profile, third row: the zoom of the density profile with the individual cells shown. The perturbed region is indicated by the grey-shaded rectangle.}
    \label{fig_Mach_10}
\end{figure}

In Fig.~\ref{fig_Mach_10}, we show the state of the simulation at $t=60$, at which the body is placed at $x=0.8,y=0.5$, for $\mathcal{M}=1.5$ (left), $\mathcal{M}=10$ (middle) and $\mathcal{M}=100$ (right). Each plot shows the map of the density using a linear scale and the profile of density along $y=0.5$. In the density profile, the initial density ($\rho=1$) is shown (bottom red solid line) and the theoretical value of $\rho_2$ given by relation~\ref{eg:rho2} (top red solid line).
While the first two panels are computed with the fiducial resolution (FR) 512 x 512, the last panel shows the simulation with a ultra-high resolution (UHR) of 2048 x 2048.

\begin{figure}
    \centering
    \includegraphics[width=0.32\columnwidth]{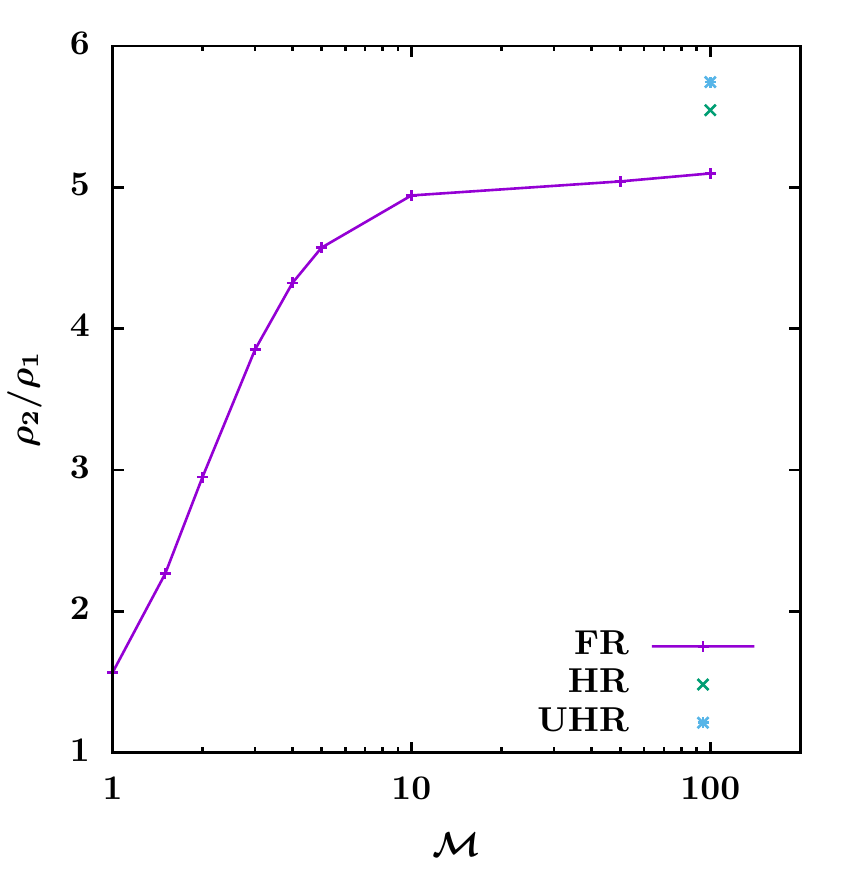}
    \includegraphics[width=0.32\columnwidth]{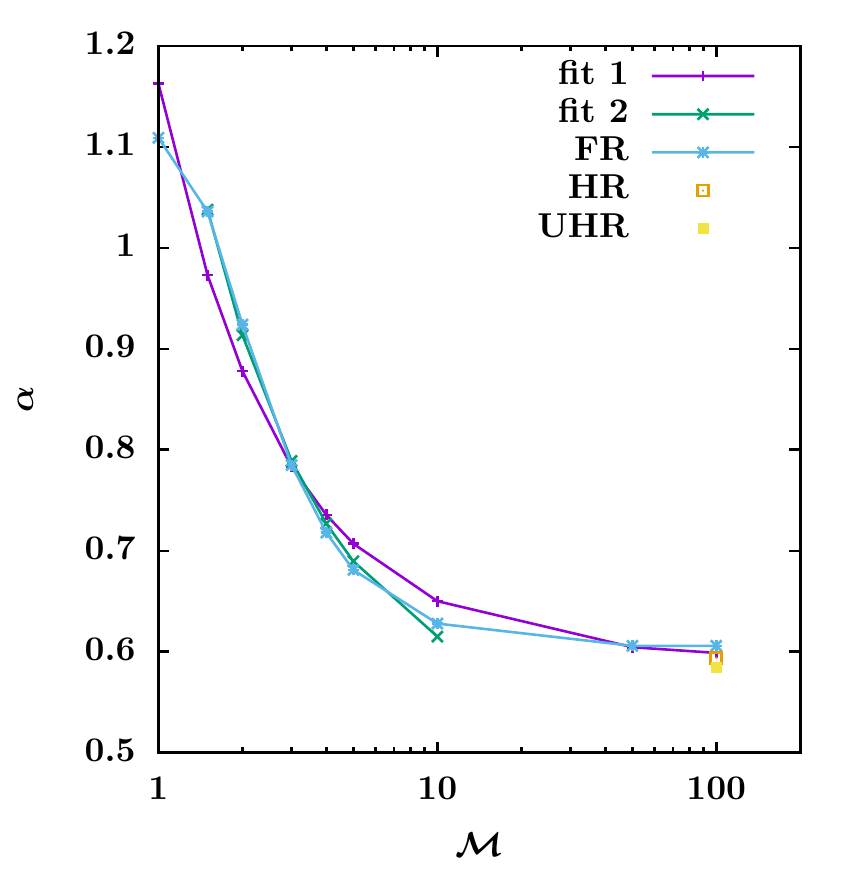}
    \includegraphics[width=0.32\columnwidth]{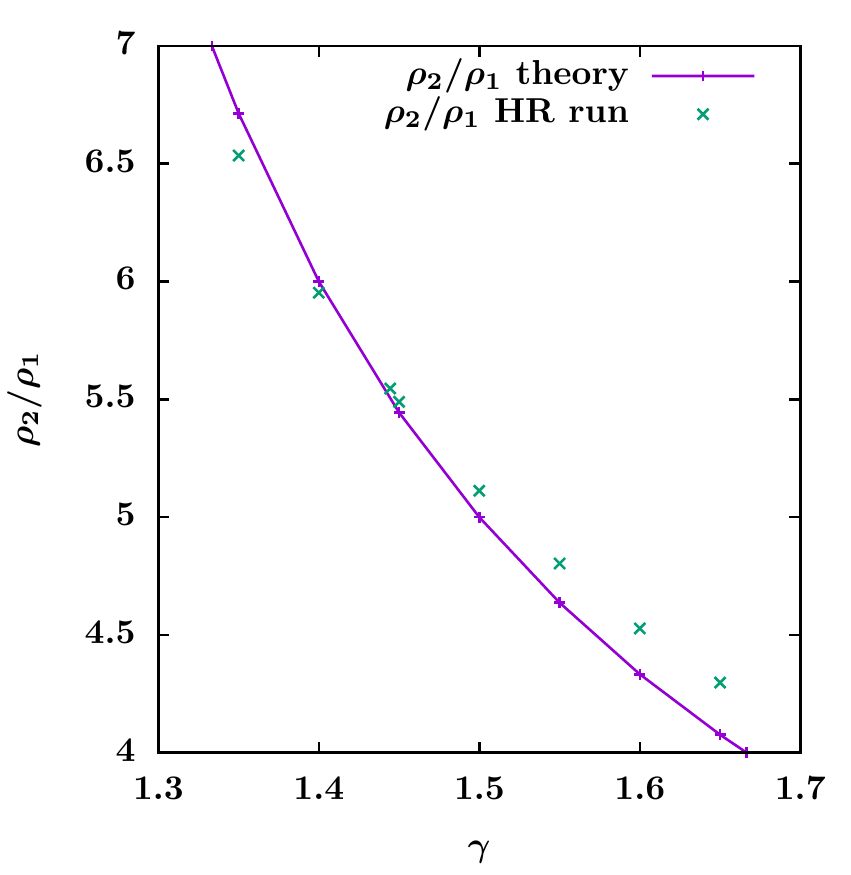}
    \caption{Dependence of the post-shock density $\rho_2/\rho_1$ (left) and the opening half-angle $\alpha$ (middle) on the Mach number $\mathcal{M}$. On the right, the dependence is shown of the post-shock density $\rho_2/\rho_1$ on adiabatic index $\gamma$ for $\mathcal{M}=100$ and HR resolution. The quantities are plotted at $t=60$ for each simulation.}
    \label{fig_Mach_zavislosti}
\end{figure}

We can notice that with the increasing Mach number, the post-shock density $\rho_2$ (which is found as the maximum value of $\rho$ between the edge of the star and the shock front) approaches the theoretical value. However, as can be seen in the first panel of Fig.~\ref{fig_Mach_zavislosti}, this increase stops around $\mathcal{M}=10$. This is caused by the fact that for higher Mach numbers, the shock front is narrower, hence it is covered by a lower number of cells, see the third row of Fig.~\ref{fig_Mach_10}. When we increase the resolution, the shock front is better captured by the grid and the $\rho_2/\rho_1$ ratio approaches and even exceeds a little bit the theoretical value for UHR (left panel of Fig.~\ref{fig_Mach_zavislosti}).

Similarly, we can also observe the decrease of the opening angle of the shock, which should depend on the Mach number as $\sim \mathcal{M}^{-1}$ \citep{1992pavi.book.....S,2014MNRAS.441..809W}.
In the second panel of Fig.~\ref{fig_Mach_zavislosti} we plot the dependence of the opening half-angle $\alpha$ on $\mathcal{M}$. 
The opening half-angle was found as the angle between the line $y=0.5$ and the line connecting the front point of the shock front, i.e. the first point along the $y=0.5$ line from the right where the density increase above $\rho_1$, and, similarly, the first point at the vertical line $(x_{\rm star}-10*\mathcal{R},y)$ (i.e. ten times the star radius behind the star) from above, where the density increase above $\rho_1$.
The construction of this angle is indicated in the three panels in the first row of Fig.~\ref{fig_Mach_10} with a red solid line.

We have fitted the obtained values by a function
\begin{equation}
    \alpha = \frac{a}{\mathcal{M}} + b
\end{equation}
and the resulting parameters of fit 1 were $a=0.5698, b=0.5931$. 
The fit is plotted in the middle panel of Fig.~\ref{fig_Mach_zavislosti} by purple points, and even though it does represent the trend, it does not capture the values very well. 
However, if we omit the points $\mathcal{M}=1, \mathcal{M}=10, \mathcal{M}=100$, the resulting fit 2 with $a=0.7462, b=0.5404$ can reproduce the data very well (green points), showing that in the regime of moderately super-sonic motion, where our results are not affected by the resolution and the finite size of the perturbing body, the resulting shock bubble behaves as expected by the theory.

\begin{figure}
    \centering
    \includegraphics[width=0.32\columnwidth]{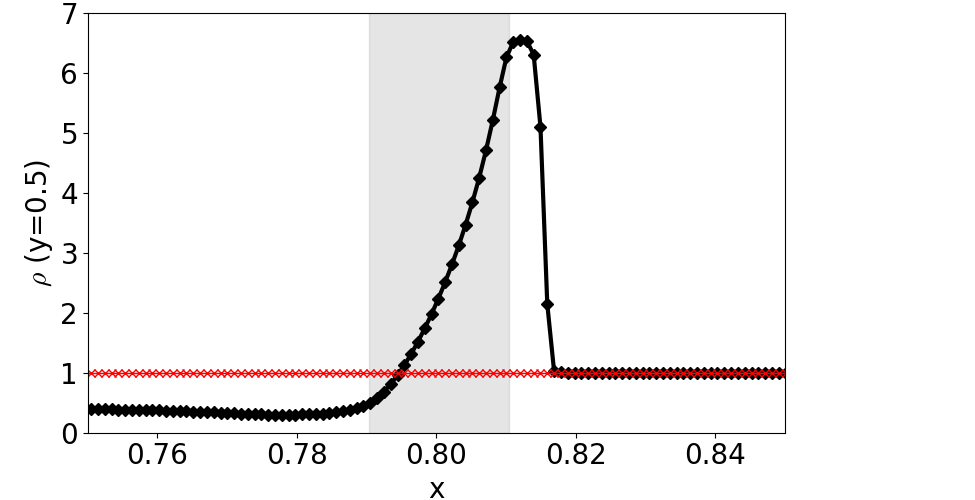}
    \includegraphics[width=0.32\columnwidth]{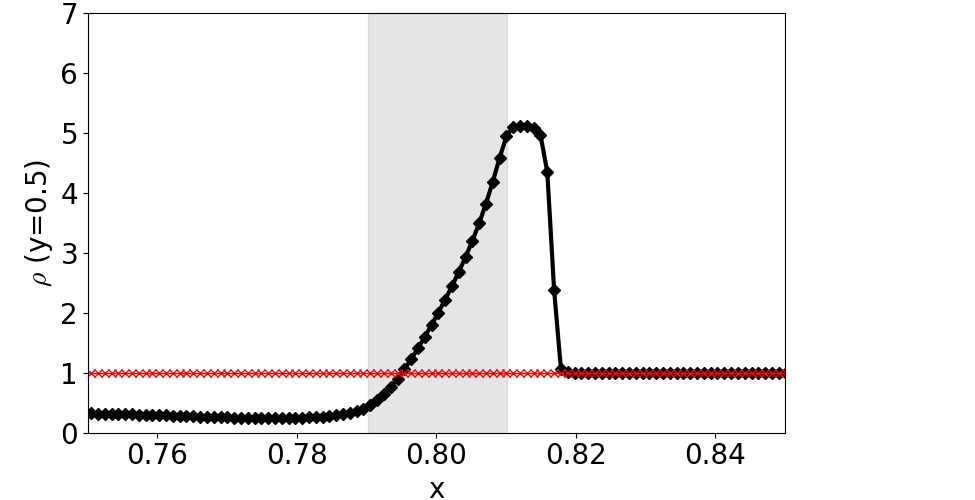}
    \includegraphics[width=0.32\columnwidth]{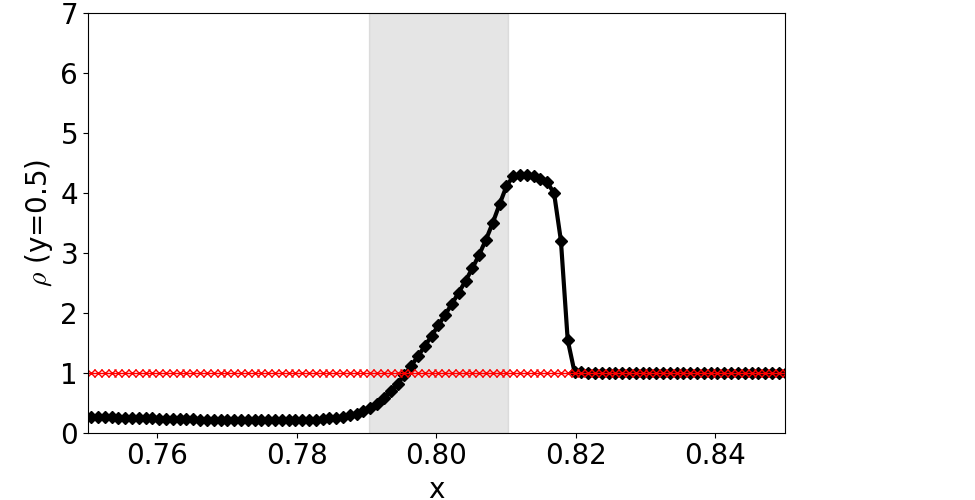}
    \caption{Density profile along $y=0.5$ for $\mathcal{M}=100$ and $\gamma=1.35$ (left), $\gamma=1.5$ (middle) and $\gamma=1.65$ (right). The perturbed region is indicated by the grey-shaded rectangle.}
    \label{fig_gamma_density}
\end{figure}

In the last panel of Fig.~\ref{fig_Mach_zavislosti}, we show the value of $\rho_2/\rho_1$ at $t=60$ for $\mathcal{M}=100$ for different values of $\gamma$. The theoretical values given by equation~\ref{eg:rho2} are plotted by a purple color, while the numerical values computed with the high resolution (HR) of 1024 x 1024 are shown as green points. The overall agreement of the values is quite good, however, we can see that the numerical values follow a little bit different trend, being smaller for lower values of $\gamma$ and higher for higher values of $\gamma$. The reason of this trend is explained in Fig.~\ref{fig_gamma_density}, where we can see that the width of the shock also depends on $\gamma$, therefore, it is better resolved for higher $\gamma$, while the resolution is quite poor for low $\gamma$. As we have seen earlier, for poorly resolved shock fronts, the post-shock density turns out to be rather smaller.

In the density maps in Fig.~\ref{fig_Mach_10} (top panel), we do not detect the development of large-scale instabilities that can be frequently seen in other hydrodynamic simulations of stellar bow shocks \citep[see e.g.][]{2012A&A...548A.113D}. In principle, stellar bow shocks are prone to the following instabilities \citep{2016MNRAS.455.1257Z}:
\begin{itemize}
    \item[(a)] Rayleigh-Taylor instability due to the centrifugal acceleration of the flow in the shock layer,
    \item[(b)] Kelvin-Helmholtz instability due to the velocity shear between the ambient medium and the denser shocked region,
    \item[(c)] radiatively cooled thin shock layers, where the cooling timescale is much shorter than the dynamical timescale, are susceptible to the non-linear thin-shell instability \citep{1994ApJ...428..186V} as well as the transverse acceleration instability \citep{1993A&A...267..155D}.
\end{itemize}

\citet{1996ApJ...461..927D} presented a linear stability analysis of stellar bow shocks. They showed that the ratio $\alpha=v_{\star}/v_{\rm w}$ between the relative stellar velocity and the stellar wind velocity determines whether the bow shock is stable ($\alpha \ll 1$) or rather unstable ($\alpha \gg 1$) with respect to short-wavelength perturbations. In our case, since we are in the limit $v_{\rm w}=0$, the simulated bow shocks are expected to be unstable.

However, there are several factors that can stabilize the bow shock and prevent the formation of hydrodynamical instabilities:
\begin{enumerate}
    \item slow cooling/warm medium \citep{2012A&A...548A.113D},
    \item ionizing radiation from an external source \citep{2014MNRAS.437..843G}, which is linked to the first point,
    \item interstellar magnetic field \citep{2014A&A...561A.152V}.
\end{enumerate}
Since we neglect a cooling term in our simulations, the first factor (no cooling) can explain the high stability of our bow shocks. A large thermal pressure can effectively compress Rayleigh-Taylor ``fingers'' and suppress their growth. In addition, the lack of cooling also diminishes the medium density contrast between the unshocked and the shocked regions due to the lack of compression via cooling, which makes the shocked layer stable. Since we include neither magnetic field nor radiation in our test runs presented in the Appendix, these do not play a role in the instability suppression. 

\end{document}